\documentclass{elsart}

 \usepackage{epsfig}

\usepackage{amssymb}
\usepackage[applemac]{inputenc}

\newcommand{\om}{\omega}

\newcommand{\epu}{\epsilon_1}
\newcommand{\epd}{\epsilon_2}
\newcommand{\muu}{\mu_1}
\newcommand{\mud}{\mu_2}

\newcommand{\E}{{\bf E}}

\newcommand{\kv}{{\bf k}}
\newcommand{\K}{{\bf K}}
\newcommand{\rv}{{\bf r}}
\newcommand{\R}{{\bf R}}

\newcommand{\Hv}{{\bf H}}
\newcommand{\G}{{\stackrel{\leftrightarrow}{\bf G}}}
\newcommand{\g}{{\stackrel{\leftrightarrow}{\bf g}}}
\newcommand{\Id}{{\stackrel{\leftrightarrow}{\bf I}}}
\newcommand{\Sv}{{\bf S}}
\newcommand{\jv}{{\bf j}}
\newcommand{\pv}{{\bf p}}
\newcommand{\nv}{{\bf n}}
\newcommand{\qv}{{\bf q}}

\newcommand{\uv}{{\bf u}}
\newcommand{\xu}{\hat{{\bf x}}}
\newcommand{\yu}{\hat{{\bf y}}}
\newcommand{\zu}{\hat{{\bf z}}}
\newcommand{\gu}{\gamma_1}
\newcommand{\gd}{\gamma_2}
\newcommand{\gt}{\gamma_3}

\newcommand{\shat}{\hat{s}}

\newcommand{\pup}{\hat{p}_1^+}
\newcommand{\pum}{\hat{p}_1^-}
\newcommand{\pdp}{\hat{p}_2^+}

\newcommand{\green}{{\stackrel{\leftrightarrow}{\bf G}}}

\begin{document}

\begin{frontmatter}

\title{Surface Electromagnetic Waves Thermally Excited: Radiative Heat Transfer,  Coherence Properties and Casimir Forces Revisited in the Near Field}
 \author[LET]{Karl Joulain,}
 \author[EM2C]{Jean-Philippe Mulet,}
  \author[EM2C]{François Marquier,}
 \author[EM2C]{R\'emi Carminati,}
 \author[EM2C]{Jean-Jacques Greffet\corauthref{cor1}} 
  \ead{greffet@em2c.ecp.fr}
 \corauth[cor1]{Corresponding author.}

\address[LET]{Laboratoire d'\'Etudes Thermiques, ENSMA, B.P 109, 86960 Futuroscope Cedex, France}
\address[EM2C]{Laboratoire d'Energ\'etique Mol\'eculaire, Macroscopique; Combustion, Ecole Centrale Paris, Grande Voie des Vignes, 92295 Ch\^atenay-Malabry Cedex, France} 





\begin{abstract}
We review in this article the influence of  surface waves on the thermally excited electromagnetic field. We study in particular the field emitted at subwalength distances of material surfaces. After reviewing the main properties of surface waves, we introduce the fluctuation-dissipation theorem that allows to model the fluctuating electromagnetic fields.  We then analyse the contribution of these waves in a variety of phenomena. They give a leading contribution to the density of electromagnetic states, they produce both temporal coherence and spatial coherence in the near field of planar thermal sources. They can be used to modify radiative properties of surfaces and to design partially spatially coherent sources. Finally, we discuss the role of surface waves in the radiative heat transfer and the theory of dispersion forces at the subwavelength scale.
\end{abstract}

\end{frontmatter}

\section{Introduction}

Many condensed matter properties are determined by surface properties. Very often, surface waves, which are electromagnetic eigenmodes of the surface, play a key role. Let us mention a few examples. It has been demonstrated \cite{Chance} that the lifetime of a molecule varies dramatically when a metallic surface is brought at a distance smaller than a micron. This effect is due to the resonant excitation of surface plasmons. The van der Waals force between a molecule and an interface is proportional to $1/\vert\epsilon+1 \vert^2$ where $\epsilon$ is the dielectric constant of the medium. There is therefore a resonance for the particular frequency such that $\epsilon=-1$. This condition coincides with a branch of the dispersion relation of a surface wave. It can be viewed as a resonant excitation of surface charge oscillations. It was shown in \cite{Bloch} that the van der Waals force between a molecule and a surface can become repulsive depending on the relative position of the molecule and the surface resonances.  Enhanced scattering due to the resonant excitation of surface charges has also been demonstrated for SiC in the infrared :
a tip brought close to a surface generates a very strong scattering signal for some particular frequencies corresponding to the excitation of surface waves \cite{Keilmannature}. Both experiments can be understood by replacing the interface by an image whose amplitude is  very large owing to the excitation of a resonance of the surface charges. Surface Enhanced Raman Scattering (SERS) is partially due to the enhancement of the electromagnetic field at the interface due to the excitation of a surface wave. The resonance of the electromagnetic (EM) field at an interface is also responsible for the enhanced transmission of a metallic film with a periodic array of holes \cite{Ebbesen,PendryOC}. The resonance of the EM field associated with the surface mode is responsible for the so-called "perfect lens" effect \cite{PendryPRL}.  A key feature of all the above examples is that they involve the interaction of a surface and an object in the near field of the structure.  As it will be explained in details in Section 2, surface waves are evanescent waves whose amplitude decreases away from the interface on a wavelength scale. In the far field, the influence of such modes is therefore negligible. In the near field on the contrary, their role is essential.

We will see in Section 3 that surface waves can be excited by thermal fluctuations inside a body. The role of surface waves in the modification of the density of EM states at the interface has a strong influence on the thermally emitted fields. Their intensity is many orders of magnitude larger in the near field than in the far field \cite{Shchegrov}. In addition, they are quasi monochromatic in the vicinity of the surface. This entails that their coherence properties are extremely different from those of the blackbody radiation \cite{Carminati}. There have been recently several experiments that have probed these thermal fields in the near-field regime : heating of trapped atoms \cite{Henkel_euro}, realization of a spatially partially coherent thermal source \cite{Greffet_nat}.  After reviewing these experiments, we will show how an EM approach with random fluctuating thermal sources can be used to describe and analyse these effects. It is based on the fluctuation-dissipation theorem. We will see that the knowledge of the electromagnetic energy density gives acces to a fundamental concept : the local density of EM states. In section 4, we study the EM coherence properties near a material supporting surface waves and held at a temperature $T$. We will see that the emitted field has very peculiar spatial coherent properties in the near field. Indeed, the field can be spatially coherent over a length larger than several tens of wavelength. We then use this property to design coherent thermal sources. In section 5 and 6 we show that the radiative heat transfer is enhanced by several orders in magnitude in the near field when two material supporting surface waves are put face to face. We will consider three cases : two nanoparticles face to face, a nanoparticle near a plane interface and two semi-infinite half-spaces separated by a narrow gap. In the last section, we will analyse the role played by the surface waves in the Casimir force i.e in the force of interaction between two semi-infinite bodies. We will see that this force is dominated in the near field by the interaction between surface waves. Finally, we review the work done to analyse the contribution of fluctuating electromagnetic fields to the friction forces.

\section{Introduction to surface electromagnetic waves}

In this section, we give a brief introduction to  the main properties of electromagnetic surface waves. This particular
type of waves exists at the interface between two different media. An electromagnetic surface wave propagates along the interface
and decreases exponentially in the perpendicular direction. Surface waves due
to a coupling between the electromagnetic field and a resonant polarization oscillation in the material are called surface polaritons. From a microscopic point of view, the surface waves at the interface of a metal is a charge density wave or plasmon. It is therefore called surface-plasmon polariton. At the interface of a dielectric, the surface wave is due to the coupling of an optical phonon with the electromagnetic field.  It is thus called surface-phonon polariton. Plasmon polaritons and phonon polaritons can also exist in the whole volume of the material and are called polaritons. More details about this subject can be found in textbooks such as Kittel \cite{Kittel}, Ashcroft and Mermin \cite{Ashcroft} and Ziman \cite{Ziman}. In what follows, we will focus our attention on surface polaritons propagating along a plane interface. Excellent reviews of the subject can be found in \cite{Raether,Agranovitch,Economou1,Boardman}.

\subsection{Surface polaritons}
\begin{figure}[h]
\begin{center}
\includegraphics[width=10cm]{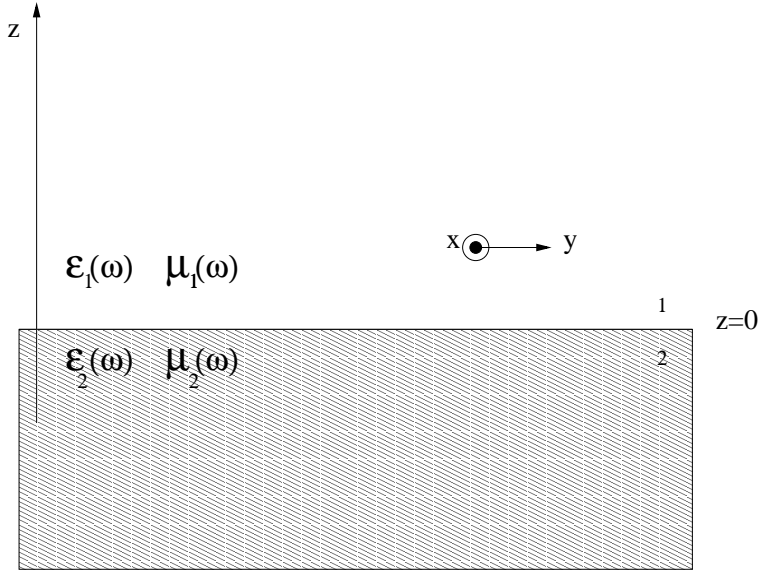}
\caption{A plane interface separating medium 1 (dielectric constant $\epu$, magnetic constant $\mu_1$) and medium 2 (dielectric constant $\epd$, magnetic constant $\mu_2$) }
\label{Fig:1}
\end{center}
\end{figure}

Let us now study the existence and the behaviour of surface polaritons in the case of a plane
interface separating two linear, homogeneous and isotropic media  with different dielectric constants. The system considered is depicted in Fig.\ref{Fig:1}.

The medium 1 (dielectric constant $\epu$ and magnetic constant $\mu_1$) fills the upper half-space $z>0$ whereas medium 2
(dielectric constant $\epd$ and magnetic constant  $\mu_2$) fills the lower half-space $z<0$. The two media are supposed to be local and dispersive so that their complex dielectric and magnetic constants only depend on $\om$. 

The three directions $x,y,z$ shown in Fig.\ref{Fig:1} are characterized by their unit vectors $\xu,\yu,\zu$. A point in space will be denoted $\rv=(x,y,z)=x\xu+y\yu+z\zu=(\R,z)$ where $\R=x\xu+y\yu$. Similarly,  a wave vector $\kv=(k_x,k_y,k_z)$ will be denoted by
$\kv=(\K,\gamma)$ where $\K$ is the component parallel to the interface and
$\gamma=k_z$ the component in the $z$ direction.

A surface wave is a particular solution of Maxwell's equations which propagates along the interface and decreases exponentially in the perpendicular directions. Because of the translational invariance of the system, it can be cast in the form:
\begin{equation}
\label{elec1}
\E_1(\rv,\om)=\left(\begin{array}{c }
      E_{x,1} \\
      E_{y,1} \\
      E_{z,1} \end{array}
     \right)\exp[i(\K.\R+\gamma_1z)]
{\rm \ (Medium 1)},
\end{equation}
\begin{equation}
\label{electric}
\E_2(\rv,\om)=\left(\begin{array}{c }
      E_{x,2} \\
      E_{y,2} \\
      E_{z,2} \end{array}
     \right)\exp[i(\K.\R-\gamma_2z)]
{\rm \ (Medium 2)},
\end{equation}
where $\gamma_1$ and $\gamma_2$ are given by,
\begin{eqnarray}
\gamma_1^2 & = & \epu\muu k_0^2-K^2 \quad {\rm \ with \ } Im(\gamma_1)>0,\label{gamma1} \\
\gamma_2^2 & = & \epd\mud k_0^2-K^2 \quad {\rm \ with \ } Im(\gamma_2)>0 \label{gamma2}.
\end{eqnarray}
Here $k_0=\om/c$ where $c$ is the speed of light in vacuum. We now look for the existence of surface waves in $s$ (TE) or $p$ (TM) polarization. In what follows, we shall assume that the wave propagates along the $y$-axis. 

\subsubsection{$s$-polarization (TE)}
In $s$-polarisation, the electric field is perpendicular to the plane $(y,z)$. The electric field $\E$ is thus parallel to the $x$ direction
\begin{equation}
\label{ }
\E_1(\rv,\om)=E_{x,1} \xu \exp[i(\K.\R+\gamma_1z)],
\end{equation}
\begin{equation}
\label{ }
\E_2(\rv,\om)=E_{x,2} \xu \exp[i(\K.\R-\gamma_2z)].
\end{equation}
The magnetic field is then derived from the Maxwell equation $\Hv=-i\nabla\times\E/(\mu(\om)\om)$. The continuity conditions of the parallel components of the fields across the interface yield the following equations :
\begin{eqnarray}
E_{x,1}-E_{x,2} & = & 0, \\
\frac{\gamma_1}{\muu}E_{x,1}+\frac{\gamma_2}{\mud}E_{x,2} & = & 0.
\end{eqnarray}
We search a mode of the system which is a solution of the homogeneous problem. The system has a non-trivial solution if and only if
\begin{equation}
\label{dets}
\mud\gamma_1+\muu\gamma_2=0.
\end{equation}
Taking into account equations (\ref{gamma1},\ref{gamma2}), one obtains from (\ref{dets}) the surface wave dispersion relation for s-polarization:
\begin{equation}
\label{disps}
K^2=\frac{\om^2}{c^2}\frac{\muu\mud[\mu_2\epsilon_1-\mu_1\epsilon_2]}{\mud^2(\om)-\muu^2(\om)}.
\end{equation}  

For the particular case where $\epsilon_1=\epsilon_2=\epsilon$, the dispersion relation takes the simple form :
\begin{equation}
\label{disps}
K^2=\frac{\om^2}{c^2}\epsilon\frac{\muu(\om)\mud(\om)}{\muu(\om)+\mud(\om)}.
\end{equation}  

\subsubsection{$p$-polarization (TM)}
For $p$-polarization, the electric field lies in the plane $(y,z)$ and can be cast in the form:
\begin{equation}
\label{ }
\E_1(\rv,\om)=\left(\begin{array}{c }
      0 \\
      E_{y,1} \\
      E_{z,1} \end{array}
     \right)\exp[i(\K.\R+\gamma_1z)],
\end{equation}
\begin{equation}
\label{ }
\E_2(\rv,\om)=\left(\begin{array}{c }
      0 \\
      E_{y,2} \\
      E_{z,2} \end{array}
     \right)\exp[i(\K.\R-\gamma_2z)].
\end{equation}
The continuity of the tangential electric field yields 
\begin{equation}
\label{Etg}
E_{y,1}-E_{y,2}=0.
\end{equation}

The Maxwell equation $\nabla.\E=0$ imposes a relation between the two components of the electric field
\begin{equation}
\label{divE}
KE_{y,2}-\gamma_2E_{z,2}=KE_{y,1}+\gamma_1E_{z,1}=0.
\end{equation}
The continuity of the $z$-component of $D$ yields:
\begin{equation}
\label{contEz}
\epu E_{z,1}=\epd E_{z,2}.
\end{equation}
Inserting  (\ref{contEz})  and (\ref{Etg}) in (\ref{divE})
yields
\begin{equation}
\label{detp}
\epu\gamma_2+\epd\gamma_1=0.
\end{equation}

Taking into account equations (\ref{gamma1},\ref{gamma2}), one obtains from (\ref{detp}) the surface wave dispersion relation for p-polarization:
\begin{equation}
\label{dispp}
K^2=\frac{\om^2}{c^2}\frac{\epu\epd[\epd\muu-\epu\mud]}{\epd^2-\epu^2}.
\end{equation}  

For the particular case where $\muu=\mud=\mu$, the dispersion relation takes the simple form :
\begin{equation}
\label{disps}
K^2=\frac{\om^2}{c^2}\mu\frac{\epu(\om)\epd(\om)}{\epu(\om)+\epd(\om)}.
\end{equation}

\subsubsection{Remarks}
\begin{itemize}
  \item When the media are non-magnetic, there are no surface waves in $s$-polarization. Indeed, the imaginary part of the $z$-components $\gamma_i$ is always positive, so that  $\gamma_1+\gamma_2$  cannot be zero.
  \item At a material-vacuum interface ($\epu=\muu=1$), the dispersion relation reads in $p$-polarization
\begin{equation}
\label{dispvac}
K=\frac{\om}{c}\sqrt{\frac{\epd(\om)}{\epd(\om)+1}}.
\end{equation}
It follows that the wave vector becomes very large for a frequency such that $\epsilon(\om)+1=0$.
  \item The conditions (\ref{dets}) and (\ref{detp}) corresponds to the poles of the Fresnel
reflection factors. To search these poles is an alternative and simple way to find the dispersion relation. This is particularly useful when searching the dispersion relation for multilayers system. 

  \item For non-lossy media, one can find a real $K$ corresponding to a real $\om$. This mode exists only if $\epd<-1$ in the case of an interface separating a vacuum from a material. 
  
  \item In the presence of losses, the dispersion relation yields two equations but both frequency and wavector can be complex so that there are four parameters. Two cases are of practical interest : i) a real frequency and a complex wavevector, ii) a complex frequency and a real wavector. These two choices leads to different shapes of the dispersion relation as discussed in \cite{Legall,Halevi,Arakawa,Kliever}. The imaginary part of $\om$ describes the finite lifetime of the mode due to losses. Conversely, for a given real $\om$, the imaginary part of $K$ yields a finite propagation length along the interface. 

  \item The dispersion relation (\ref{dispvac}) shows that for a real dielectric constant $\epu<-1$, $K>\om/c$.  This mode cannot be excited by a plane wave whose wavevector is such that $K<\om/c$. In order to excite this mode, it is necessary to increase the wavevector. One can use a prism \cite{Raether,Otto,Kretschmann} or a grating \cite{Legall}. A scatterer can also generate a wave with the required wavevector.

\end{itemize}

\subsection{Dispersion relation}

\begin{figure}[h]
\begin{center}
\includegraphics[width=10cm]{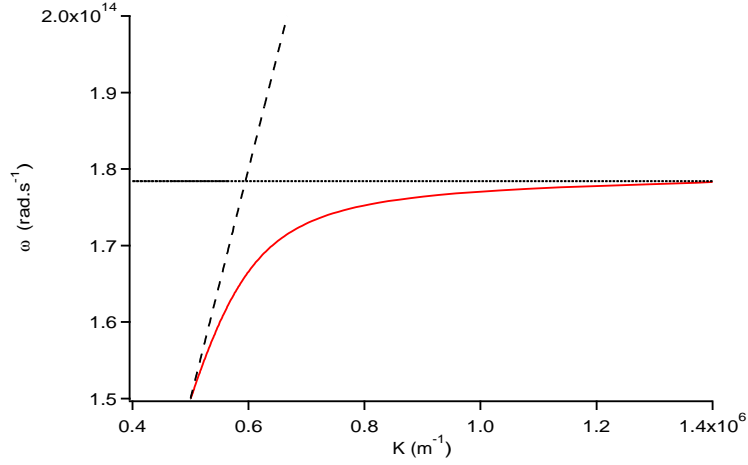}
\caption{Dispersion relation for surface phonon-polariton at a SiC/Vacuum interface. The flat asymptote is situated at $\om_{asym}=1.784$ 10$^{14}$ rad s$^{-1}$. The slanting dashed line represents the light cone above which a wave is propagating and below which a wave is evanescent.}
\label{Fig:2}
\end{center}
\end{figure}

In this subsection, we will consider two types of surface waves: surface-plasmon polaritons and surface-phonon polaritons. Surface-plasmon polaritons are observed at surfaces separating a dielectric from a medium with a gas of free electrons such as a metal or a doped semiconductor. The dielectric constant of the latter can be modelled by a Drude model : 
\begin{equation}
\label{Drude1}
\epsilon(\omega)=\epsilon_{\infty}-\frac{\omega_p^2}{\omega^2+i\Gamma\omega},
\end{equation}
where $\omega_p$ is the plasma frequency and $\Gamma$ accounts for the losses. Using this model and neglecting the losses, we find that the resonance condition $\epsilon(\om)+1=0$ yields $\omega=\omega_p/\sqrt{2}$. For most metals, this frequency lies in the near UV so that these surface waves are more difficult to excite thermally. By contrast, surface-phonon polaritons can be excited thermally because they exist in the infrared. They have been studied through measurements of emission and reflectivity spectra by Zhizhin and Vinogradov \cite{Vinogradov}. Let us study the dispersion relation of surface-phonon polaritons at a vacuum/Silicon Carbide (SiC) interface. SiC is a non-magnetic material whose dielectric constant is well described by an oscillator model in the [2-22 $\mu$m] range\cite{Spitzer}:
\begin{equation}
\label{oscilsic}
\epsilon(\om)=\epsilon_\infty\left(1+\frac{\om_L^2-\om_T^2}{\om_T^2-\om^2-i\Gamma\omega}\right),
\end{equation}
with $\om_L=969$ cm$^{-1}$, $\om_T=793$ cm$^{-1}$, $\Gamma=4.76$ cm$^{-1}$ and
$\epsilon_\infty=6.7$. The dispersion relation at a SiC/vacuum interface is represented in Fig.\ref{Fig:2}. This dispersion relation has been derived by assuming that the frequency $\om$ is complex and the parallel wavevector $K$ is real. This choice is well suited to analyse experimental measurements of spectra for fixed angles. The width of the resonance peaks observed is related to the imaginary part of the frequency of the mode. We note that the curve is situated below the light cone $\om=cK$ so that the surface wave is evanescent. We also observe a horizontal asymptote for $\om_{asym}=1.784$ 10$^{14}$ rad s$^{-1}$ so that there is a peak in the density of electromagnetic states.  We will see in the next sections, that the existence of surface modes at a particular frequency plays a key role in many phenomena.

\begin{figure}[h]
\begin{center}
\includegraphics[width=10cm]{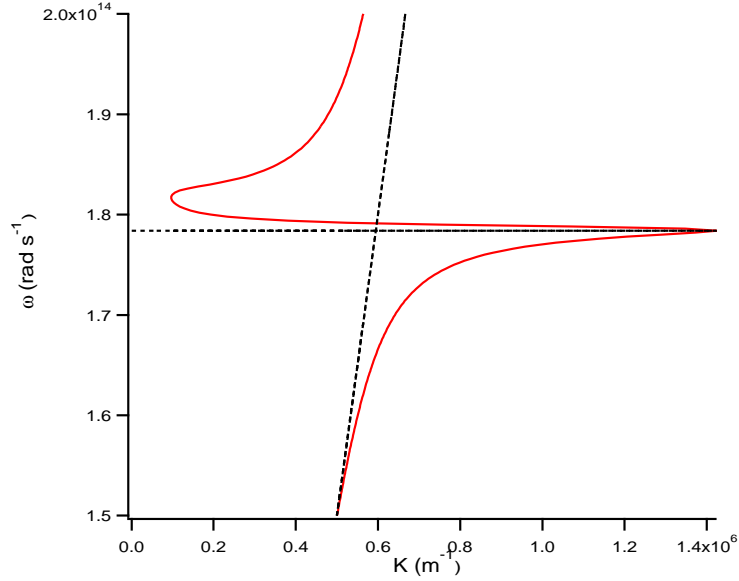}
\caption{Dispersion relation for surface phonon-polariton at a SiC/Vacuum interface. Real $\om$ chosen to obtain a complex $K$. The real part of $K$ is represented. The horizontal asymptote is situated at $\om_{asym}=1.784$ 10$^{14}$ rad s$^{-1}$. The slanting dashed line represents the light line above which a wave is propagating and below which a wave is evanescent.}
\label{Fig.2bis}
\end{center}
\end{figure}

In Fig.\ref{Fig.2bis}, we have shown the dispersion relation obtained when choosing a real frequency $\om$ and a complex wavevector $K$. The real part of the complex wavevector is represented. It is seen that the shape of the dispersion relation is significantly changed. A backbending of the curve is observed. This type of behaviour is observed experimentally when performing measurements at a fixed frequency and varying the angle. Observed resonances in reflection or emission experiments have an angular width which is related to the imaginary part of the  complex wavevector. 

\begin{figure}[h]
\begin{center}
\includegraphics[width=10cm]{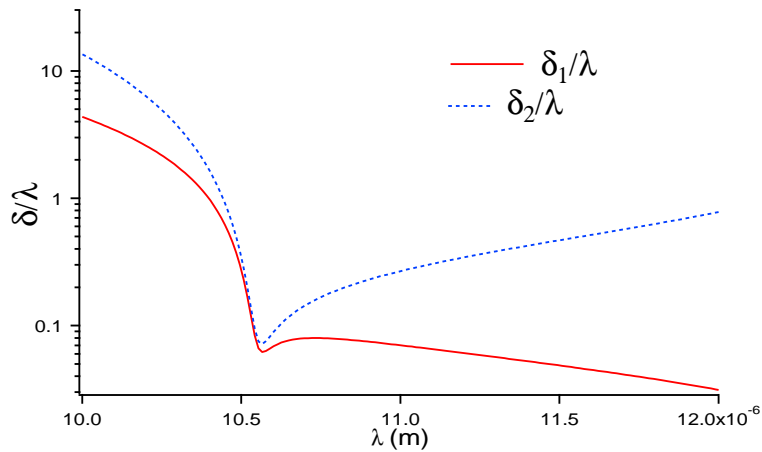}
\caption{Surface wave decay length along the $z$-direction in medium 1 and 2 versus the wavelength for a SiC-vacuum interface.}
\label{Fig:3}
\end{center}
\end{figure}

We have plotted in Fig.\ref{Fig:3} the surface wave decay length in the 
direction perpendicular to the interface versus the wavelength. From Eqs (\ref{elec1}) and (\ref{electric}), it is seen that the amplitude of the electromagnetic field decreases exponentially in the $z$ direction with a decay length $\delta_1=1/Im(\gamma_1)$ in medium 1 and $\delta_2=1/Im(\gamma_2)$ in medium 2.
We note that the smallest penetration depth in SiC is obtained for the frequency $\om_{asym}$. At this frequency, losses are very large. 

\begin{figure}[h]
\begin{center}
\includegraphics[width=10cm]{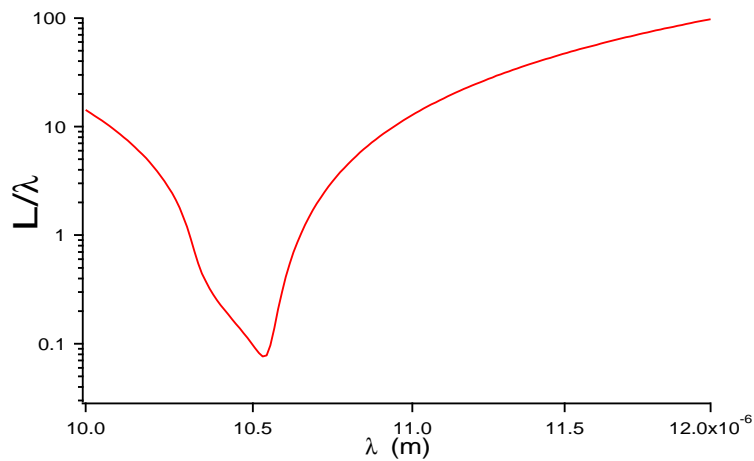}
\caption{Surface wave propagation length along the interface versus the wavelength $\lambda$.}
\label{Fig:4}
\end{center}
\end{figure}

We study in Fig.\ref{Fig:4} the surface wave propagation length along a SiC-vacuum interface. It is given by the inverse of the imaginary part of the parallel wavevector $L=1/Im(K)$. Around $\om_{asym}$, $L$ is minimum. It can  be as large as several tens of wavelengths.

It will be seen below that the existence of these surface modes is responsible for a long coherence time and a long coherence length of the electromagnetic field in the near field. They are essentially given by the lifetime and the decay length of the surface wave.

\section{Fluctuation-dissipation theorem. Cross-spectral density.}

In this section, we introduce the tools and methods that are useful to derive the field radiated by a body in thermal equilibrium at temperature $T$ both in the near field and in the far field. Whereas the phenomelogical theory of radiometry based on geometrical optics describes correctly the field emitted in the far field, it fails to predict the behaviour of the emitted radiation in the near field. Indeed, geometrical optics does not include evanescent waves. A new framework to describe thermal radiation is thus needed. Such a framework has been introduced by Rytov \cite{Rytov1,Rytov2} and is known as fluctuational electrodynamics. The key idea is that for any material in thermal equilibrium,  charges such as electrons in metals, or ions in polar crystals undergo a random thermal motion. This generates  fluctuating currents which radiate an electromagnetic field. A body at temperature $T$ is thus viewed as a medium with random currents that radiate the thermal field. The statistical properties of this field can be determined provided that i) the statistical properties of the random currents are known, ii) the radiation of a volume element below an interface is known. The first information is given by the Fluctuation-Dissipation Theorem (FDT), the second is given by the Green's tensor of the system. 

This approach is very similar to the Langevin model for Brownian motion. Langevin \cite{Langevin08} introduced a random force as a source for the dynamical equations of the particles. This allows to derive the statistical properties of their random motion. An important feature of the model is that the random force is not arbitrary. Its correlation function is related to the losses of the system by the FDT. In the case of random electromagnetic fields, the dynamic equations for the fields are Maxwell equations. We need to introduce an external random source to model the fluctuations of the field. These external sources are random currents. The key issue now is to know the statistical properties of these random sources. They are given by the FDT. The remaining of this section introduces the technical tools needed to derive the radiated field. The first part is devoted to the statistical properties of the random currents given by the FDT, the second part deals with the FDT applied to the EM fields. The Green's tensor are given in appendix A.

\subsection{Cross-spectral density}

The spectral analysis of a signal is usually done using its Fourier transform. In the case of a stationary stochastic signal, the Fourier transform cannot be computed in the sense of a function because the integral is not square integrable. However, it is possible to compute the Fourier transform of the time-correlation function of the random signal. In what follows, we will be interested in the space-time correlation function of the electromagnetic fields $\left<E_k(\rv,t)E_l(\rv',t')\right>$. For a stationary field, this correlation function depends only on $t-t'$. Its Fourier transform $\mathcal{E}_{kl}(\rv,\rv',\omega)$ is called cross-spectral density:
\begin{equation}
\mathcal{E}_{kl}(\rv,\rv',\omega)=\int_{-\infty}^{\infty}\left<E_k(\rv,t)E_l(\rv',t')\right> e^{i\omega (t-t')}d(t-t').
\label{csdensity}
\end{equation}
Note that for $\rv=\rv'$, the above equation reduces to the Wiener-Khinchin theorem that relates the power spectral density of a random stationary signal to the Fourier transform of its time correlation function \cite{Mandel}. It is convenient to introduce a correlation function of the Fourier transforms using generalized functions :
\begin{equation}
\left<E_k(\rv,\omega)E_l^*(\rv',\omega')\right>=2\pi\delta(\omega-\omega')\mathcal{E}_{kl}(\rv,\rv',\omega).
\label{correlomega}
\end{equation}

\subsection{Fluctuation-dissipation theorem for the current density}

The FDT derived by Callen and Welton \cite{Callen} yields a general form of the symmetrized correlation function of a vector $\mathbf{X}(\rv,\rv',\om)$. Whereas for classical quantities, the symmetrization does not change the results, it plays an important role in quantum mechanics for non-commuting observables. If the cross-spectral density of $\mathbf{X}$ is defined by $\mathcal{X}_{kl}$, we define the symmetrised correlation function of $\mathbf{X}$  by:
\begin{equation}
\label{symcorr}
\mathcal{X}^{(S)}_{kl}=\frac{1}{2}[\mathcal{X}_{kl}+\mathcal{X}_{lk}].
\end{equation}

The symmetrised correlation function of the dipole moment of a particle in thermodynamic equilibrium with polarisability $\alpha$ defined by $p_i=\epsilon_0\alpha_{ij} E_j$ can be written as \cite{Callen,Agarwal}:
\begin{eqnarray}
\label{FDTP}
\left<p_k(\om)p_l^*(\om')\right>_S&=&2\pi\delta(\om-\om') \mathcal{P}^{(S)}_{kl}(\omega)
\nonumber \\
&=& \hbar\coth(\hbar\omega/2k_{B} T) Im[ \epsilon_0\alpha_{kl}(\om)] 2\pi\delta(\om-\om'),
\end{eqnarray}
where the brackets denote an ensemble average. In the preceding expression $\hbar$ is the reduced Planck constant and $k_{B}$ is the Boltzmann's constant. 

For a bulk in thermodynamic equilibrium at temperature $T$, the symmetrised correlation function of the polarization density can be written as \cite{Callen,Agarwal}:
\begin{eqnarray}
\label{FDTP}
\left<P_k(\rv,\om)P_l^*(\rv',\om')\right>_S&=&2\pi\delta(\om-\om') \mathcal{P}^{(S)}_{kl}(\rv,\rv',\omega)
\nonumber \\
&=& \hbar\coth(\hbar\omega/2k_{B} T) Im[ \epsilon_0\epsilon_{kl}(\rv,\rv',\om)]2\pi\delta(\om-\om'),
\end{eqnarray}
 where the spatial dependence of the dielectric constant accounts for a possible non-locality. From this equation, we can easily derive the correlation function of the current density $\jv=-i\om\mathbf{P}$:
\begin{equation}
\label{FDTj}
\left<j_k(\rv,\om)j_l^*(\rv',\om')\right>_S
= \hbar\om^2\coth(\hbar\omega/2k_{B} T) Im[ \epsilon_0\epsilon_{kl}(\rv,\rv',\om)]2\pi\delta(\om-\om').
\end{equation}

Note that if the medium is isotropic and local, the quantity $Im[\epsilon_{kl}(\rv,\rv',\om)]$ becomes $Im[\epsilon(\rv,\om)]\delta_{kl}\delta(\rv-\rv')$. We also note that:
\begin{equation}
\label{Theta}
\frac{\hbar\om}{2} \coth(\hbar\omega/2 k_{B} T) =\hbar \omega \left[\frac{1}{2}+\frac{1}{\exp(\hbar\om/k_{B} T)-1}\right]
\end{equation}
is the mean energy of a harmonic oscillator in thermal equilibrium. In the following we will also use the compact notation
\begin{equation}
\Theta(\om,T)=\frac{\hbar\om}{\exp(\hbar\om/k_{B}T)-1}
\end{equation}
for the mean energy of the harmonic oscillator without the zero point energy  $\hbar \omega/2$.

\subsection{Fluctuation-dissipation for the fields.}

Another very useful application of the fluctuation-dissipation theorem yields a relation between the cross-spectral density of the fluctuating fields at equilibrium and the Green's tensor of the system \cite{Agarwal}. The Green's tensor appears as the linear response coefficient relating the fields to their sources. Note that these quantities are defined in classical electrodynamics.  In what follows, we will use three different Green's tensors defined by:
\begin{equation}
\label{etoj}
\E(\rv,\om)=i\mu_0\om\int d^3\rv' \; \G^{EE}(\rv,\rv',\om)\jv(\rv',\om),
\end{equation} 

\begin{equation}
\label{htoj}
\Hv(\rv,\om)=\int d^3\rv' \; \G^{HE}(\rv,\rv',\om)\jv(\rv',\om),
\end{equation}
and 
\begin{equation}
\label{htom}
\Hv(\rv,\om)=\int d^3\rv' \; \G^{HH}(\rv,\rv',\om)\mathbf{M}(\rv',\om).
\end{equation}
In this last equation $G_{kl}^{HH}$ is the Green tensor relating the magnetic field to the {\it magnetization} $M$. Note that both $\G^{HE}$ and $\G^{HH}$ are related to $\G^{EE}$ through the Maxwell equations \cite{Tai}. 
\begin{eqnarray}
 \G^{HE}(\rv,\rv',\om)& = & \frac{\mu_0}{\mu(\rv)}\nabla_{\rv}\times\G^{EE}(\rv,\rv',\om), \\
\G^{HH}(\rv,\rv',\om) & = & \frac{\mu_0}{\mu(\rv)} \nabla_{\rv}\times\G^{EE}(\rv,\rv',\om)\times\overleftarrow{\nabla}_{\rv'}.
\end{eqnarray}
Here
$$
G^{EE}(\rv,\rv',\om)\times\overleftarrow{\nabla}_{\rv'}=^T\![\nabla_{\rv'}\times^T\!G^{EE}(\rv,\rv',\om)]
$$
where the symbol $T$ denotes a transposition of a tensor.
 Explicit forms of the Green's tensors are given in Appendix \ref{AppGT}.

The cross-spectral correlation function of the electromagnetic field at equilibrium or blackbody radiation then reads \cite{Agarwal}:
\begin{equation}
\label{FDTE}
\mathcal{E}^{(S)}_{kl}(\rv,\rv',\omega)=\mu_0\hbar\om^2 \coth[\hbar\omega/2k_{B}T] Im[G_{kl}^{EE}(\rv,\rv',\om)].
\end{equation}
This last equation is the FDT for the electric field. The cross-spectral correlation function of the magnetic field obeys a similar relation
\begin{equation}
\label{FDTH}
\mathcal{H}^{(S)}_{kl}(\rv,\rv',\omega)=\epsilon_0\hbar\om^2 \coth[\hbar\omega/2k_{B}T] Im[G_{kl}^{HH}(\rv,\rv',\om)].
\end{equation}

These relations yield the coherence properties of the equilibrium field provided that the Green's tensor of the system are known. Note that the Green's tensor is a classical object so that the vacuum fluctuations are already included in the above formalism. It is also important to note that the Green's tensor can be computed including the losses of the system. 

\subsection{Relation between symmetrized correlation function and observables}

In what follows, we will use the FDT either with $\Theta(\om,T)$ or with $(\hbar\om/2) \coth[\hbar\omega/2k_{B}T]=\Theta(\om,T)+\hbar\om/2$. The former amounts to drop the vacuum energy $\hbar\om/2$. This choice can be justified using a heuristic argument \cite{Rytov2}. When it comes to the derivation of fluxes, the vacuum energy cancels when taking the difference between emission and absorption\cite{Rytov2}. Instead, when computing the Casimir force, one has first to compute the energy variation of the electromagnetic field in the space between two parallel plates. In that case, the vacuum fluctuation energy $\hbar\omega/2$ plays a fundamental role and cannot be ignored so that  it is kept in the calculation. This procedure may seem arbitrary. A more rigorous approach to the choice of the relevant form of the FDT can be derived from quantum electrodynamics as discussed by Agarwal \cite{Agarwal}. It can be shown that when the process studied involves an absorption measurement, the relevant correlation function is the normally ordered correlation function \cite{Mandel}. If the measurement involves a quantum counter, then one needs to calculate the antinormally ordered correlation function \cite{Mandel}. The relevant forms of the FDT are given in the paper by Agarwal \cite{Agarwal}.  We show in Appendix \ref{AppFD}  that one can end up with an \textit{effective} cross-spectral density defined for positive frequencies only. This effective cross-spectral density depends on the type of measurement. The time-correlation function can be written as :
\begin{equation}
\left<E_k(\rv,t+\tau)E_l(\rv',t)\right>=Re [\int_0^{\infty}
\frac{d\om}{2\pi}\exp(i\om\tau)\mathcal{E}^{eff}_{kl}(\rv,\rv',\om) ],
\end{equation}

where the effective cross-spectral density $\mathcal{E}^{eff}_{kl}(\rv,\rv',\om)$ that should be used for an absorption measurement is given by: 
\begin{equation}
\label{csdabs}
\mathcal{E}^{(N)}_{kl}(\rv,\rv',\om)=4\om\mu_0Im[G^{EE}_{kl}(\rv,\rv',\om)]\Theta(\om,T),
\end{equation}

and the cross-spectral density appropriate for a quantum-counter measurement is given by:
\begin{equation}
\label{csdqc}
\mathcal{E}^{(A)}_{kl}(\rv,\rv',\om)=4\om\mu_0Im[G^{EE}_{kl}(\rv,\rv',\om)][\Theta(\om,T)+\hbar\om].
\end{equation}

Only the latter includes the energy of vacuum fluctuations.  For the sake of comparison, we also report the symmetrised form apropriate for positive frequencies only :
\begin{equation}
\label{csdqc}
\mathcal{E}^{(S)}_{kl}(\rv,\rv',\om)=4\om\mu_0Im[G^{E}_{kl}(\rv,\rv',\om)][\Theta(\om,T)+\frac{\hbar\om}{2}].
\end{equation}

 A simple rule can  thus be used when starting with the usual symmetrised FDT as given by (\ref{FDTE}) in order to get the relevant correlation function for a process involving absorption: 1) restrict the spectrum to positive frequencies, 2) multiply the spectrum by 2, 3) remove the energy fluctuation contribution, 4) take the real part. 

\subsection{Fluctuational electrodynamics out of equilibrium}

In the previous section, we have given the form of the cross-spectral densities of the fields and current densities at equilibrium. However, it is possible to derive the fields radiated by a system out of equilibrium. The approach is based on the FDT for the current density. Assuming local thermal equilibrium, we can derive the statistical properties of the currents. We can thus derive the fields radiated by a system with an inhomogeneous temperature field. Although the mean values of the fields are zero, their correlation are non-zero.  Let us consider for instance the symmetrized cross-spectral correlation function of the electric field
\begin{eqnarray}
\label{ }
&\left<E_k(\rv,\om)E_l^*(\rv',\om')\right>_S= \nonumber \\
&\left<\mu_0^2\om^2\int d^3\rv_1d^3\rv_2
G_{km}^{EE}(\rv,\rv_1,\om)G_{ln}^{EE*}(\rv',\rv_2,\om)j_m(\rv_1,\om)j_n^*(\rv_2,\om')\right>.
\end{eqnarray} 
Using the FDT for the fluctuating currents (\ref{FDTj}), we obtain 
\begin{eqnarray}
\label{Wklnoeq}
&\left<E_k(\rv,\om)E_l^*(\rv',\om')\right>_S=
\frac{\mu_0\om^3}{c^2}
\int d^3\rv_1Im[\epsilon(\rv_1)]   \nonumber \\
&\times [\Theta[\om,T(\rv_1)]+\hbar\om/2]G_{km}^{EE}(\rv,\rv_1,\om)G_{lm}^{EE*}(\rv',\rv_1,\om')
2\pi\delta(\om-\om').
\end{eqnarray}

With the help of the FDT, we have seen that it is possible to calculate all kind of cross-spectral spatial correlation  functions involving the electric and the magnetic fields. With these functions, we are now able to calculate other quantities such as the energy density, the Poynting vector or the Maxwell stress tensor. In the case of thermal equilibrium situations, we will use the application of the FDT for the fields which give simpler expressions. Nevertheless, in non-equilibrium situation such as the study of heat transfer between materials held at different temperature, these expressions are no longer valid. It is however still possible to use the fluctuation-dissipation theorem for the currents by assuming local thermal equilibrium. It will thus be possible to derive the fluxes for non-equilibrium situations. 

\section{Electromagnetic energy density and Local Density Of States (LDOS)}

In this section, we will study how the electromagnetic energy density is modified by the presence of material media. We shall first examine the amount of electromagnetic energy emitted by a half-space at temperature $T$. It will be shown that the density of energy is dramatically different in the near field and in the far field when surface waves are excited. The second point that we address is the general problem of the definition of the local density of electromagnetic states. Whereas it is possible to derive the density of electromagnetic states for a non-lossy system by searching the eigenmodes, the lossy case is more difficult. An alternative approach was introduced by Agarwal \cite{Agarwal3} based on the FDT. Using this approach, we will discuss the role of surface waves.

\subsection{Density of emitted electromagnetic energy}

Let us return to the configuration described in Fig.\ref{Fig:1}. We calculate the electromagnetic energy density in the vacuum above a material (medium 2) at temperature $T$ due to the emission of this material.  We do not take into account the energy incident on the medium. In order to retrieve the density of energy at equilibrium, we should include the energy incident on the surface. The density
of energy in vacuum reads  (See Jackson \cite{Jackson} p.242)
\begin{equation}
\label{NRJvide}
\left<U\right>=\frac{\epsilon_0}{2}\left<|\E(\rv,t)|^2\right>+\frac{\mu_0}{2}\left<|\mathbf{H}(\rv,t)|^2\right>
=\int_0^{\infty}\frac{d\om}{2\pi} u_{tot}(z,\om),
\end{equation}

where we have introduced a spectral density of energy $u_{tot}$. The details of the derivation are given in \cite{Shchegrov,Henkel00}. The basic procedure amounts to derive the field radiated by the random currents in the lower half-space. Adding the electric and magnetic contributions, the total electromagnetic energy above a medium at temperature $T$ in a vacuum at $T=0$ K is
\begin{eqnarray}
\label{utotnoneq}
u_{tot}(z,\om)&=&\frac{\Theta(\om,T)\om^2}{2\pi^2 c^3}  
\left\{
\int_0^{\om/c} \frac{KdK}{k_0\vert \gamma_1\vert} \frac{\left(1-|r_{12}^s|^2\right)
+\left(1-|r_{12}^p|^2\right)}{2}    \right. \nonumber \\
&+& \left. \int_{\om/c}^\infty \frac{4K^3dK}{k_0^3|\gamma_1|}
 \frac{Im(r_{12}^s)+Im(r_{12}^p)}{2} e^{-2Im(\gamma_1)z}
\right\}, 
\end{eqnarray}

where the Fresnel reflection factors are given in the appendix. 

\subsection{Discussion}

\begin{figure}[h!]
\begin{center}
\includegraphics[width=10cm]{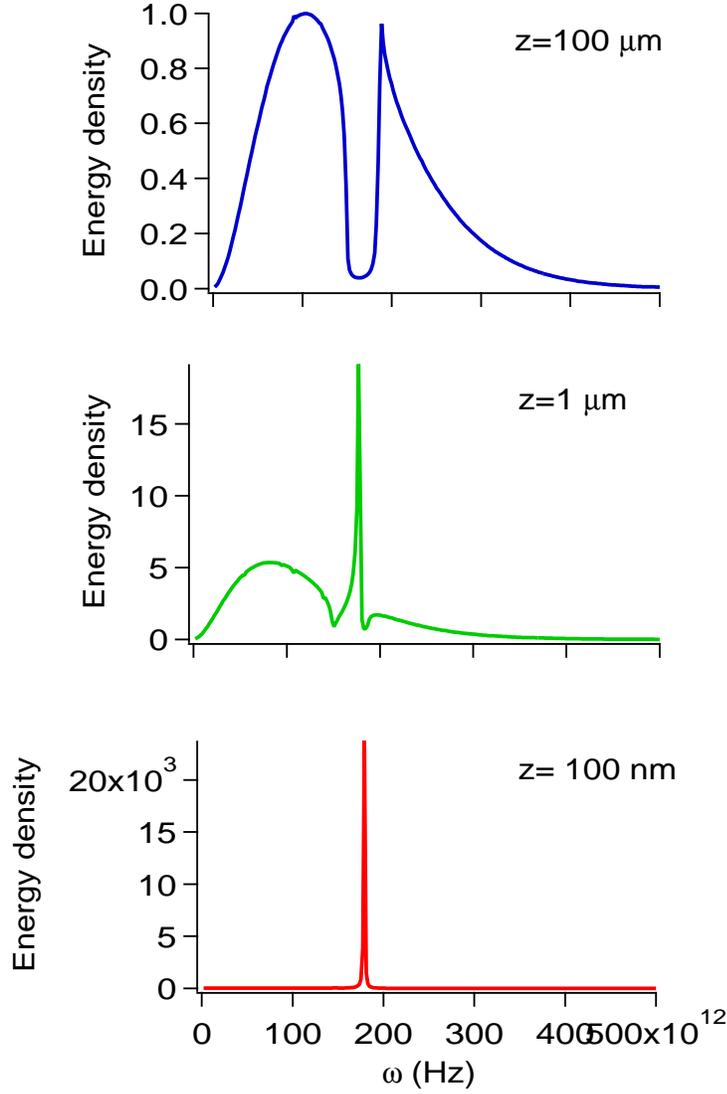}
\caption{Total electromagnetic energy density above a plane interface separating SiC at $T=300$ K from vacuum at $T=0$ K.}
\label{Fig:spectresic}
\end{center}
\end{figure}

In order to illustrate this discussion, we study the density of electromagnetic energy above some specific materials.  Let us first consider a material supporting surface waves in the infrared such as
SiC. In Fig.\ref{Fig:spectresic}, we plot the energy density $u_{tot}(z,\om)$ versus the frequency at different distances of a semi-infinite interface of SiC. The semi-infinite medium is at temperature $T=300$ K whereas the vacuum is at $T=0$ K. Note that at $T=300$ K, Wien's law gives a peak wavelength for thermal radiation $\lambda_{Wien}=10 \mu$m. In the far field, i.e. for distances $d$ larger than $\lambda_{Wien}/2\pi$, the energy density spectrum resembles that of a blackbody. The difference with a Planck spectrum comes from the fact that SiC is a very reflecting material around $\lambda=10\mu$m or $\om=1.7$ 10$^{14}$ rad s$^{-1}$. Thus, its emissivity is small in this frequency interval. 
This property is easily recovered from the electromagnetic energy due to propagating waves only (First term in (\ref{utotnoneq}))
\begin{equation}
\label{utotprop}
u_{tot}^{prop}(z,\om)= u^0(\om,T) \int \frac{d\Omega}{4\pi}\frac{1-|r_{12}^s|^2+1-|r_{12}^p|^2}{2},
\end{equation}
where we have used $2\pi KdK=k_0^2 \cos(\theta)d\Omega$, $\theta$ is the angle between the emission direction and the nomal of the surface.  The integral is performed over a half-space and
\begin{equation}
\label{ }
u^0(\om,T)=\frac{\hbar\om^3}{\pi^2c^3}\frac{1}{\exp\frac{\hbar\om}{kT}-1}=\frac{\omega^2}{\pi^2c^3}\Theta(\omega,T)
\end{equation}
is the electromagnetic energy density in a cavity at thermal equilibrium $T$. In the far field, the evanescent waves do not contribute to the energy density because of the exponential decay  ($\rm{e}^{-2Im(\gamma_1)z})$.  We note that if medium 2 is totally absorbing ($r_{12}^{s,p}=0$), the energy density due to propagating waves is half the energy calculated in a vacuum at thermal equilibrium. This is not surprising since we are computing only the emitted part of the radiation. In the case of equilibrium radiation, there is also the contribution of the radiation coming from the upper half space. We note that  in (\ref{utotprop}), the emissivity appears to be $(1-|r_{12}^s(\om,\Omega)|^2+1-|r_{12}^p(\om,\Omega)|^2)/2$. It is thus the half sum of the energy transmission factors for both polarizations. Thermal emission by a half-space can be viewed as a transmission process of a blackbody radiation in the material medium through an interface. This point of view yields insight in  Kirchhoff's law. Indeed, the equality between emissivity and absorptivity appears to be a consequence of the equality of the transmission factor when interchanging source and detector. 

\begin{figure}[h!]
\begin{center}
\includegraphics[width=10cm]{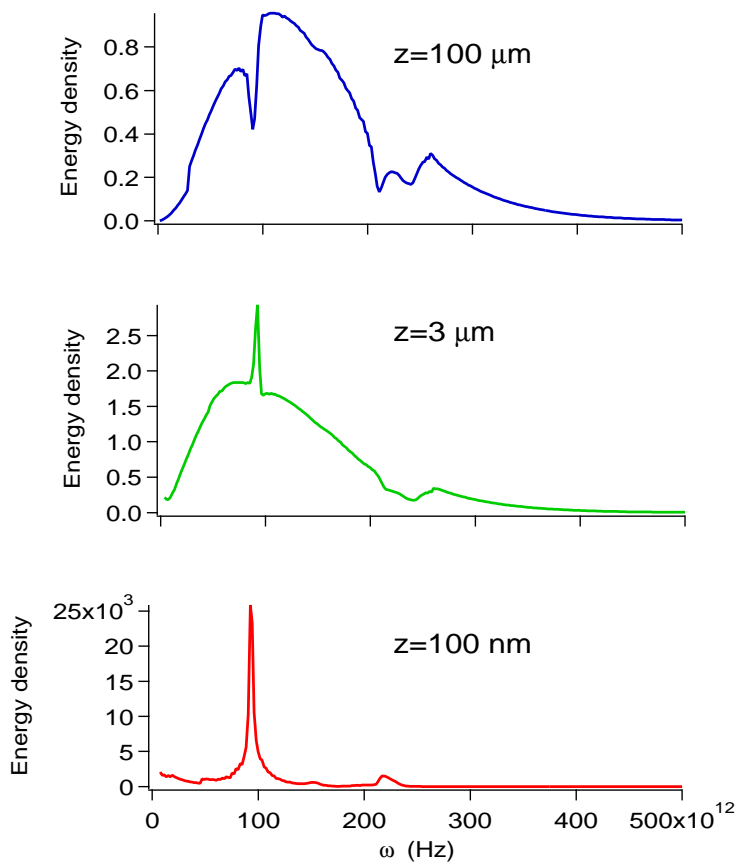}
\caption{Electromagnetic energy density above a plane interface separating glass at $T=300$ K from vacuum at $T=0$ K.}
\label{spectreverre}
\end{center}
\end{figure}

At a distance $z=3\mu$m  which is slightly larger than $\lambda_{Wien}/2\pi$, the energy density spectrum changes drastically and a strong peak emerges. At 100 nm from the interface, one observes that the thermal emission is almost monochromatic around $\om=1.78$ 10$^{14}$rad s$^{-1}$. At this frequency the energy density has increased by more than four orders of magnitude. The peak corresponds to the excitation of a surface wave. This distance is in agreement with the decay length of the surface waves as discussed in Section 2.  At distances much smaller than the wavelength, we enter a regime that we call extreme near field. The leading contribution comes from the very large wavevectors $K$ in the energy density integral. At large $K$, it can be shown that $\gamma_1\sim i K$, so that $Im(r_{12}^s)$ tends to zero and $Im(r_{12}^p)$ tends to its electrostatic limit $(\epd-1)/(\epd+1)$. We finally obtain a very simple asymptotic form of the energy density
\begin{equation}
\label{uasymp}
u_{tot}=\frac{\om^2}{4\pi^2c^3}\frac{Im[\epd(\om)]}{|\epd(\om)+1|^2}\frac{1}{(k_0z)^3}\Theta (\om,T),
\end{equation}
showing that the energy density will diverge at the frequencies where $\epu=-1$. These are the frequencies where the dispersion relation has horizontal asymptotes. For lossy materials, a resonance will occur at the frequency where $Re(\epd)=-1$ provided that absorption (i.e $Im[\epd]$) is not too large. Let us mention that the resonance of the reflection factor in the electrostatic limit has been observed experimentally by Hillenbrand \textit{et al.}\cite{Keilmannature}. A spectrum of the light scattered by a tip very close to a surface shows a peak for the resonance frequency. 
This peak is due to the field emitted by the image of the tip. Indeed, its amplitude is proportional to the reflection factor $(\epd-1)/(\epd+1)$ and is therefore resonant when $Re(\epd)=-1$.

In order to prove that this surprising behaviour is not specific to SiC, we plot in Fig.\ref{spectreverre}, the energy density spectrum above a flat interface of glass. This material is amorphous so that the optical phonons are poorly defined. 
Here again, the energy density spectrum resembles that of a blackbody in the far field whereas peaks emerge while approaching the surface. The strongest peak is at $\om=9.2$ 10$^{13}$ rad s$^{-1}$ and the weakest one at $\om=2.2$ 10$^{14}$ rad s$^{-1}$.
These frequencies are solution of  $Re[\epd(\om)]=-1$. Yet, the corresponding surface waves have a very short propagation length. 

Not all materials supporting surface waves exhibit strong peaks in their near-field thermal energy density spectrum. Indeed, as it can be seen in (\ref{uasymp}), a peak is exhibited if the frequency where $\epsilon_2(\om)$ approaches -1 corresponds to a frequency range where $\Theta(\om,T)$ is not too small. For example, metals exhibit surface-plasmon polariton in the UV or visible range where $\Theta(\om,T)$ is exponentially small at ambient temperature. Thus metals do not exhibit strong peak in their thermal energy density spectrum in the near field.

\subsection{Local Density Of States}
The  density of states (DOS) is a fundamental quantity from which many macroscopic quantities can be derived. In statistical physics, the DOS allows to calculate the partition function of a system from which all the macroscopic properties follow.  The local density of states (LDOS) is useful to study a non-uniform system. The local density of electronic states is widely used in solid state physics. It has been shown \cite{Tersoff} for instance that a scanning tunneling microscope images the electronic LDOS. The local character of the LDOS clearly describes the spatial distribution of electrons in the solid. A similar spatial dependence is also relevant for electromagnetic waves. Whereas the intensity is uniform in a vacuum in equilibrium, this is not the case in a waveguide or above an interface. The distribution of the energy is no longer uniform. Whereas LDOS is well defined for electrons in solid state physics \cite{Economou2}, its electromagnetic counterpart is not very well defined in the literature. As compared to electronic systems, two differences must be taken into account : the vectorial nature of the fields and the existence of losses.

 In electrodynamics, the LDOS is used in different contexts. From Fermi's golden rule, it is known that the DOS determines the radiation rate. It can be shown that the lifetime of an atom with an electric  dipole along a unit vector $\mathbf{u}$ is inversely proportional to $Im[\mathbf{u}\cdot G^{EE}\cdot\mathbf{u}]$. This is often refered to as the LDOS. To avoid confusion, we shall refer to this quantity as the projected LDOS. It is the relevant quantity that one needs to study when designing a microcavity or a photonic crystal to tailor emission properties. Yet, note that only those states that can be coupled to the dipole are taken into account. Thus it is not a good definition if one is interested in the total energy of the system. Such a quantity is required when computing dispersion forces \cite{vankampen,Gerlach} or shear forces \cite{Pendry97} for instance. Those forces depend on the energy stored in all available modes. In a vacuum, the LDOS can be shown to be given by the imaginary part of the trace of the Green's tensor $\G^{EE}$. This seems to be a straightforward extension of the scalar result which is proportional to the imaginary part of the Green function. The vacuum form is thus usually assumed to be valid for any other situations \cite{Pendryldos,Chicanne,Colas}. In what follows, we will summarize a recent analysis of the LDOS \cite{Joulain} that follows the original approach by Agarwal \cite{Agarwal3}. It will be seen that the LDOS is not given by the imaginary part of the trace of the Green's tensor $\G^{EE}$. It will also appear that surface waves dominate the LDOS close to an interface.

We consider a system in thermal equilibrium at temperature $T$. In a vacuum, one can define the electromagnetic energy $U(\om)$ by the product of the DOS $\rho(\om)$ by the mean energy of each state at temperature $T$ :
\begin{equation}
\label{ }
U(\om)=\rho(\om) \frac{\hbar\om}{\exp(\hbar\om/k_{B}T)-1}.
\end{equation}
We can now introduce \cite{Shchegrov,Joulain} a \textit{local} density of states by using as a starting point the \textit{local} density of electromagnetic energy $U(\rv,\omega)$ 
at a given point $\rv$  in space, and  at a 
given angular frequency $\omega$. This can be written by definition of the 
LDOS $\rho(\rv,\omega)$ as

\begin{equation}
    U(\rv,\omega)=\rho(\rv,\omega) \ \frac{\hbar \omega}{\exp(\hbar
    \omega/k_{B}T) - 1}.
    \label{eq:defLDOS}
\end{equation}

The density of electromagnetic energy is the sum of the density of electric 
energy and
of the density of magnetic energy given in (\ref{NRJvide}). In equilibrium, it can be calculated using the
system Green's function and the fluctuation-dissipation theorem. 
We start from the electric and magnetic field correlation functions for a stationary system
\begin{eqnarray}
 \left<E_k(\rv,t)E_l(\rv',t')\right> & = &Re\left[\int_0^{\infty} \frac{d\omega}{2\pi} \mathcal{E}^{(N)}_{kl}(\rv,\rv',\omega)
e^{-i\omega(t-t')}\right], \\
\left<H_k(\rv,t)H_l(\rv',t')\right> & = & Re\left[\int_0^{\infty}  \frac{d\omega}{2\pi} \mathcal{H}^{(N)}_{kl}(\rv,\rv',\omega)
e^{-i\omega(t-t')}\right] ,
\end{eqnarray}
with $t=t'$.
 The cross-spectral density for normally ordered fields is given by (\ref{csdabs}). It follows that the energy per unit volume can be cast in the form \cite{Agarwal3,Joulain}:

\begin{equation}
    U(\rv,\omega)=\frac{\hbar \omega}{\left[\exp(\hbar
    \omega/k_{B}T) - 1\right]}\frac{\omega}{\pi c^2}Im 
Tr\left[\green^{EE}(\rv,\rv,\om)+\green^{HH}(\rv,\rv,\om)\right].
    \label{eq:imtr}
\end{equation}

A comparison of Eqs.~(\ref{eq:defLDOS}) and (\ref{eq:imtr}) shows 
that the LDOS is the sum of an electric contribution $\rho^{E}$ and a magnetic contribution $\rho^{H}$:
\begin{equation}
\rho(\rv,\omega)=\frac{\omega}{\pi c^2}Im 
Tr\left[\green^{EE}(\rv,\rv,\om)+\green^{HH}(\rv,\rv,\om)\right]
= \rho^{E}(\rv,\om)+\rho^{H}(\rv,\om)
    \label{eq:ldosimtr}
\end{equation} 
In what follows, we shall discuss a few examples to illustrate the modification of the LDOS. It will be seen that in some cases, the LDOS is accurately given by the trace of the electric Green's dyadic but it can also be very different. 

\subsection{Electromagnetic LDOS in simple geometries}
\subsubsection{Vacuum}
In the vacuum, the Green's tensors $\G^{EE}$ and $\G^{HH}$ obey the same equation and have the same boundary conditions. Therefore, their contribution to the electromagnetic energy density are equal:
\begin{equation}
\label{ }
Im[\green^{EE}(\rv,\rv,\om)]=Im[\green^{HH}(\rv,\rv,\om)]=\frac{\om}{6\pi c}{\stackrel{\leftrightarrow}{\bf I}}.
\end{equation}
The LDOS is thus obtained by multiplying the electric field contribution by 2. After taking the trace, the usual result for a vacuum is retrieved
\begin{equation}
\label{ }
\rho_v(\rv,\om)=\rho_v(\om)=\frac{\om^2}{\pi^2c^3}.
\end{equation}
As expected, we note that the LDOS is homogeneous and isotropic.

\subsubsection{Plane interface}
We now consider a plane interface separating a vacuum (medium 
1 in the upper half-space) from a semi-infinite material 
(medium 2, in the lower half-space) characterised by its complex dielectric 
constant $\epsilon_2(\om)$. The material is assumed to be homogeneous, linear, 
isotropic and non-magnetic. The expression of the LDOS at a given 
frequency and at a given height $z$ above the interface in vacuum is obtained
by inserting the expressions of the electric and magnetic Green's tensors for this geometry \cite{Sipe} into equation (\ref{eq:ldosimtr}).
Note that in the presence of an interface, the magnetic and electric Green's tensors are no longer the 
same. Indeed, the boundary counditions at the interface
are different for the electric and magnetic fields.

Let us consider some specific examples for real materials like metals and 
dielectrics. We first calculate $\rho(z,\om)$ for aluminum at different 
heights. Aluminum is a metal whose dielectric constant is well 
described by a Drude model for near-UV, visible and 
near-IR frequencies \cite{Palik}:
\begin{equation}
\label{ }
\epsilon(\omega)=1-\frac{\omega_p^2}{\omega(\omega+i\Gamma)},
\end{equation}
with $\omega_p=1.747$ 10$^{16}$ rad.s$^{-1}$ and $\Gamma=7.596$ 
10$^{13}$ rad.s$^{-1}$.
\begin{figure}
\begin{center}
\includegraphics[width=10cm]{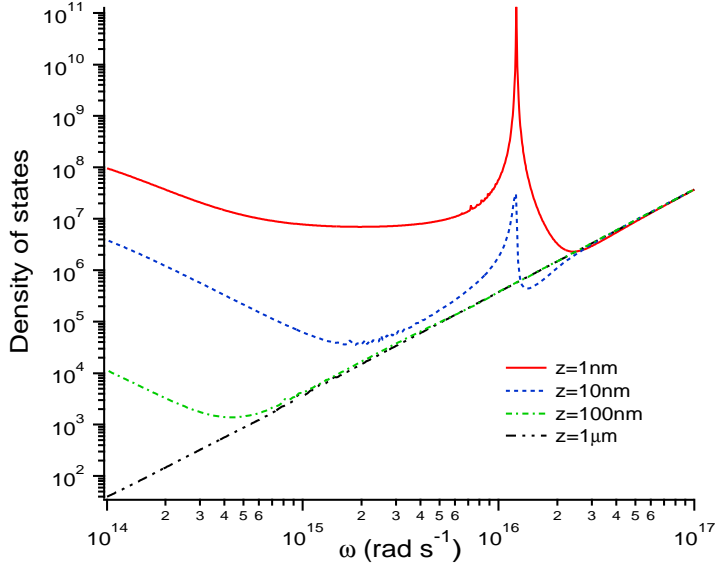}
\caption{LDOS versus frequency at different heights above a 
semi-infinite sample of aluminum. From \cite{Joulain}}
\label{ldosz}
\end{center}
\end{figure} 
We plotted in Fig.\ref{ldosz} the LDOS $\rho(\rv,\om)$ in the near 
UV-near IR frequency domain at four different heights. We first note 
that the LDOS increases drastically when the distance to the material 
is reduced. As discussed in the previous paragraph, at larger distances 
from the material, one 
retrieves the vacuum density of states. Note that at a given 
distance, it is always possible to find a sufficiently high frequency 
for which the corresponding wavelength is small compared to the 
distance so that a far-field situation is retrieved. This is clearly seen  when looking at the curve for $z=1 \mu m$ which coincides with the vacuum LDOS. When the 
distance to the material is decreased, additional modes are present: 
these are the evanescent modes which are confined close to the 
interface and which cannot be seen in the far field. Moreover, 
aluminum exhibits a resonance around $\omega=\omega_p/\sqrt{2}$. 
Below this frequency, the material supports surface-plasmon polaritons so that 
these additional modes are seen in the near field. 
This produces an increase of the LDOS close to the interface.
The enhancement is particularly important at the 
resonant frequency  which corresponds to $Re[\epsilon_2(\omega)]=-1$. 
This behaviour is analogous to that 
previously described in the electromagnetic energy density paragraph for a glass surface 
supporting surface-phonon polaritons. Also note that in the low 
frequency regime, the LDOS increases. Finally,  Fig.\ref{ldosz} shows 
that it is possible to have a LDOS smaller than that of vacuum at 
some particular distances and frequencies.
\begin{figure}
\begin{center}
\includegraphics[width=10cm]{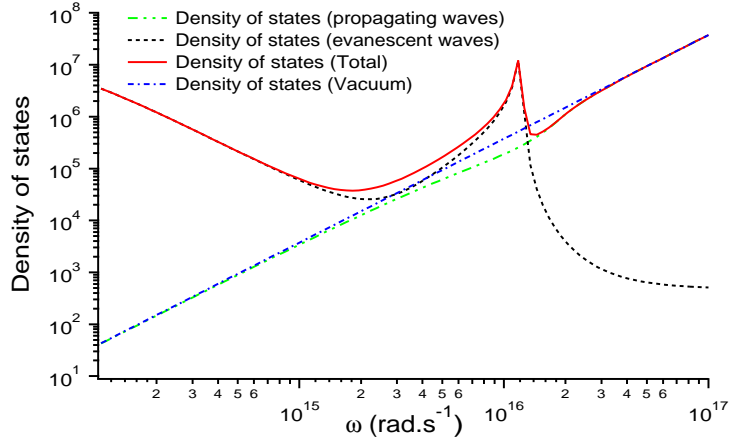}
\caption{Density of states contributions due to the propagating and 
evanescent waves compared to the total density of states and the 
vacuum density of states. These quantities are calculated above an 
aluminum sample at a distance of 10 nm. From \cite{Joulain}}
\label{fig2}
\end{center}
\end{figure}
Fig.\ref{fig2} shows the propagating and evanescent waves 
contributions to the LDOS above an aluminum sample at a distance of 
10 nm. The propagating contribution is very similar to that of the vacuum 
LDOS. As expected, the evanescent contribution dominates at 
low frequency and around the surface-plasmon polariton resonance, 
where pure near-field contributions dominate.

\begin{figure}
\begin{center}
\includegraphics[width=10cm]{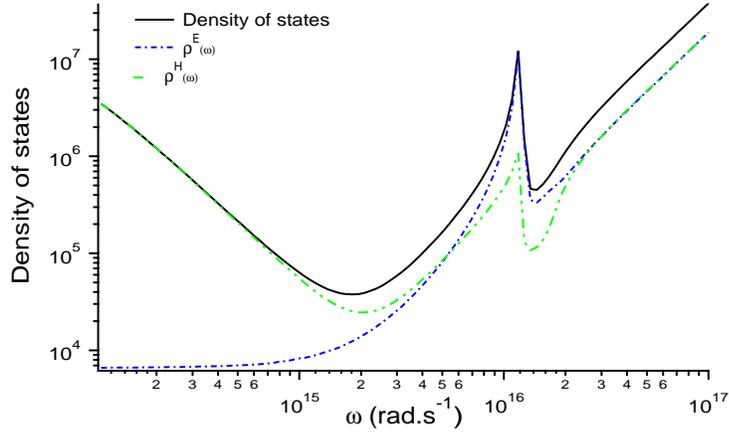}
\caption{LDOS at a distance $z=10$ nm above a semi-infinite aluminum 
sample. Comparison with $\rho^E(z,\omega)$ and $\rho^H(z,\omega)$. From \cite{Joulain}.}
\label{fig3}
\end{center}
\end{figure}

Let us now turn to the comparison of $\rho(z,\om)$ with the usual 
definition often encountered in the literature\cite{Pendryldos,Chicanne,Colas}, which corresponds to 
$\rho^E(z,\omega)$. We plot in Fig.\ref{fig3}, $\rho$, $\rho^E$ and $\rho^H$ above 
an aluminum surface at a distance $z=10$~nm. In 
this figure, it is possible to identify three different domains for 
the LDOS behaviour. We note again that in the far-field situation 
(corresponding here to high frequencies i.e. $\lambda/2\pi\ll z$), the LDOS reduces to the vacuum situation. In this case 
$\rho(z,\omega)=2\rho^E(z,\omega)=2\rho^H(z,\omega)$. Around the 
resonance, the LDOS is dominated by the electric 
contribution $\rho^{E}$. Conversely, at low frequencies, $\rho^H(z,\om)$ 
dominates. Thus, Fig.\ref{fig3} shows that we have to be very careful 
when using the approximation $\rho(z,\om)=\rho^E(z,\omega)$. Above aluminum 
and at a distance $z=10$ nm, it is only valid in a 
narrow range  between $\om=10^{16}$ rad s$^{-1}$  and $\om=1.5\times
10^{16}$ rad s$^{-1}$, i.e. around the frequency where the surface wave exists.

\subsubsection{Asymptotic form of the LDOS in the near-field}

In order to get more physical insight, we have calculated the 
asymptotic LDOS behaviour in the three regimes mentioned above. As we 
have already seen, the far-field regime ($\lambda/2\pi\ll z$) corresponds to the vacuum 
case. To study the near-field situation  ($\lambda/2\pi\gg z$), we focus on the evanescent 
contribution due to the large wavevectors $K$ as suggested by the results in Fig.\ref{fig2}.  In this 
(quasi-static) limit, the Fresnel reflection factors reduce to
\begin{eqnarray}
\lim_{K\rightarrow\infty}r^s_{12} & = &  \frac{\epsilon_2 
-1}{4 (K/k_0)^2} ,\\
\lim_{K\rightarrow\infty}r^p_{12} & = &  \frac{\epsilon_2 
-1}{\epsilon_2+1}\label{asympr}.
\end{eqnarray}
Asymptotically, the expressions of  $\rho^E(z,\omega)$ and 
$\rho^H(z,\omega)$ are\cite{Joulain}:
\begin{eqnarray}
\rho^E(z,\omega) & = & 
\frac{\rho_v}{|\epsilon_2+1|^2}\frac{\epsilon_2^{''}}{4k_0^3z^3} ,
\label{asympe}\\
\rho^H(z,\omega) & = & \rho_v\left[\frac{\epsilon_2^{''}}{8k_0 
z}+\frac{\epsilon_2^{''}}{2
|\epsilon_2+1|^2k_0z} \right] .
\label{asymph}
\end{eqnarray}
At a distance $z=10$ nm above an aluminum surface, these asymptotic 
expressions matches almost perfectly with the evanescent 
contributions ($K>k_0$) of $\rho^E$ and $\rho^H$. These 
expressions also show that for a given frequency, one can always find  a distance $z$ to the 
interface below which the dominant contribution to the LDOS will be 
the one due to the imaginary part of the electric-field Green 
function that varies like $(k_0z)^{-3}$. But for aluminum at a distance $z=10$~nm, this is not the 
case for all frequencies. As we mentioned before, this is only true 
around the resonance. For example, at low frequencies, and for 
$z=10$~nm,  the LDOS is actually dominated by $\rho_v\epsilon_2^{''}/(16k_0 
z)$.

\subsubsection{Spatial oscillations of the LDOS}
\begin{figure}[h!]
\begin{center}
\includegraphics[width=10cm]{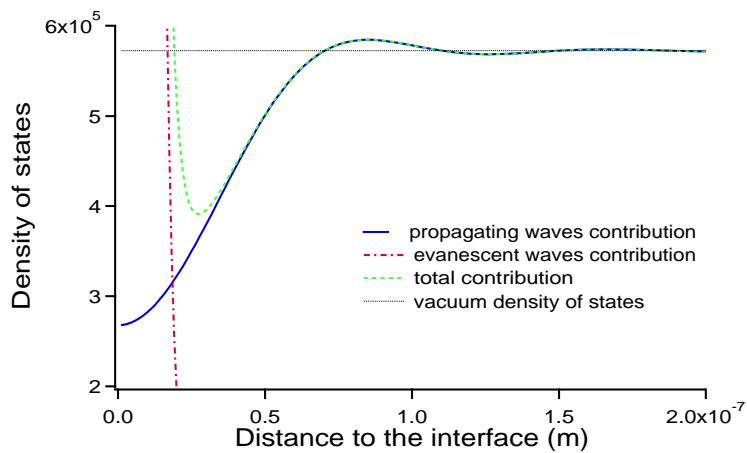}
\caption{LDOS versus the distance $z$ from an aluminum-vacuum 
interface at the aluminum resonant frequency. From \cite{Joulain}}
\label{fig4}
\end{center}
\end{figure}

Let us now focus on the LDOS variations at a given frequency $\om$ versus 
the distance $z$ to the interface. There are essentially three regimes. First, for distances much larger than the wavelength, the LDOS is 
given by the vacuum expression $\rho_v$.
The second regime is observed close to the interface where oscillations are observed. Indeed, at a given frequency, each incident plane wave on the interface can interfere with its reflected counterpart. This generates an interference pattern with a fringe spacing that depends on the angle and the frequency. Upon adding the contributions of all the plane waves over angles, the oscillating structure disappears except close to the interface.
This leads to oscillations around distances on the 
order of the wavelength. This phenomenon is the 
electromagnetic analog of Friedel oscillations which can be 
observed in the electronic density of states near interfaces~\cite{Ashcroft,Harrisson}. For a highly reflecting 
material, the real part of the reflection coefficients are negative 
so that the LDOS decreases while approaching the surface. These 
two regimes are clearly observed for aluminum in 
Fig.~\ref{fig4}. The third regime is observed at small distances as seen in Fig.~\ref{fig4}. It is due to the contribution of surface waves. Its behaviour is thus dependent on the frequency. Let us first consider the particular case of the frequency corresponding to the asymptote of the dispersion relation. It is seen that the evanescent 
contribution dominates and ultimately the LDOS always increases as 
$1/z^3$, following the behaviour found in (\ref{asympe}). This is 
the usual quasi-static contribution that is always found at short 
distance \cite{Henkel00}. 
At a frequency slightly lower than the resonance frequency, surface 
waves are still excited on the surface. These modes increase the 
LDOS according to an exponential law, a 
behaviour which was already found for thermally emitted 
fields~\cite{Carminati,Henkel00}.

\subsubsection{Conclusion about the LDOS}
The main results of this section can be summarized as follows.
The LDOS of 
the electromagnetic field can be unambiguously and properly 
defined from the 
local density of electromagnetic energy in a vacuum above a sample at 
temperature $T$ in equilibrium. The LDOS can still be written as a function of the 
electric-field Green's tensor only, but it is in general not 
proportional to the trace of its imaginary part. An additional term 
proportional to the trace of the imaginary part of the magnetic-field 
Green's tensor is present in the far field and at low frequencies. At 
short distance from the surface of a material supporting surface 
modes (plasmon or phonon-polaritons), the LDOS has a resonance 
at frequencies such that $Re[\epsilon(\om)]=-1$. Close to this 
resonance, the  approximation $\rho(z,\omega)=\rho^E(z,\omega)$ 
holds. If it is possible to measure the near-field thermal emission spectrum of a material,  the local density of states can be retrieved \cite{Joulain}.

\section{Coherence properties of  planar thermal sources in the near-field}

In this section, we examine the second order coherence properties of the fields due to thermal excitation in the presence of surface waves. We have shown that the density of energy is completely dominated by the contribution of surface waves in the near field. We shall see that they are also responsible for a deep modification of coherence properties. In what follows, we restrict ourselves to second-order coherence properties. 

\subsection{Spatial coherence in the near field}

The spatial coherence of the electromagnetic field is characterized by its cross-spectral density $\mathcal{E}_{kl}(\rv,\rv',\om)$. Roughly speaking, we study the correlation function of the electromagnetic field at two different points for a particular frequency. For a system in thermal equilibrium, this quantity is readily given by the fluctuation-dissipation theorem (\ref{FDTE}). The spatial dependence is thus included in the spatial dependence of the imaginary part of the Green's tensor. For the particular case of a vacuum, one finds the properties of the blackbody radiation. The Green's tensor is given in appendix A.  
 When taking its imaginary part, it is seen that the spatial dependence of the cross-spectral density is proportional to $\sin(kr)/kr$. The typical length scale of the correlation function for the blackbody radiation is thus the wavelength $\lambda=2\pi/k$. 

\begin{figure}
\begin{center}
\includegraphics[width=10cm]{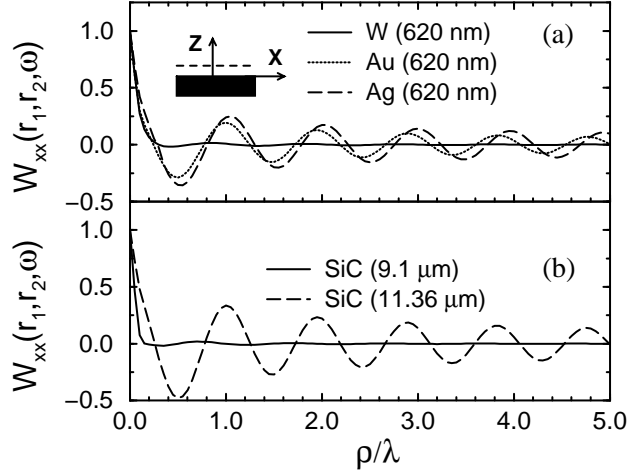}
\caption{Cross spectral density function $\mathcal{E}_{xx}(\rv_1,\rv_2,\om)$ (denoted $W_{xx}$ in the label of the figure) of the thermally emitted $x$ component of the electric field versus $\rho/\lambda$ where $\rho=|\rv_1-\rv_2|$ for different metals (a) and for SiC at different wavelengths (b). The long range correlation is due to surface plasmon polaritons for metals and to surface phonon polaritons for SiC.  From \cite{Carminati}}
\label{Figcoh1}
\end{center}
\end{figure}

It is more interesting to analyse what happens in the presence of an interface. It turns out that the coherence length may be either much larger or much smaller than the wavelength. This problem has been studied in \cite{Carminati,Henkel00} for a slightly different case. The authors considered the coherence properties of the field emitted by a solid. The difference with the above result is that in equilibrium, one has to consider two contributions : i) the blackbody radiation illuminating the surface and reflected by the surface, ii) the radiation emitted by the surface. In what follows, we focus only on the emitted part of the radiation so that we do not consider the equilibrium situation. The correlation is given by  (\ref{Wklnoeq}). This equation is valid inasmuch as the temperature $T$ can be defined everywhere in a half-space. It requires a local thermal equilibrium.  We have represented in Fig.\ref{Figcoh1} the cross-spectral density of the electric field for different metallic surfaces at a given distance $z=0.05\lambda$ to the interface. It is seen that the correlation oscillates and has an exponentially decaying enveloppe. The decay length is much larger han the wavelength indicating that the fields are coherent over large distances. This surprising phenomenon is due to the excitation of surface waves along the interface. The  physical mechanism is based on the fact that a small volume element contains random currents that excite a surface wave. This surface wave propagates along the interface over distances larger than the wavelength. It follows that different points may be illuminated by the same random source so that they are correlated. Accordingly, one does not expect any correlation between the $s$-polarized field since no surface wave exists for $s$-polarization. If one uses a material with a real part of the dielectric constant larger than $-1$, no surface wave can propagate so that no correlation should be observed. We have shown in Fig.\ref{Figcoh1} the case of tungsten in the visible that does not support surface waves. It is seen that the coherence length is smaller than a wavelength so that the radiation field appears to be more incoherent than blackbody radiation. It has been shown in  \cite{Henkel00} that in the short distance regime, the coherence length is given by~$z$.

A similar behaviour is observed for SiC, a polar material that supports surface-phonon polaritons in a frequency band. Within this band, at a wavelength $11.36  \mu$m the correlation is seen to be a long-range correlation whereas the correlation decays very rapidly for a wavelength ($9.1 \mu$m) that is not in the band where surface waves exist. 

Let us discuss in simple terms the physical origin of these unusual coherence properties. The long-range coherence is unexpected because the fluctuating currents are delta correlated as shown by the FDT. This is the reason why the fields are usually assumed to be delta-correlated in space. However, the fluctuating currents excite \textit{induced currents} in the material. In the case of a metal, a surface plasmon can be excited. In the case of a polar crystal, a surface-phonon polariton can be excited. Both surface waves are extended modes along the surface. The induced currents associated with these extended modes are therefore coherent over large distances. More precisely, the coherence length is expected to be given by the decay length of these surface modes.  This has been confirmed by a detailed asymptotic analysis in \cite{Henkel00}.

The other surprising property shown in Fig.\ref{Figcoh1} is that the coherence length defined as the FWMH of the cross-spectral density can be smaller than the wavelength. In other words, a source can be more spatially incoherent than the blackbody radiation. The key idea is that close to an interface, the field contains evanescent waves so that features smaller than the wavelength can exist. This is not the case in a vacuum so that the field has a minimum coherence length. Since the amplitude of evanescent waves of large wave vector $K$ decays as $\exp(-Kz)$, it is clear that the distance $z$ appears as a cutoff frequency. This explains that the coherence length increases as $z$ in the near-field regime.

\subsection{Temporal coherence in the near field}

The temporal coherence of the electromagnetic field is characterized by the same point time-correlation function of the electromagnetic field :

\begin{equation}
\left<E_k(\rv,t+\tau)E_l(\rv,t)\right>.
\end{equation}
This correlation function is a measurement of the memory of  the random field. It is useful to introduce a typical decay time $t_{coh}$ of the correlation function called coherence time. A Michelson interferometer with aligned mirrors performs a measurement of the correlation function. Indeed, the interference term of the signal can be written as $E(\rv,t+\tau)E(\rv,t)$ where $\tau$ is the flight time corresponding to the optical path length difference $\delta_{opt}$ between the two paths $\delta_{opt}=c \tau$. If the path-length difference is larger than the longitudinal coherence length $c t_{coh}$, no interferences can be observed. 

The temporal coherence of the EM field is related to its power spectral density. This is clearly seen by using the Wiener-Khinchin theorem \cite{Mandel,Goodman} which shows that the power spectral density is the Fourier transform of the correlation function. Alternatively, we can start from (\ref{FDTE}). It follows that 
\begin{equation}
\label{WKT}
\left<E_k(\rv,t+\tau)E_l(\rv,t)\right>=Re\left[ \int_0^{\infty}  4 \mu_0 \omega \Theta(\omega,T)Im[G_{kl}^{EE}(\rv,\rv,\omega)]e^{i \omega \tau} \frac{d \omega}{2\pi}\right].
\end{equation}

 Let us first consider the temporal coherence of the field in a vacuum. The imaginary part of the Green's tensor does not diverge and yields zero for non diagonal terms and $\omega/6 \pi c$ for diagonal terms. It follows that the time correlation function of the blackbody radiation is given by :
\begin{equation}
\left<E_k(\rv,t+\tau)E_l(\rv,t)\right>=\delta_{kl} ~Re\left[ \int _{0}^{\infty}4 \mu_0 \omega \Theta(\omega,T) \frac{\omega}{6\pi c}
e^{i \omega \tau} \frac{d \omega}{2\pi}\right].
    \label{cohtemp}
\end{equation}
 Since the integrand has a large spectral width, it appears that the coherence time is on the order of the peak radiation period.

 If we now consider the case of an interface, we know that the spectrum can be very different in the near field. We have seen previously that the contribution of the surface wave modifies dramatically the density of electromagnetic energy. In particular, we have seen that the density of energy becomes quasi-monochromatic which suggests a large coherence time. More specifically, in the extreme near field, we have seen in  (\ref{asympr}) that the Green's function has a resonant denominator $\epsilon+1$. Close to the resonance where $Re[\epsilon(\om_0)]=-1$, we can expand the dielectric constant as :
 \begin{equation}
 \epsilon({\om})=-1+i\epsilon''(\om_0)+(\om-\om_0) \frac{\d \epsilon}{\d \om},
 \end{equation}
 where we have used the notation $\epsilon=\epsilon'+i\epsilon''$. The denominator $\epsilon+1$ can be cast in the form:
 \begin{equation}
\epsilon(\omega)+1=\left (\frac{\d \epsilon'}{\d \om}\right ) [\om-\om_0+i\Gamma],
\end{equation} 
 where $\Gamma=\epsilon''(\om_0)/\frac{\d \epsilon'}{\d \om}$. It is seen that the Green's dyadic has a pole at the frequency corresponding to the asymptote of the dispersion relation of the surface wave.  Its contribution to the integral (\ref{cohtemp}) yields an exponential decay of the form $\exp[i\omega_0 t-\Gamma t]$. It follows that in the extreme near field, the thermally emitted field is partially temporally coherent with a coherence time given by $\Gamma^{-1}$. The origin of the temporal coherence of the electromagnetic field can thus be assigned to the very large density of states due to the asymptote of the surface wave. It follows that whereas the plane interface of a hot metallic surface is a temporally incoherent source for an atom located in the far field, it is a partially temporally coherent source for an atom located within a nanometric distance from the interface. 
 
\subsection{Polarization coherence in the near field}

We have seen that the excitation of surface waves by delta-correlated currents produces both a spatial and a temporal coherence. The correlation of the field is characterized by the decay time and the decay length of the surface wave that propagates along the interface. There is a further interesting coherence effect that has been studied by Set\"al\"a \textit{et al.}  \cite{Friberg}. The basic idea is that surface waves are $p$-polarized waves. As shown in section 2, a surface wave that propagates along the $y$ axis has an electric field with two components along the $y$ and $z$ axis. It follows that the $y$ and $z$ components of the field are correlated. 

The study of correlation of the fields in the near field of the source cannot be performed using the standard formalism. Indeed, when dealing with a beam, it is usually possible to work with the two components of the field perpendicular to the propagation axis. This is no longer possible in the near field of a thermal source. A generalization of the degree of polarization has been introduced in \cite{Friberg}. It was found that the degree of polarization varies as a function of the distance to the interface. Like the coherence time, it increases when approaching the interface from the far field because the surface wave creates a correlation. When reaching the very near-field regime, the degree of polarization decays and tends to $1/4$ for all the materials.

\section{Spatially partially coherent thermal sources in the far field}

\subsection{Introduction}

In this section, we will discuss the possibility of designing a source that is spatially partially coherent. In simple terms, a  spatially partially coherent source is a source that radiates a field which has a narrow angular aperture at a given wavelength. The typical examples of coherent sources are lasers and antennas. These sources have well defined emission angular lobes. In what follows, we will show that a narrow angular emission lobe is a signature of the spatial coherence of the field  in the plane of the source. 

 We first introduce an analogue of the Wiener Khinchin theorem (WKT)  (\ref{WKT}) to analyse the spatial coherence of the field. In simple terms, WKT states that a quasi monochromatic source with bandwidth $\Delta \nu$ has a coherence time roughly equal to $1/\Delta \nu$. Similarly, a quasi parallel beam with a spatial frequency bandwidth $\Delta k_x$ has a transverse coherence length $2\pi/\Delta k_x$. A formal proof is based on two properties : the relationship between the cross-spectral density of the field in the plane of the source $z=0$ and the power spectral density of the field, the relationship between the power radiated in far field and the power spectral density. For a translationally invariant system, the Fourier transform of the field does not converge in the sense of a function. Yet, one can define a field equal to the random field in a square of area $A$ and null outside. We can now define the Fourier transform of the field in the plane $z=0$ as :

\begin{equation}
\E_A(\rv_{\parallel},\om)= \int\frac{\d^2\K}{4\pi^2} ~ \E_A(\K,\om)\exp(i\K\cdot\rv_{\parallel})
\end{equation}

It can be shown \cite{Goodman} that the WKT yields a relation between the cross-spectral density and the power spectral density of the spectrum of the field :
\begin{equation}
\left<E_l(\rv_{\parallel},\om)E_l^{*}(\rv_{\parallel}+\rv '_{\parallel},\om)\right>=
\int\frac{d^2\K}{4\pi^2}
\lim_{A\rightarrow\infty}\frac{1}{A}\left<E_{l,A}(\K,\om)E_{l,A}^{*}(\K,\om)\right>
\exp(i\K\cdot\rv '_{\parallel})
\end{equation}

This relation implies that if the spectrum has a  bandwidth smaller than $2\pi/\lambda$, the coherence length is larger than $\lambda$. The second step is to show that the bandwidth of the spectrum power density $\left<\E_{l,A}(\K,\om)\E_{l,A}^{*}(\K,\om)\right>/A$ is given by the emission pattern in the far field. Indeed, the field can be written everywhere as \cite{Nietolivre,Mandel} :
\begin{equation}
\E_A(\rv,\om)=\int \frac{\d^2\K}{4\pi^2}~ \E_A(\K,\om)\exp(i\K\cdot\rv_{\parallel}+i\gamma z),
\end{equation}
where $\mathbf{k}=(\K,\gamma)$ and $\gamma$ is given by $\gamma^2=k_0^2-\K^2$. This field can be evaluated asymptotically in the far field using the stationary phase approximation \cite{Mandel}. It can be cast in the form :
\begin{equation}
\E_A(\rv,\om)=\frac{K\exp(ik_0r)}{r}~\E_A\left(\K=\frac{2\pi}{\lambda}\hat{\rv}_{\parallel},\om\right),
\end{equation}
where $K$ is a constant. The power $d P$ flowing through an element of surface $d S=r^2d\Omega$ is given by the flux of the Poynting vector. In far field, the Poynting vector has locally a plane wave structure so that its time averaged amplitude is given by $\epsilon_0 c \vert \E\vert^2/2$.
\begin{equation}
d P=\frac{\epsilon_0c}{2} \vert K\vert^2 
\left\vert \E_A\left(\K=\frac{2\pi}{\lambda}\hat{\rv}_{\parallel},\om\right)\right\vert^2d\Omega
\end{equation}
where $\hat{\rv}$ is the unit vector $\rv/\vert \rv\vert$. This relation completes the discussion of the link between the directivity of the emitted field and the coherence of the field in the plane of the source. It is clear that a directional source implies a narrow spectrum and therefore a large correlation length. 

\subsection{Design of coherent thermal sources}

We have seen in the previous section that a source which supports a surface wave is partially spatially coherent along the surface. However, because these waves cannot propagate in a vacuum, the coherence remains confined in the vicinity of the surface. The question that we address now is whether it is possible to export the near-field coherence to the far field. In essence, that amounts to couple the surface waves to the propagating waves. This can be done in several ways. A practical way is to rule a grating on the surface. The grating can then diffract the surface wave.  By properly choosing the period of the grating, it is possible to control the angle of propagation of the diffracted light. This was first observed in \cite{Heskethnature,HeskethPRB1,HeskethPRB2} for a very deep grating ruled on a doped silicon surface. Such a material supports surface-plasmon polaritons in the infrared. A more effective source was realized using a gold grating by Kreiter \textit{et al.}  \cite{Kreiter}.  Heinzel \textit{et al.} \cite{Heinzel} have also realized a source using surface plasmons on tungsten in the near infrared. A source based on the use of surface-phonon polaritons on SiC was reported by Le Gall \textit{et al.} \cite{Legall}. The first discussion of the spatial coherence of these sources was reported in \cite{Greffet_nat}. An extended discussion of these properties has been recently reported \cite{FrancoisPRB} where transverse coherence lengths have been derived from experimental measurements of angular widths of emission peaks. Angular peaks as narrow as 1° can be obtained. 

\begin{figure}[h!]
\begin{center}
\includegraphics[width=10cm]{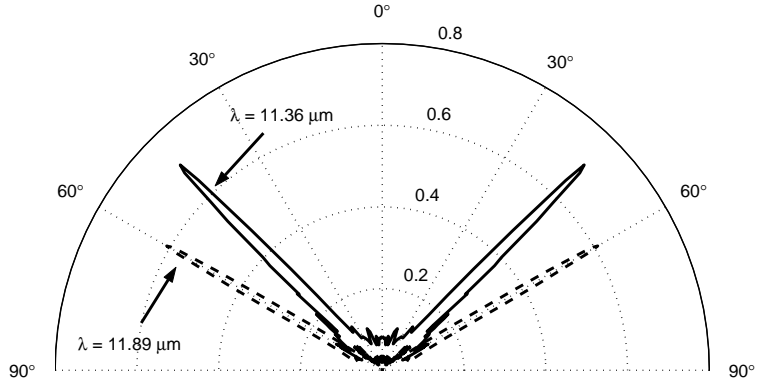}
\caption{Experimental angular emission of a SiC grating at two different wavelengths. The angular pattern has the characteristic shape of an antenna. It demonstrates the spatial coherence of the thermal source. Measurements are taken at 800 K.  From \cite{FrancoisPRB}}
\label{figemiscoh}
\end{center}
\end{figure}

 In order to have an efficient source of light, it is necessary to make sure that the coupling of the surface wave into a propagating wave is optimum. This can be accomplished by designing a surface with total absorption. From Kirchhoff's law, it follows that if absorption is unity, then emissivity is also unity. Another condition must be satisfied : the emission wavelength should lie in a region where Planck's function takes large values. The optimum choice of the wavelength thus depends on the temperature of the source. Fig.\ref{figemiscoh} shows the angular emission pattern of a SiC grating. It is clearly seen that the angular aperture is very narrow indicating a large coherence length \cite{FrancoisPRB}. Let us emphasize that the coherence is not due to to the grating but only to the surface wave. The role of the grating is merely to couple the surface wave into a propagating wave. 
 
 Different schemes have been proposed to produce partially coherent thermal sources. They are based on a filtering of the emission pattern in order to reduce the angular width of the emission pattern. The scheme that we have described so far is based on the use of a grating that couples a surface wave to propagating wave can be viewed as a device that increases the absorptivity or emissivity to 1 for a narrow set of angles. It can thus be viewed as a filtering process. A different type of filter can be designed using multilayers systems. Several authors \cite{Kollyukh,Ben-abd} have reported a narrow angular pattern emission obtained by interferences between several layers. This mechanism leads to angular widths on the order of 10°.

\subsection{Engineering radiative properties of surfaces}

For many applications, it is desirable to modify radiative properties of surfaces. An introduction to radiative properties of surfaces can be found in a review paper by Z. Zhang \cite{Zhang}. Roughness has often been used to increase the emissivity. An analysis of the different mechanisms involved can be found in \cite{Ghmari}. Further references on scattering by rough surfaces can be found in several reviews and monographs \cite{Nietolivre,Beckmann,Desanto,Ogilvy,Tsang,Voronovich}. Microstructures can be used to design efficient selective absorbers and sources. The decay of reflectivity of a shallow rough surface due to the excitation of surface plasmons is addressed in \cite{Baylard}. Hava and coworkers have examined silicon microstructured surfaces \cite{Hava1,Hava2,Hava3,Hava4}. Sai \textit{et al.} \cite{Sai} have designed silicon microstructured surfaces for thermophotovoltaics applications. Marquier \textit{et al.}\cite{MarquierOC} have studied the effect of surface plasmon on highly doped silicon showing that the peak absorption frequency can be tuned by varying the doping. Kusunoki \textit{et al.} \cite{Kusunoki} have reported emissivity measurements on tantalum surfaces with two-dimensional periodic structures. They observed peaks of emission due to the excitation of surface plasmons. Pralle \textit{et al.} \cite{Pralle} have designed selective infrared emitters using periodic structures on silicon wafers coated with gold. 

\begin{figure}[h!]
\begin{center}
\includegraphics[width=10cm]{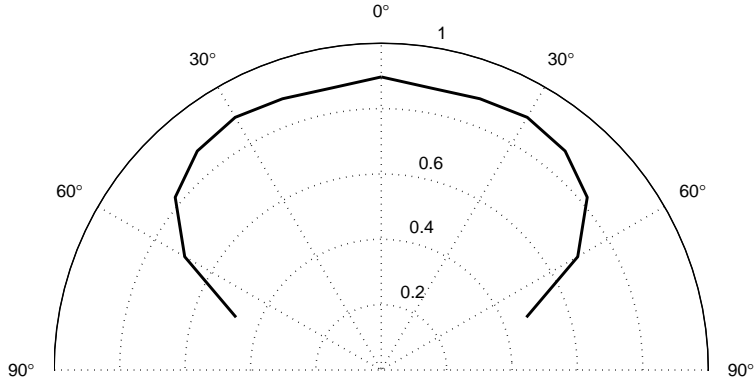}
\caption{Angular emissivity measurements of a SiC grating at the surface phonon polariton frequency. The emissivity is above 0.9 and quasi-isotropic in the plane of incidence.  The sample temperature is 800K. From \cite{FrancoisPRB}.}
\label{figsourceiso}
\end{center}
\end{figure}

An interesting application of surface waves to the enhancement of light emission has been demonstrated by Marquier \textit{et al.} \cite{FrancoisPRB,MarquierOC}. The idea is to emit light in all directions at a given frequency. To this aim, it is necessary to be able to couple light in all directions. When using a grating of period $d$, the emission angle $\theta$ is given by $\frac{\omega}{c}\sin(\theta)=K(\omega)+p\frac{2\pi}{d}$ where $p$ is an integer. The key idea is to work at the resonance frequency taking advantage that the dispersion relation is flat. At that particular frequency, surface plasmons exist for a very large set of wavevectors $K$.  It follows that one can design a source emitting in all directions. This has been demonstrated for a SiC surface on which a grating was ruled. Fig.\ref{figsourceiso} represents the angular emission pattern. It is seen that the emissivity is close to its maximum value 1 and almost isotropic. The mechanism of emission can be again viewed as a two steps process. First, each volume element is equivalent to a random dipole that can either emit a photon or excite a surface wave. The latter is a very efficient process so that usually, most of the desexcitation goes into surface waves and eventually into heat. The second step is the introduction of a grating that converts the surface wave into a propagating wave by diffraction. Thus, the excitation of a surface wave which usually tends to decrease the emission of light becomes a factor that enhances the emission of light. A similar mechanism has been proposed recently to use the surface waves in order to enhance the emission of light by quantum wells placed close to a metallic layer \cite{BarnesNatMat,Okamoto}. The idea is that the high density of states due to the surface plasmon enhances the emission. If the conversion of a surface wave into a propagating wave is efficient, the process enhances the emission. 

\section{Radiative heat transfer in the near field}

We have shown previously that the density of electromagnetic energy increases in the near field due to the contribution of surface waves. We now address the question of heat transfer between bodies separated by distances smaller than the wavelength. In that case, it is expected to observe contributions of the surface waves to the radative heat transfer. This topic has already a long history.  Anomalous radiative heat transfer was observed in the sixties. Cravalho \textit{et al.} \cite{Cravalho}, Boehm and Tien \cite{Boehm} studied that problem and took into account waves that are propagative in the materials and evanescent in the vacuum gap. Yet, this correction does not take into account all the evanescent waves. Waves that are evanescent on both sides of the interface (i.e. surface modes) were not taken into account in these early works. The first correct derivation of the flux between two plane parallel plates was reported by Polder and Van Hove \cite{Polder} in 1971. Their method allowed them to analyse the contributions of different polarizations and to compute  separately the contributions of evanescent and propagating waves. Similar works were reported later on by different authors \cite{Levin,loom,Pendry99,Volokitin01,Chen1,Mulet02}. It is found in all these works that the flux diverges as the distance decreases. This point was considered to be unphysical in \cite{Pan,Pan2} where the increase in the density of states due to surface waves was not included. It is pointed out in \cite{Maradudin,Mulet01b} that when keeping the contribution of all the wavevectors, the flux diverges if the temperature difference is assumed to be kept to a constant value. This amounts to say that the thermal resistance goes to zero which does not violate any physical law. Only a few experiments \cite{Heargraves1,Dransfeld,Xu} were reported on measurements of heat transfer due to near-field radiation. 

Heat transfer between a plane and a small particle was first discussed by Dorofeyev \cite{Dorofeyev98} and later by Pendry \cite{Pendry99} and Volokitin and Persson \cite{Volokitin01}.  Practical results were derived for a metal using a Drude model and making the additional assumption that $\vert\epsilon\vert>>1$. Mulet \textit{et al.} \cite{Mulet01} pointed out the resonant contribution of surface waves to the heat transfer. It was shown that the heat transfer is quasimonochromatic at the frequency of the optical phonons for a polar crystal given by $Re[\epsilon(\omega)]+1=0$. A similar effect is observed between metallic parallel surfaces \cite{Mulet02} and for doped semiconductors \cite{MarquierOC}.  For metals, this resonance does not play a significant role because the plasma frequency is in the UV domain so that the Bose-Einstein factor takes low values at usual temperatures. It was later suggested \cite{Greffet_nat,ChenAPL} that a quasi-monochromatic enhanced heat transfer could be used to increase the efficiency of thermophotovoltaics devices by matching the energy of the emitted photons with the absorption band gap of the photovoltaics cell. That might reduce the loss of excess energy of ultraviolet photons. The heat transfer between two small particles has been studied in \cite{Volokitin01}. It has been shown that the dipole-dipole interaction yields a large contribution to the heat transfer whereas the contribution of the photon emission and absorption process is negligible. This near-field heat transfer between nanoparticles is analogous to the energy transfer between molecules due to the dipole-dipole coupling known as Förster transfer \cite{Forster}. It may also be resonant for surface plasmon resonances. In what follows, we shall derive explicitly the heat transfer between nanoparticles and the emission by a surface plane.

\subsection{Radiative power exchanged between two spherical nanoparticles}
We now calculate the heat transfer between two spherical nanoparticles held at different temperature and separated in the vacuum. Such a calculation was first reported by Volokitin and Persson \cite{Volokitin01}.
Let us consider two nanoparticles whose dielectric constant are $\epu$ and $\epd$ and whose temperatures are $T_1$ and $T_2$. We first calculate the power dissipated in particle 2 by the electromagnetic field induced by particle 1 using the dipolar appoximation:
\begin{equation}
\label{eq:47}
P_{1\rightarrow2}(\om)=\epsilon_0\frac{\om}{2}Im(\alpha_2)|\E_{inc}(\rv_2,\om)|^2,
\end{equation}
where $\rv_2$ denotes the position of the particule 2 and $\alpha_2$ is the polarisability of a sphere of radius $a$ \cite{Bohren}:
\begin{equation}
\label{polar}
\alpha_2=4\pi a^3\frac{\epd-1}{\epd+2}.
\end{equation}
 The field incident on the particle 2 created by the thermal fluctuating dipole of  particle 1 located at $\rv_1$ at temperature $T_1$ is given by:
 \begin{equation}
\label{einc}
\E_{inc}(\rv_2,\om)=\mu_0\om^2\G(\rv_2,\rv_1,\om)\cdot\pv
\end{equation}
where $\G(\rv_2,\rv_1,\om)$ is the vacuum Green's tensor given in appendix A. To proceed, we need the correlation function of the dipole given by the fluctuation-dissipation theorem whose symmetrised form is given by (\ref{FDTP}). We finally obtain the heat exchange between two spherical nanoparticles at temperature $T_1$ and $T_2$:
\begin{equation}
\label{nano-nano}
P_{1\leftrightarrow2}=\frac{3}{4\pi^3}\frac{Im[\alpha_1(\om)]Im[\alpha_2(\om)]}{|\rv_2-\rv_1|^6}
\left[\Theta(\om,T_1)-\Theta(\om,T_2)\right].
\end{equation}
Let us note the $1/r^6$ spatial dependance of the heat transfer. This dependence is typical of the dipole-dipole interaction. It is actually a van der Waals type interactions that can be interpreted in the following way: fluctuations (thermal or not) distort the charge distribution of a nanoparticle producing a fluctuating dipole. This fluctuating dipole induces in turn an electromagnetic field on the other nanoparticle initiating a second dipole. This dipole interaction causes both an energy transfer and a momentum transfer or force. For molecules this energy transfer is known as Förster transfer and the force is called van der Waals force. We find that nanoparticles follow a similar behaviour with a resonance at the surface polariton resonance. Indeed, in the case of spherical particle with radius $a$, the polarisability has a resonance when the dielectric constant approaches $-2$ provided that the imaginary part of the dielectric constant is not too high. The particle resonances appear in the visible part of the spectrum for metals and in the infrared for polar materials.

\subsection{Thermal emitted flux by a planar interface}
In this section, we analyse the emission by a plane interface. Let us consider the situation of a planar interface ($z=0$) separating a dielectric ($z<0$) at temperature $T$ from a vacuum ($z>0$). We shall derive the flux emitted from the interface.
\subsubsection{Classical theory of radiation.}
\begin{figure}[h]
\begin{center}
\includegraphics[width=7cm]{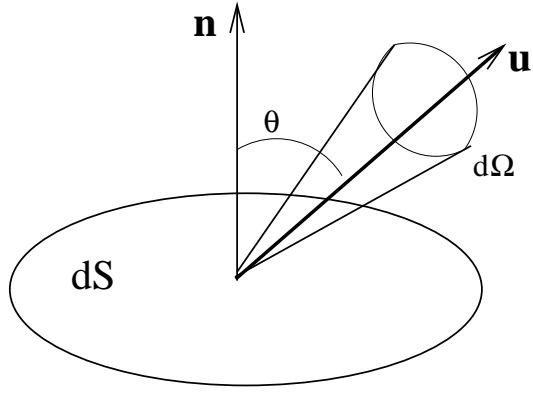}
\caption{Solid angle $d\Omega$ around a direction $\uv$ making an angle $\theta$ with the normal of an elementary surface $dS$.}
\label{emislum}
\end{center}
\end{figure}

In the classical theory of radiation, the power $d^2 Q$ emitted by an elementary opaque surface $dS$ at temperature $T$, in a solid angle $d\Omega$ around a direction $\uv$ making an angle $\theta$ with the normal to the surface (Fig.\ref{emislum}), whose monochromatic emissivity is $\epsilon'(\theta)$ is
\begin{equation}
\label{ }
d^2 Q(\om,\theta)=\varepsilon'_\om(\theta)I^0_\om(T) \,d\Omega \, dS 
\end{equation}
where
$$
I^0_\om(T)=\frac{\hbar\om^3}{4\pi^3c^2}\frac{1}{\e^{\hbar\om/k_{B}T}-1}
$$ 
is the blackbody specific intensity.
The power $d Q$ emitted by the elementary surface is thus
\begin{equation}
\label{ }
d Q(\om)=\int \varepsilon'_\om(\theta)I^0_\om(T) \cos\theta d\Omega dS =\mathbf{q}\cdot\nv \, dS
\end{equation}
where the integral is performed over a half-space. We have introduced the radiative vector defined by 
\begin{equation}
\label{ }
\qv(\om)=\int \varepsilon'_\om(\theta)I^0_\om(T)  \, \uv \,d\Omega
\end{equation}
In the case of a blackbody, the integration over the angles is straightforward. The monochromatic heat flux is thus equal to $\pi L^0_\om(T)$ and the total heat flux is equal to $\sigma T^4$ where  $\sigma$ is the Stefan constant
$$
\sigma=\frac{\pi^2k_{B}^4}{60c^2\hbar^3}=5.67\ 10^{-8} Wm^{-2}K^{-4}
$$
\subsubsection{Fluctuational electrodynamics method}

 We will follow  Polder and van Hove \cite{Polder} to show that the phenomenological form of the emitted flux can be derived in the framework of fluctuational electrodynamics. In the following sections, we use this approach to derive the expressions valid in the near field. Let us consider the situation of a planar interface ($z=0$) separating a dielectric ($z<0$) at temperature $T$ from a vacuum ($z>0$). The flux emitted by the half-space is given by the Poynting vector $\Sv=\E\times\Hv$. In the case of monochromatic quantities, the time average Poynting vector reads $\Sv(\rv,\om)=\frac{1}{2}Re\left(\E(\rv,\om)\times\Hv^*(\rv,\om)\right)$. This quantity can be derived using the fluctuation-dissipation theorem. Thus, the electric and magnetic Green's tensor are needed. In this geometry, these tensors are given by (\ref{gelec}) and (\ref{gmagn}).  The Poynting vector reduces to its $z$ component $\left<S_z(\rv,\om)\right>=(1/2)Re[\left<E_xH_y^*-E_yH_x^*\right>]$. In order to obtain the Poynting vector, one calculates quantities like
$\left<E_i(\rv,\om)H_j^*(\rv,\om)\right>$. Using (\ref{etoj},\ref{htoj})
\begin{equation}
\label{ }
\left<E_i(\rv,\om)H_j^*(\rv,\om)\right>=\left<i\mu_0\om\int G^{EE}_{ik}(\rv,\rv')G^{HE*}_{jl}(\rv,\rv'')j_k(\rv')
j_l^*(\rv'')d^3\rv'd^3\rv''\right>
\end{equation}
Using the effective FDT for the currents (\ref{FDTj}) defined for positive frequencies only, the preceding equation reduces to
\begin{equation}
\label{ }
\left<E_i(\rv,\om)H_j^*(\rv,\om)\right>=\frac{i\Theta(\om,T)\om^2}{\pi c^2}
\int \epsilon''(\rv')G^{EE}_{ik}(\rv,\rv')G^{HE*}_{jk}(\rv,\rv')d^3\rv'
\end{equation}
Using the Green functions expressions (\ref{gelec}) and (\ref{gmagn})
and the identities $\epsilon''=0$ in the upper half space and that $\epsilon''_2\om^2/c^2=2Re(\gamma_2)Im(\gamma_2)$, the Poynting vector can be cast in the form
\begin{eqnarray}
\label{ }
\left<S_z(\rv,\om)\right>&=&\frac{\om\Theta(\om,T)}{16\pi^3c}Re\left\{\int d^2\K e^{-2Im(\gamma_1)z}
\frac{\gamma_1Re(\gamma_2)}{k_0|\gamma_2|^2}\right.\nonumber\\
&\times&\left.\left[|t_{21}^s|^2+|t_{21}^p|^2\frac{|\gamma_2|^2+K^2}{|n_2|^2k_0^2}\right]\right\}
\end{eqnarray}
Let us note that only $\gamma_1$ may not be real in the preceding expression. In fact, as the upper half space is vacuum (medium 1), $\gamma_1$ can only be real ($K<\om/c$) or pure imaginary ($K>\om/c$). Therefore, the contribution of evanescent waves vanishes in the radiative flux expression. Using the identities (\ref{rel1}-\ref{rel4}),

\begin{eqnarray}
Re(\epsilon^*\gamma) & = & Re(\gamma)\frac{|\gamma|^2+K^2}{k_0^2} \label{rel1} \\
Im(\epsilon^*\gamma) & = & Im(\gamma)\frac{-|\gamma|^2+K^2}{k_0^2} \label{rel2}\\
Re(\gamma_2)|t_{21}^s|^2\frac{|\gamma_1|^2}{|\gamma_2|^2}&=&Re(\gamma_1)(1-|r_{21}^s|^2)-2Im(\gamma_1)Im(r_{21}^s) \label{rel3}\\
\frac{|n_1|^2}{|n_2|^2}\frac{|\gamma_1|^2}{|\gamma_2|^2}Re(\epd^*\gamma_2)|t_{21}^p|^2&=&\left[Re(\epu^*\gamma_1)(1-|r_{21}^p|^2)-2Im(\epu^*\gamma_1)Im(r_{21}^p)\right]. 
\label{rel4},
\end{eqnarray}

one finally obtains:
\begin{equation}
\label{ }
\left<S_z(\rv,\om)\right>=\frac{\hbar\om^3}{2\pi^2c^2}\frac{1}{\e^{\hbar\om/k_{B}T}-1}
\int_0^{\om/c}\frac{KdK}{k_0^2}\;\frac{1-|r_{12}^s|^2+1-|r_{12}^p|^2}{2}
\end{equation}
As already mentioned, only propagating waves $(K<\om/c)$ contribute to this expression. This is not surprising because no waves come from the positive $z$ direction. Moreover, there is a revolution symmetry around the $z$ axis. Introducing $d\Omega$ the elementary solid angle, we have the relation $KdK/k_0^2=d\Omega\cos\theta/(2\pi)$. The Poynting vector is then given by
\begin{equation}
\label{ }
\left<S_z(\rv,\om)\right>=\frac{\hbar\om^3}{2\pi^2c^2}\frac{1}{\rm{e}^{\hbar\om/k_{B}T}-1}
\int_{\Omega=2\pi}\frac{\cos\theta d\Omega}{2\pi}\;\frac{1-|r_{12}^s|^2+1-|r_{12}^p|^2}{2}
\end{equation}
In the case of a blackbody, i.e. a body for which the reflection factors are null, the usual expression of the radiative flux  $\pi I^0_\om(T)$ is recovered. When the dielectric situated below the interface does not behave as a blackbody, the flux takes the usual form
\begin{equation}
\label{ }
q(\om)=\left<S_z(\rv,\om)\right>=\int d\om~ \varepsilon'_{\om}(\theta)I^0_\om(T)\cos\theta d\Omega
\end{equation}
where we have identified the emissivity $\varepsilon'_\om(\theta)=(1-|r_{12}^s|^2+1-|r_{12}^p|^2)/2$. In presence of a single interface, we note that the radiation emitted is not different from the usual one. The near field does not play any role in this situation.

\subsection{Heat transfer between two semi-infinite polar materials. Interference effects}
We now focus on the heat transfer between two semi-infinite half-spaces separated by a vacuum gap and whose temperature $T_1$ and $T_2$ are uniform. The main changes that occur at small distance is the fact that evanescent waves can contribute to the heat transfer through tunneling. 

We summarize in the next sections the results. Detailed derivations can be found in \cite{Mulet02} for instance. \begin{figure}[h]
\begin{center}
\includegraphics[width=8cm]{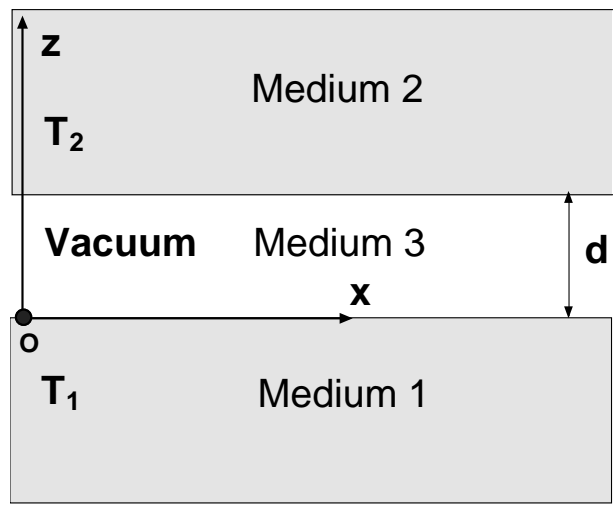}
\caption{Two semi-infinite half-spaces separated by a vacuum (distance $d$)}
\label{fig_sys_2}
\end{center}
\end{figure}
The radiative transfer is characterised by the radiative flux. In the phenomenological theory, this flux is given by
\begin{equation}
\label{ }
q(\om)=\int_0^{2\pi}\cos\theta d\Omega\int_0^\infty d\om\frac{\varepsilon'_{1\om}\varepsilon'_{2\om}}
{1-\rho'_{1\om}\rho'_{2\om}}\times\left[I^0_\om(T_1)-I^0_\om(T_2)\right]
\end{equation}
where the $\epsilon'_{i\om}$ are the directional monochromatic emissivities and the $\rho'_{i\om}$ are the directional monochromatic reflectivities.

 Using a fluctuational electrodynamics model, the flux can be written as the sum of two terms $q(\om)=q^{prop}(\om)+q^{evan}(\om)$.

The first term $q^{prop}(\om)$ is the propagating waves contribution.
\begin{equation}
\label{ }
q^{prop}(\om)=\sum_{q=s,p}\int \, d\om \, d\Omega\, \cos\theta\, \left[\frac{(1-|r_{31}^{q}|^2)(1-|r_{31}^{q}|^2)}{|1-r_{31}^{q}r_{32}^{q}e^{2i\gt d}|^2}\right]\left[I^0_\om(T_1)-I^0_\om(T_2)\right]
\end{equation}
Let us note that $1-|r_{31}^{s,p}|^2$ and $1-|r_{32}^{s,p}|^2$ are the transmission energy coefficients between media 1 and 3 and 2 and 3 for the $s$ or $p$-polarization. These coefficients can be identified as an emissivity in the same way that it has been defined for a single interface.
Let us remark that this expression for the propagating waves contribution to the radiative flux between two semi-infinite media is very close to the usual one. Only the denominators are different because interferences are not taken into account in the phenomenological model.  Nevertheless, if one considers a frequency interval small in comparison with the frequency but larger than $c/d$, the variation of $e^{i\gt d}$ with $\om$ is much faster than the Fresnel reflection coefficient  variations. The integration over this interval yields an average value of  $|1-r_{31}^{i}r_{32}^{i}e^{2i\gt d}|^2$ which is exactly $1-|r_{31}^{i}|^2|r_{32}^i|^2$. Matching the reflectivity with the Fresnel reflection energy coefficient, one can then identify this expression for the radiative flux with the classical one.

\subsubsection{Tunneling of evanescent waves}
The second term $q^{evan}(\om)$ is the contribution of the evanescent waves. It reads:
\begin{eqnarray}
\label{ }
&q^{evan}(\om)=&\nonumber\\
&\sum_{q=s,p}\int d\om \int_{\om/c}^\infty \frac{2KdK}{k_0^2}e^{-2Im(\gt)d}\left[\frac{Im(r_{31}^{q})Im(r_{32}^{q})}{|1-r_{31}^{q}r_{32}^{q}e^{2i\gt d}|^2}\right] \left[I^0_\om(T_1)-I^0_\om(T_2)\right]&
\end{eqnarray}
Contrary to the single interface case, this contribution does not vanish because of the existence of both upward and downward evanescent waves in the space between the two media\cite{CarminatiPRA}. When the distance reduces, this term is more and more important due to the presence of the exponential $e^{-2Im(\gt)d}$. When the material involved are supporting surface waves, the imaginary part of the reflection coefficient in $p$-polarisation becomes important around the resonant frequency, when the dielectric constant approaches -1. If the two media are sufficiently close to allow the interaction between the exponentially decaying surface waves bound to each interface, a transfer occurs due to the tunneling of evanescent waves.

Let us define a radiative heat transfer coefficient as the limit of the ratio of the radiative flux on the temperature difference between the two media when this temperature tends to zero:
\begin{equation}
\label{ }
h^R(\om)=\lim_{(T_1-T_2)\rightarrow 0}\frac{q(\om)}{T_1-T_2}
\end{equation}
\begin{figure}[t]
\begin{center}
\includegraphics[width=8cm]{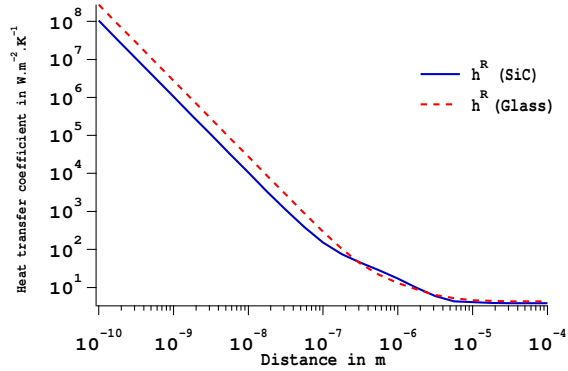}
\caption{Radiative heat transfer coefficient versus distance for semi-infinite media of temperature $T=300$ K. From \cite{Mulet02}}
\label{fig:50}
\end{center}
\end{figure}
\begin{figure}[t]
\begin{center}
\includegraphics[width=8cm]{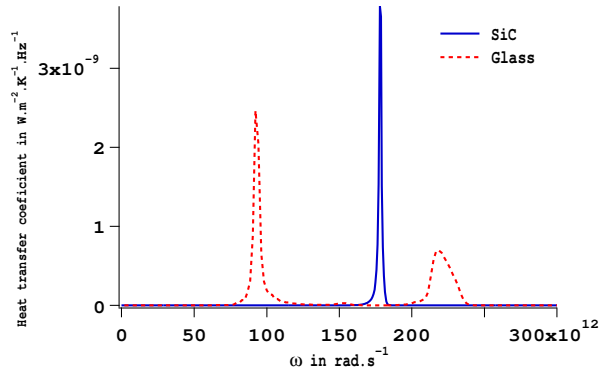}
\caption{Monochromatic heat transfer coefficient for a distance $d=10$ nm and a temperature $T=300$ K. From \cite{Mulet02}}
\label{fig:60}
\end{center}
\end{figure}
In Fig.\ref{fig:50},  $h^R(\om)$ is represented versus the distance between  two semi-infinite media of glass or SiC. For a distance larger than the thermal radiation wavelength given by the Wien law, i.e. for $d>10\ \mu$m, the transfer does not depend on the distance. We are then in the classical regime where the transfer occurs through the radiation of propagating waves. At shorter distances, the transfer increases as $1/d^2$. For a distance of 10 nm, the radiative heat transfer coefficient has increased by 4 orders of magnitude compared to its far field value. If we now focus on the spectral dependance of the heat transfer coefficient at a 10 nm distance (Fig.\ref{fig:60}), we note that the heat transfer is important for the frequencies corresponding to resonant surface waves. The heat transfer is therefore practically monochromatic in the near field. We can also expand asymptotically the radiative heat transfer coefficient for short distances:
\begin{equation}
\label{ }
h^R(\om)\sim\frac{1}{d^2}\frac{Im(\epu)Im(\epd)}{|1+\epu|^2|1+\epd|^2}\times
k_{B}\left(\frac{\hbar\om}{k_{B}T}\right)^2\frac{e^{\hbar\om/(k_{B}T)}}{(e^{\hbar\om/(k_{B}T)}-1)^2}
\end{equation}
This expression yields the $1/d^2$ dependance of the transfer coefficient and its strong frequency dependance. Indeed, when the dielectric constant approaches~-1, the radiative heat transfer coefficient exhibits a peak as well as the Fresnel reflection factor. This is the signature of the presence of a surface wave. The validity of the $1/d^2$ dependence has been questionned in \cite{Pan} on the basis that an infinite flux is not physical. As a matter of fact, the flux is infinite if one assumes that the temperature difference is kept constant. This problem is analogous to the problem of an electron flux or intensity that goes to infinity if the resistance goes to zero at fixed voltage. This raises another question: in the case of heat transfer, the resistance across a vacuum gap on the order of 10 nm usually remains much larger than the bulk conduction resistance of solids over distances on the order of 100 nm. It is thus safe to assume that the temperature is uniform in the bulk over a skin depth so that the calculation is valid. 

\subsection{Calculation of the heat transfer between a dielectric sphere and a half-space}
\subsubsection{Introduction}
\begin{figure}[h]
\begin{center}
\includegraphics[width=8cm]{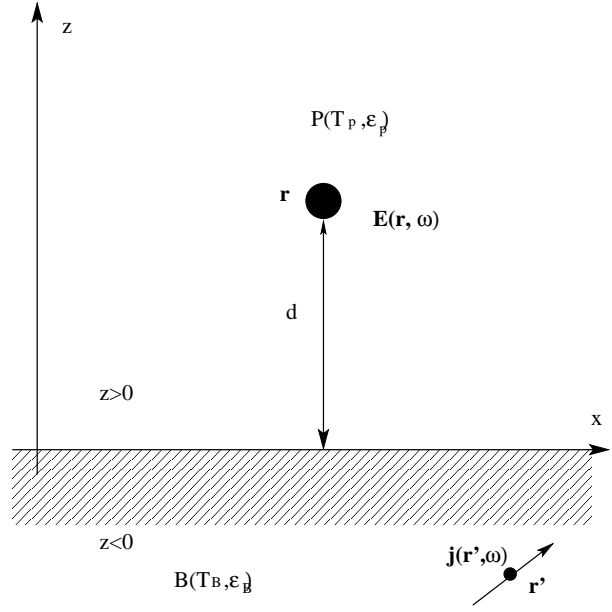}
\caption{Geometry of the system}
\label{systapl}
\end{center}
\end{figure}

We calculate in this part the radiative power exchanged between a small spherical particle and a semi-infinite medium. To this aim, we first calculate the power absorbed by the dielectric sphere placed above a heated half-space. We then calculate the power dissipated by the half-space situated below a heated sphere from reciprocity \cite{Mulet01}. The geometry of the problem is presented in Fig.\ref{systapl} : the upper medium $z>0$ is vacuum. A particle $(P)$ of radius $a$ and dielectric constant $\epsilon_P(\om)=\epsilon'_P(\om)+i\epsilon^{''}_P(\om)$ is held at temperature $T_P$. The lower medium is filled by a homogeneous, isotropic material (bulk) of dielectric constant $\epsilon_B(\om)=\epsilon'_B(\om)+i\epsilon^{''}_B(\om)$ and held at temperature $T_B$. The center of the particle is at a distance $d$ above the interface.

\subsubsection{Power absorbed by the bulk. Near-field limit}
A calculation following the procedures already introduced (see e.g.\cite{Volokitin01,Mulet01}), yields the power absorbed by the particle when illuminated by the field radiated by a half-space: 
\begin{equation}
\label{ }
P_{abs}^{B\rightarrow P}(\om)=\frac{2}{\pi}\frac{\om^4}{c^4}Im[\epsilon_B(\om)]Im[\alpha(\om)]
\Theta(\om,T_B)\sum_{n,m}\int_B|G_{nm}(\rv_P,\rv',\om)|^2d^3\rv'
\end{equation} 

If we consider the fluctuating currents inside the particle that radiates into the bulk and dissipates, one can calculate by the same formalism the power locally dissipated per unit volume at a point $\rv$ inside the bulk. It reads
\begin{equation}
\label{PtoB}
P_{abs}^{P\rightarrow B}(\rv,\om)=\frac{2}{\pi}\frac{\om^4}{c^4}Im[\epsilon_B(\om)]Im[\alpha(\om)]
\Theta(\om,T_P))\sum_{n,m}|G_{nm}(\rv,\rv_P,\om)|^2
\end{equation}
From the expression of the one-interface Green's tensor, it is possible to expand asymptotically the expression of the power absorbed by the particle. This quantity behaves as $1/d^3$ and reads
\begin{equation}
\label{ }
P_{abs}^{B\rightarrow P}(d,\om)\sim\frac{1}{4\pi^2d^3}\; 4\pi a^3\frac{3\epsilon_P^{''}(\om)}{|\epsilon_P(\om)+2|^2}
\frac{\epsilon_B^{''}(\om)}{|\epsilon_B(\om)+1|^2}\; \Theta(\om,T_B)
\end{equation}
From this expression, we see that there is an enhancement of the power absorbed if the denominators vanish or approach zero. We have seen that it is the case if the material support resonant surface waves so that the dielectric constant of the material can take negative values. We study in the next section the case of SiC.
\subsubsection{Example of SiC}
\begin{figure}[h]
\begin{center}
\includegraphics[width=3in]{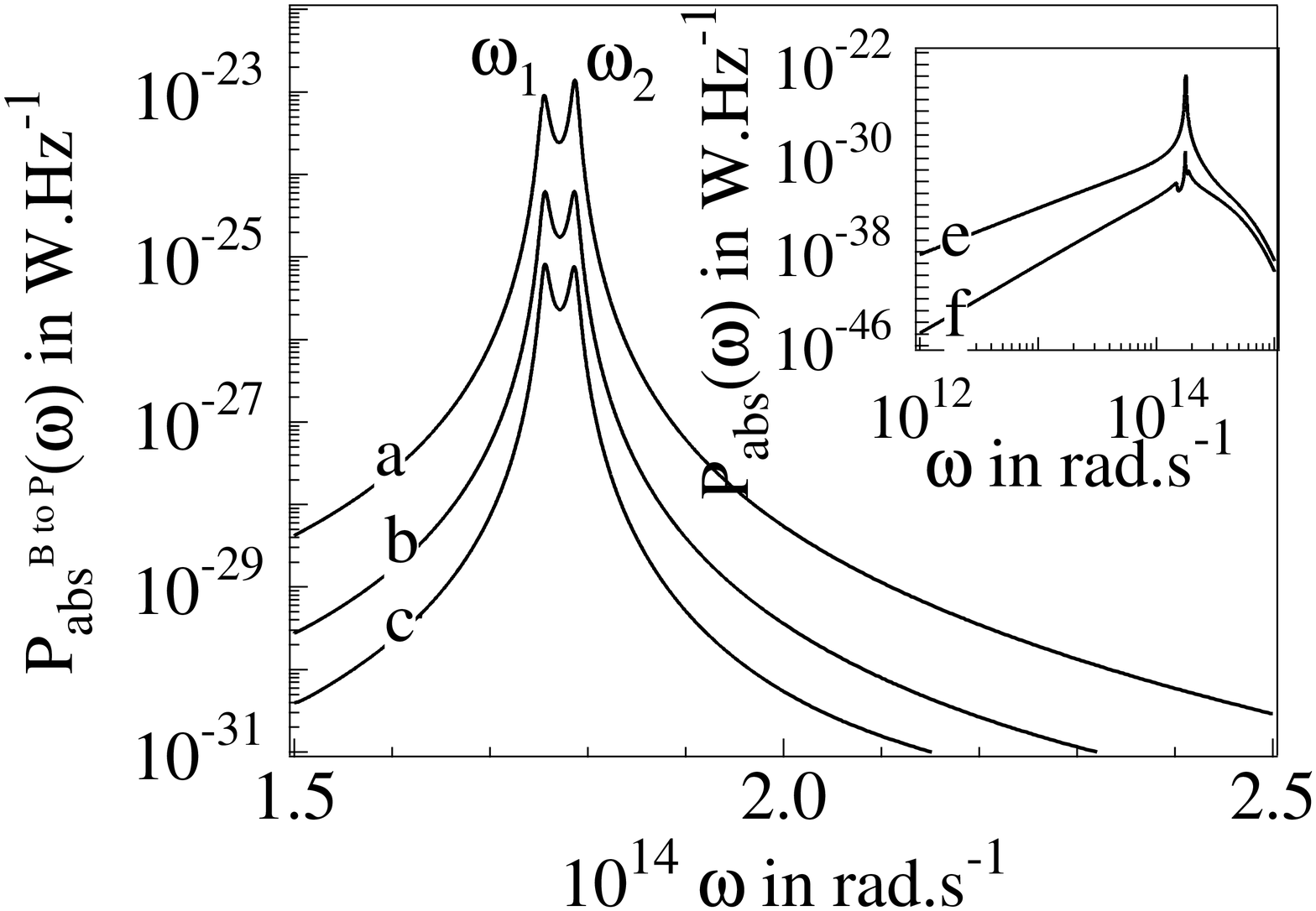}
\caption{Mean power radiated by the bulk (at $T_B=300$ K) and absorbed by the particle (of radius $a$=5 nm) versus frequency. (a) $d$=20 nm; (b) $d$=50 nm; (c) $d$=100 nm. The insert (log-log scale) shows the spectrum of the absorbed power between $10^{12}$ rad s$^{-1}$ and $10^{15}$ rad s$^{-1}$; (e) $d$=20 nm; (f) $d$=1mm. From \cite{Mulet01}
}
\label{fig:2apl}
\end{center}
\end{figure}

\begin{figure}[h]
\begin{center}
\includegraphics[width=3in]{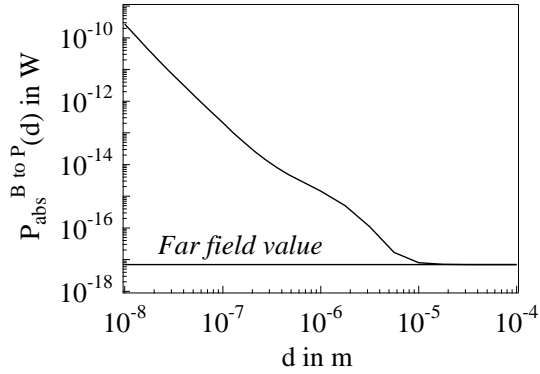}
\caption{Total power radiated by the bulk (at $T_B=300$ K) and absorbed by the particle (of radius $a=$ 5 nm) versus distance. From \cite{Mulet01}}
\label{fig:3apl}
\end{center}
\end{figure}

\begin{figure}[h]
\begin{center}
\includegraphics[width=3in]{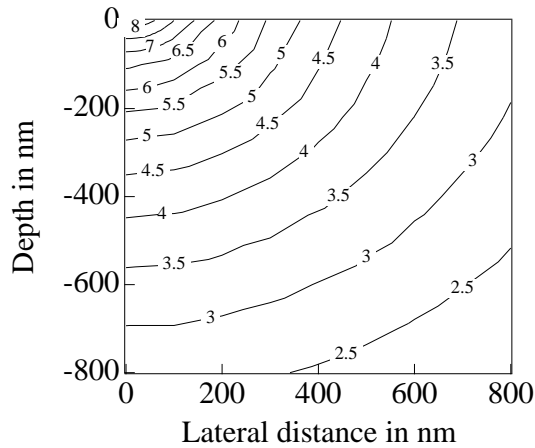}
\caption{Deposited power per unit volume inside the bulk. The particle has a radius $a=5$ nm and is held at temperature $T_P=300$ K. From \cite{Mulet01}}
\label{powerdep}
\end{center}
\end{figure}

As it has been said in the first part of the article, SiC is a polar material that can be described by an oscillator model (\ref{oscilsic}).
In Fig.\ref{fig:2apl}, we plot $P_{abs}^{B\rightarrow P}(\om)$ for a spherical particle held at temperature $T_P=300$ K, of radius $a=5$ nm at different distances above the surface. We note that the figure displays two remarkable peaks at frequency $\om_1\approx1.756\times10^{14}$ rad.s$^{-1}$ and $\om_2\approx1.787\times10^{14}$ rad.s$^{-1}$. These two peaks correspond to the resonances of the system. The first one corresponds to a frequency where $\epsilon_P(\om)$ approaches -2 : a volume phonon polariton is excited in the particle inducing a large electric dipole and a large dissipation. The second one is related to the resonant surface wave corresponding to a large increase of the electromagnetic LDOS. Thus, here too, the radiative heat transfer in the near field can be considered as monochromatic.
The electromagnetic waves associated with the resonant surface waves are evanescent. The energy transfer, which finds its origin in the presence of these waves, is important because the particle lies in the region (up to many micrometers) where the evanescent field is large, so that there is an efficient coupling between the field and the particle. In the far-field, evanescent waves are negligible and usual results are retrieved.
Fig.\ref{fig:3apl} shows the integrated power absorbed by the same particle versus the distance $d$. The near-field radiative heat transfer increases as $1/d^3$ (as it was suggested by the asymptotic behavior) and is larger at small distances by several orders of magnitude than the far field one. This enhancement comes from the contribution of evanescent waves.
Reciprocity requires that the same enhanced radiative heat transfer appears when the particle illuminates the surface. This situation may help to understand the radiative heat exchange between a nano-tip (like those used in near-field microscopy) and a sample. To answer this question, we calculated from (\ref{PtoB}) the total power (integrated over frequencies) dissipated per unit volume for different points in the sample. Fig.\ref{powerdep} displays a map, in log-scale, of the dissipation rate in the case of a 10nm-diameter sphere of SiC at $T_P=300$ K situated 100 nm above a sample of SiC. It is seen that the energy is dissipated on a scale comparable to the tip-sample distance. The dissipation per unit volume decreases very fast ($1/r^6$) with the distance $r$ between the source and the point of the sample where the dissipation is considered (the isocontour labeled with a '6' corresponds to the points where the dissipation per unit volume is 10$^6$ W m$^{-3}$. The amount of energy locally deposited is as large as 100 MW m$^{-3}$.
We note that the dependance of the heat deposited follows the same regime as in the two particles exchange in the vacuum. This is not surprising. Here also the phenomenon is due to an interaction between induced dipoles. This phenomenon is also at the origin of the force between macroscopic bodies at nanometric distances which is the subject of the next section.

\section{Role of surface electromagnetic waves on the Casimir force}
\subsection{Introduction}
After having considered the energy exchange due to the interaction in the near field between surface waves, it is natural to wonder what happens in terms of momentum exchange in the near field when two semi-infinite bodies are approached face to face. This situation is actually well known since 1948, when Casimir \cite{Casimir,Caspol} first showed the existence of an attracting force between two parallel perfect conductors. A large body of litterature has been devoted to this effect and several reviews are available \cite{Mostepanenko,Milonnibook,Plunien,Bordag,Lamoreaux,Milton}. The seminal paper of Lifshitz \cite{Lifshitz56} occupies a special place because it was the first calculation of this force by means of the FDT for the currents. Agarwal \cite{Agarwal2} reported a similar calculation using the FDT for the fields. There has been an increasing interest in the Casimir force since it has been shown that this force could be measured with high accuracy \cite{Lamoreaux97,Mohideen,Harris,Ederth,Bressi,Decca} and that it should be considered in the design of micro electromechanical systems (MEMS) \cite{Chan01a,Chan01b}.  Various correction to this force have been studied such as finite conductivity \cite{Lambrecht00} or temperature corrections \cite{Genet00,Klim01}. In what follows, we discuss the role of surface waves in the Casimir force. We will show that they play a key role in the short distance regime. 

To get a simple picture of Casimir force, we recall that a system of two plane parallel reflecting planes is a waveguide. The number of electromagnetic modes in the waveguide is discrete and depends on the thickness of the waveguide. From quantum electrodynamics, it is known that each mode with frequency $\omega$  has a minimum energy $\hbar \omega/2$ refered to as vacuum fluctuations. If the thickness decreases, the number of modes decreases so that the electromagnetic energy decreases.
Hence, the existence of vacuum fluctuations entails that there is an attractive force between the two plates. This phenomenon is clearly a macroscopic manifestation of the electromagnetic energy of the vacuum which is a pure quantum effect. Yet, its computation amounts to count the number of electromagnetic modes available and this is a pure classical problem. As has been discussed in the previous sections, the local density of electromagnetic states is completely dominated by the existence of surface modes. It follows that the role of surface waves is essential in the physics of the Casimir force in the short distance regime. 

In classical electrodynamics, the momentum transfer is given by the Maxwell stress tensor $T_{ij}$ \cite{Jackson}. The fields can be derived using the FDT \cite{Henkel04}. Using the system Green tensors for the electric and the magnetic field, it is possible to obtain this quantity. In the case of two semi-infinite bodies separated along the $z$ axis by a vacuum gap, the momentum flux reduces to the $zz$ component $T_{zz}$ of the Maxwell stress tensor given by
\begin{equation}
\label{ }
T_{zz}=\frac{\epsilon_0}{2}\left[|E_x|^2+|E_y|^2-|E_z|^2\right]
+\frac{\mu_0}{2}\left[|H_x|^2+|H_y|^2-|H_z|^2\right].
\end{equation}
One has to substract the infinite contribution to the force in the absence of bodies \cite{Schwinger78}. The force can be attractive or repulsive depending on the materials properties \cite{Milonni88,Hushwater}. One obtains an attractive force in the case of dielectrics \cite{Genet03b} and a force that might be repulsive in some configurations implying magnetic materials \cite{Kenneth,Buks02}.

\subsection{Spectrum of the force}

Lifshitz \cite{Lifshitz56} obtained a force per unit area given by
\begin{eqnarray}
     F &=&
     \int_{0}^{\infty}\!\frac{ {\rm d}\omega }{ 2\pi }
     \int_{0}^{\infty}\!\frac{ {\rm d} u }{ 2\pi } \,
     F(u,\omega)
     \label{eq:1a}
     \\
     F(u, \omega ) &=&
	- \frac{2\hbar
     \omega^3}{c^3}
     \,{\rm Im}\ 
     u v \sum_{\mu \,=\,{\rm s},\, {\rm p}}
     \frac{ r_{\mu}^2( u, \omega) \, {\rm e}^{ -2
     \frac{\omega}{c} v d } }{
     1 - r_{\mu}^2( u ,\omega ) \, {\rm e}^{ -2
     \frac{\omega}{c} v d } }
     ,
\label{eq:1a}
\end{eqnarray}
where $v = (u^2 - 1)^{1/2}$ (${\rm Im}\,v \le 0$), and $r_{\mu}$ is
the Fresnel reflection coefficient for a plane wave with polarization
$\mu$ and wavevector $K = \frac{\omega}{c} u$ parallel to the
vacuum-medium interface. We use he convention that an attractive force
corresponds to $F < 0$. The force appears as the contribution of elementary plane waves which angular frequency is $\om$ and which wavevector parallel to the interface is $u\om/c$. In his paper, Lifshitz used a deformation contour  in the complex plane  of frequencies to obtain a final formula where the summation over the wavevector is replaced by an integral over $v$ and the summation over the frequencies is replaced by an integral over the imaginary frequencies $\om=i\xi$. This approach has the advantage to replace the oscillating exponentials with smooth real functions  that make the integral easy to integrate numerically. Nevertheless, by doing such a deformation contour, one is losing the spectral information contained in the expression (\ref{eq:1a}). What we are going to show in the following is that the main contribution to the force in the near field is coming from the coupled polaritons of both interfaces. Therefore, we will see that there is a complete analogy in the interpretation of the momentum exchange with the interpretation of the energy exchange in terms of interaction of surface-polaritons. 

\begin{figure}[h]
     \vspace*{10mm}
\begin{center}
\includegraphics[width=6cm,angle=90]{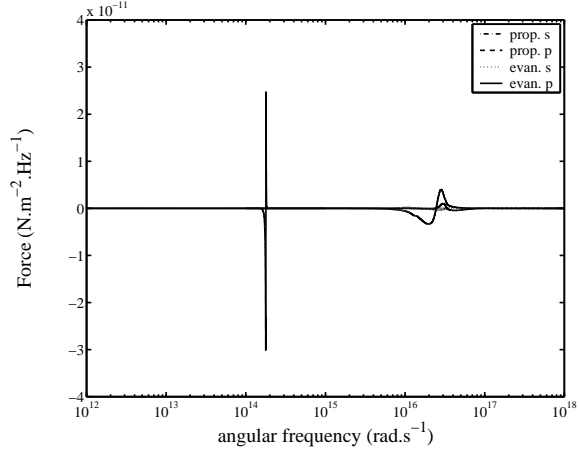}
\end{center}
     \caption[]{Contributions of $s$ and $p$-polarized, propagating and
     evanescent modes to the force spectrum (as given by ~(\ref{eq:1a})
    after integration over the wavevector $u$).  Distance $d =
     10$~nm. Material: SiC, dielectric function with two resonances.
     The angular frequencies of the corresponding surface resonances
     are $1.78\times 10^{14}\,{\rm s}^{-1}$ in the IR and $2.45\times
     10^{16}\,{\rm s}^{-1}$ in the UV~\cite{Palik}. From  \cite{Henkel04}}
\label{fig:1a}
\end{figure}

Let us study then the force spectrum in the case of two real materials. In the case of SiC (Fig.\ref{fig:1a}), the force is dominated by the UV and IR contributions. Actually, due to the presence of the $\om^3$ in the expression (\ref{eq:1a}), the UV contribution is much more important than the IR one.

\begin{figure}[h]
     \vspace*{10mm}
\begin{center}
\includegraphics[width=6cm,angle=90]{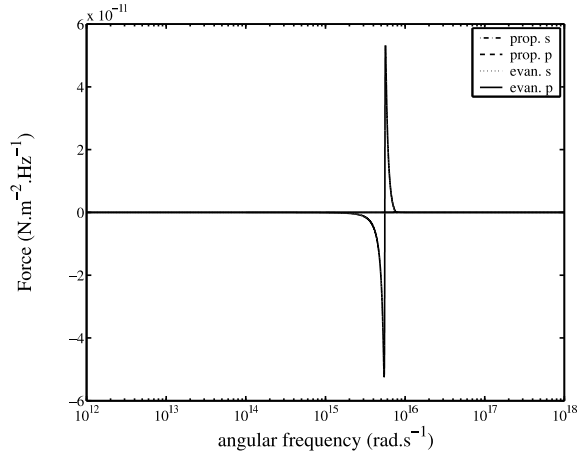}
\end{center}
     \caption[]{Contributions of $s$ and $p$-polarized, propagating and
     evanescent modes to the force spectrum (~(\ref{eq:1a})
     integrated on the wavevector $u$).  Distance $d =
     10$~nm. Material: Aluminum. described by tabulated optical data
\cite{Palik}. From  \cite{Henkel04}}
     \label{fig:1b}
\end{figure}
 Fig.\ref{fig:1b} represents the force spectrum between two aluminum half-spaces. It is seen that all the contributions are coming from frequencies very close to plasmon-polariton resonances. From expression (\ref{eq:1a}), it is easy to separate the polarisation contributions, the propagative contribution $(0\leq u\leq1)$ and the evanescent one $(u>1)$. We see in Figs.\ref{fig:1a},\ref{fig:1b} that the peaks exist only in $p$-polarisation and for evanescent waves. This fact is an additional argument in favor of an interpretation of the force being caused by the interaction between polaritons which are not propagating and only appear in $p$-polarisation.
An interpretation of the phenomenon is the following. At frequencies close to the system resonances i.e. close to the surface waves frequencies, the density of electromagnetic states increases. Thus, the amount of momentum carried in the gap increases too\footnote{As for energy transfer case, momentum is carried by evanescent waves if and only if an upward and a downward wave are present.}. This interpretation is only valid at distances of the order of the wavelength of the surface wave. At larger distances, the force is dominated by the propagative contribution as the original perfect conductors situation first pointed out by Casimir. We note also that the contribution to the force at such small distances has a different  sign whether the frequency is lower or higher than the resonant frequency. However, the total Casimir force between dielectrics is always positive and smaller than the one for perfect conductors.

\subsection{Binding and antibinding resonances}

\begin{figure}
\begin{center}
\includegraphics[width=3in]{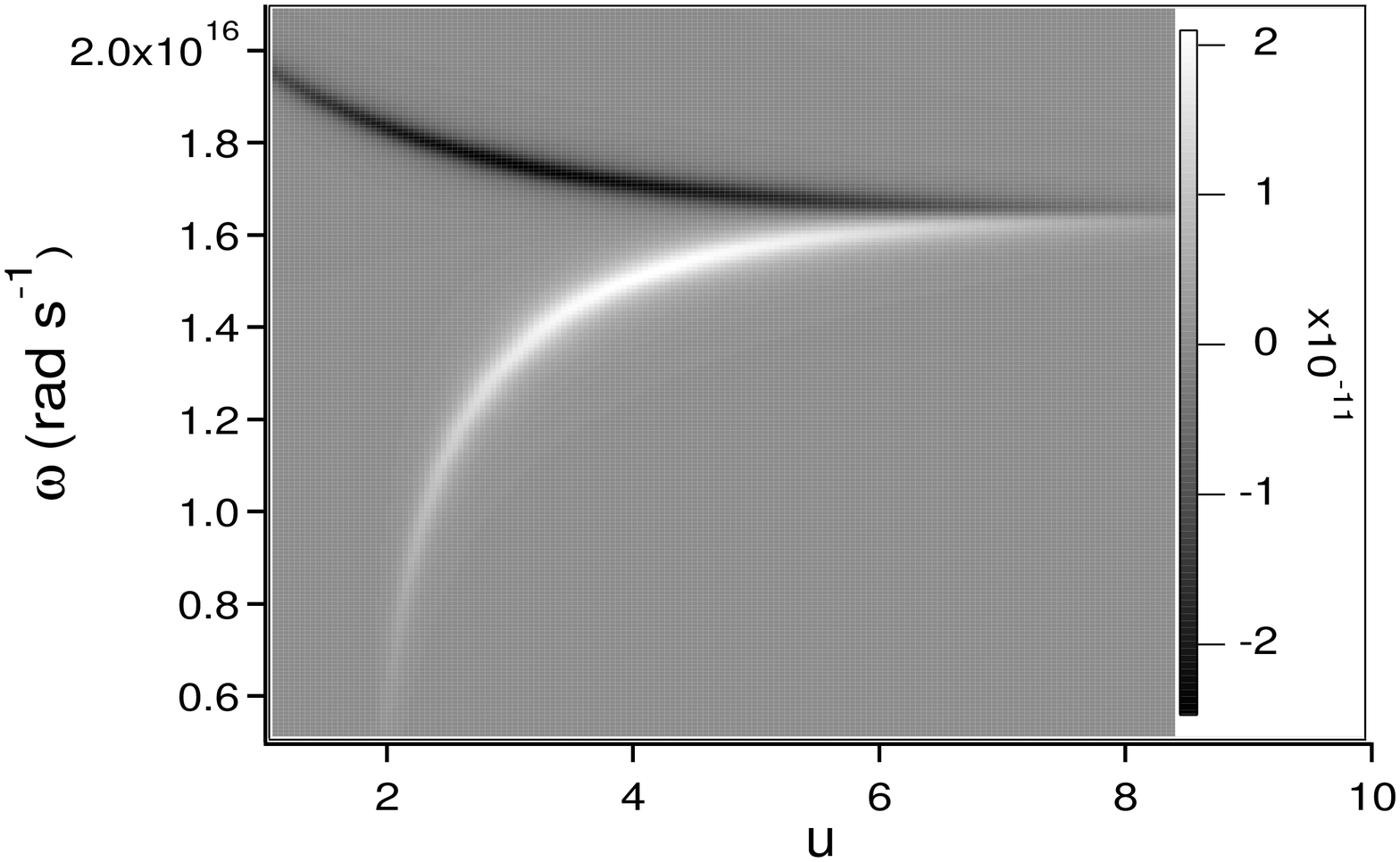}
\includegraphics[width=3in]{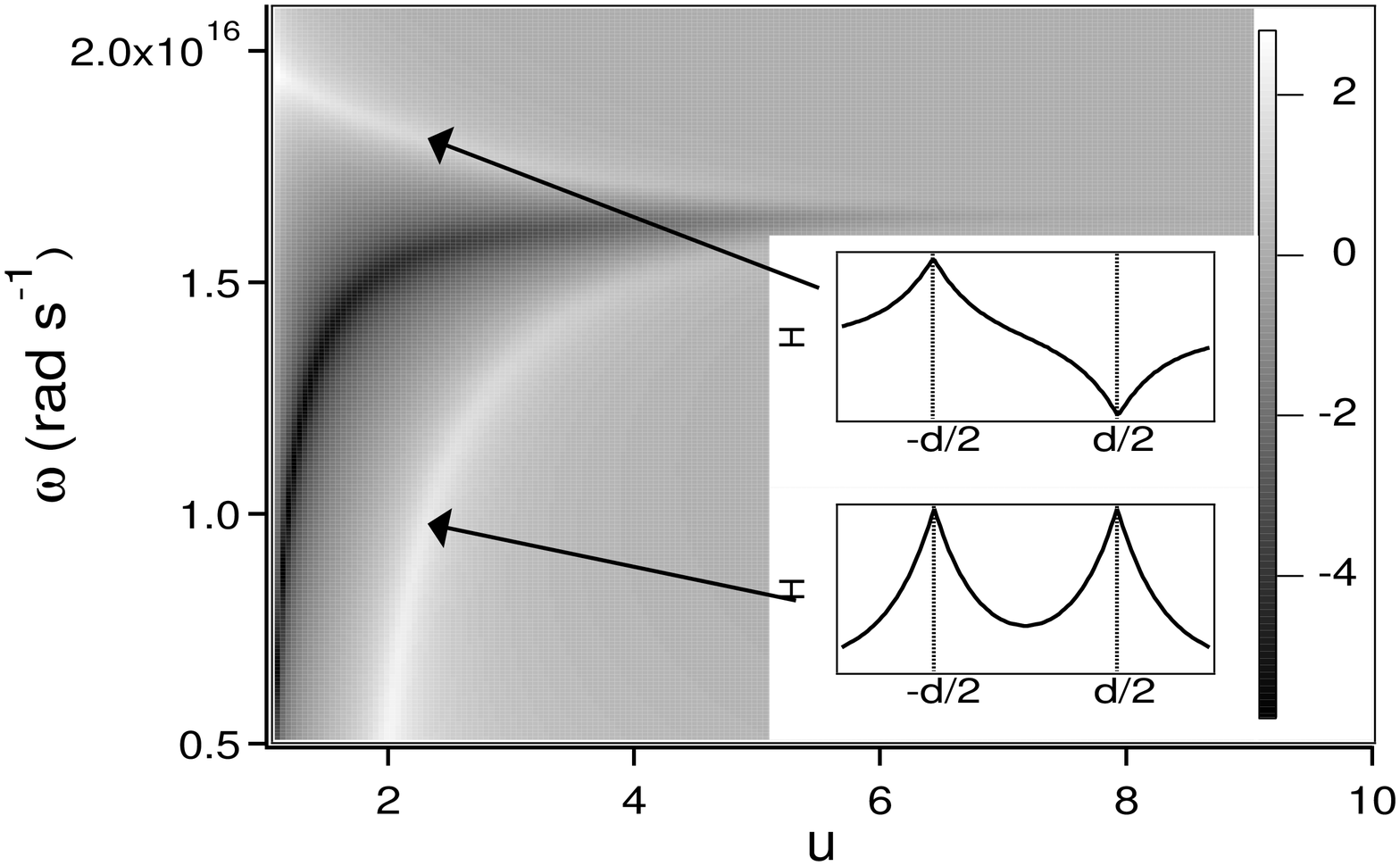}
\caption{(a) Wave-vector resolved spectrum of the Casimir Force  (\ref{eq:1a}) in the $(u,\om)$
 plane between two aluminum half-spaces separated by a distance of 10 nm. The frequency of the flat asymptote corresponds to the peaks of the force spectrum (Fig.\ref{fig:1b}). Light (dark) areas : attractive (repulsive) force. (b) Resonant denominator $|1/1-r_{\rm p}^{2}e^{-2 \frac{\omega}{c} v d}|^2$ in the $(u,\om)$ plane, the grayscale giving the logarithm to base 10. The dispersion relation of the coupled surface resonnance correspond to light areas; dark area : dispersion for a single interface (\ref{dispvac}). The dielectric function is extracted from tabulated data \cite{Palik}. The inset sketches the magnetic field of the coupled surface resonances (antisymmetric and symmetric combinations). From \cite{Henkel04}}
 \label{fig:4a}
\end{center}
\end{figure}

The role of SPP can be further analysed by studying the variation of the integrand of the force $F(u,\om)$ in the plane $(u,\om)$. Close to the resonant surface-wave frequency $\om_{SW}$, the contributions to the force come from evanescent waves. We therefore limit our study to the case $u>1$ and close to $\om\sim\om_{SW}$. In Fig.\ref{fig:4a}a, we plot the integrand $F(u,\om)$ for two aluminum half-spaces separated by a distance $d=10$ nm. Two branches mainly contribute, the higher frequency branch yielding a repulsive contribution whereas the lower one gives an attractive contribution. These two branches actually follow the two-interfaces system dispersion relation given by
\begin{equation}
1-r_{\rm p}^{2}e^{-2 \frac{\omega}{c} v d}=0
\label{dispslab}
\end{equation}

\begin{figure}
\begin{center}
\includegraphics[width=3 in]{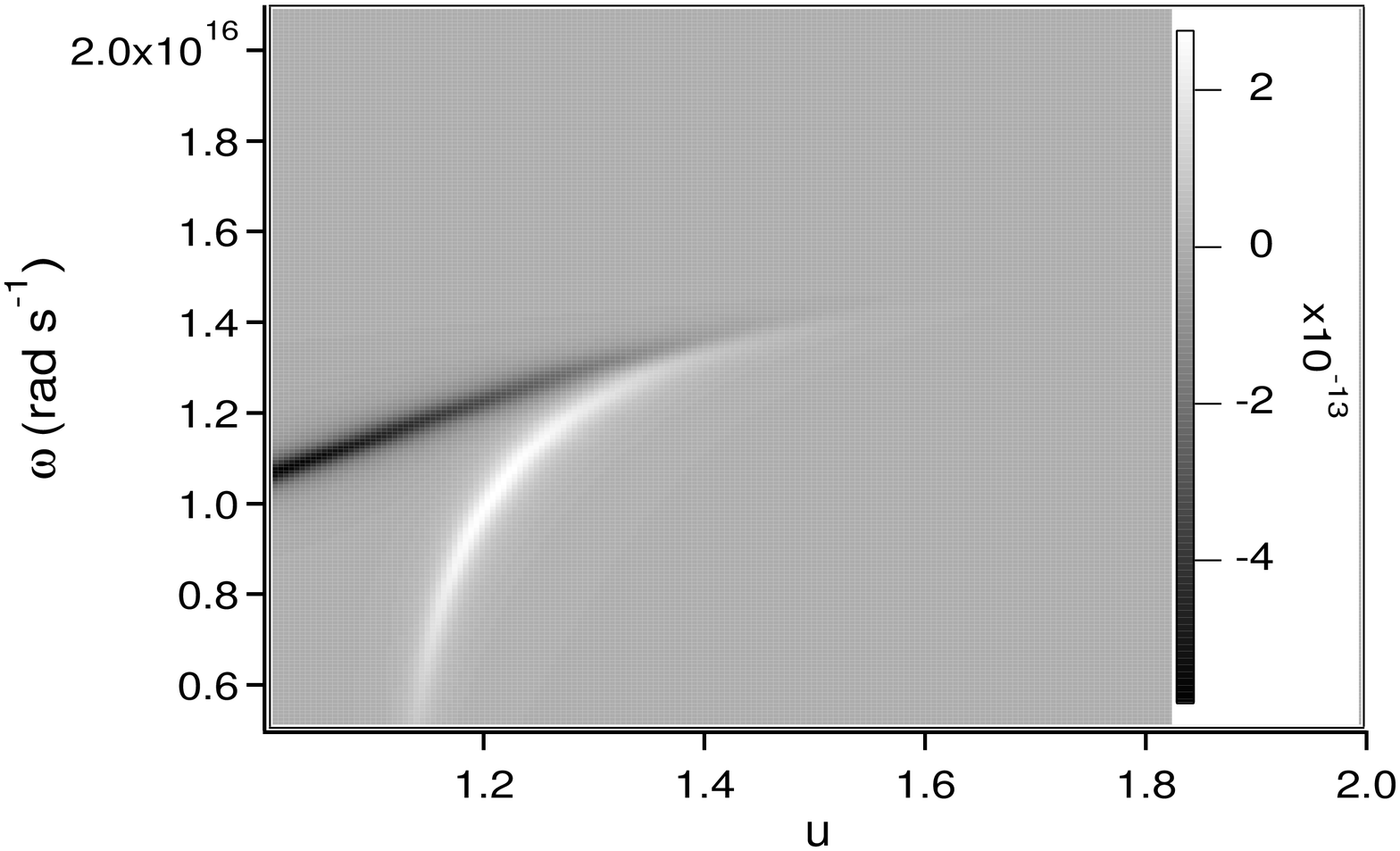}
\includegraphics[width=3in]{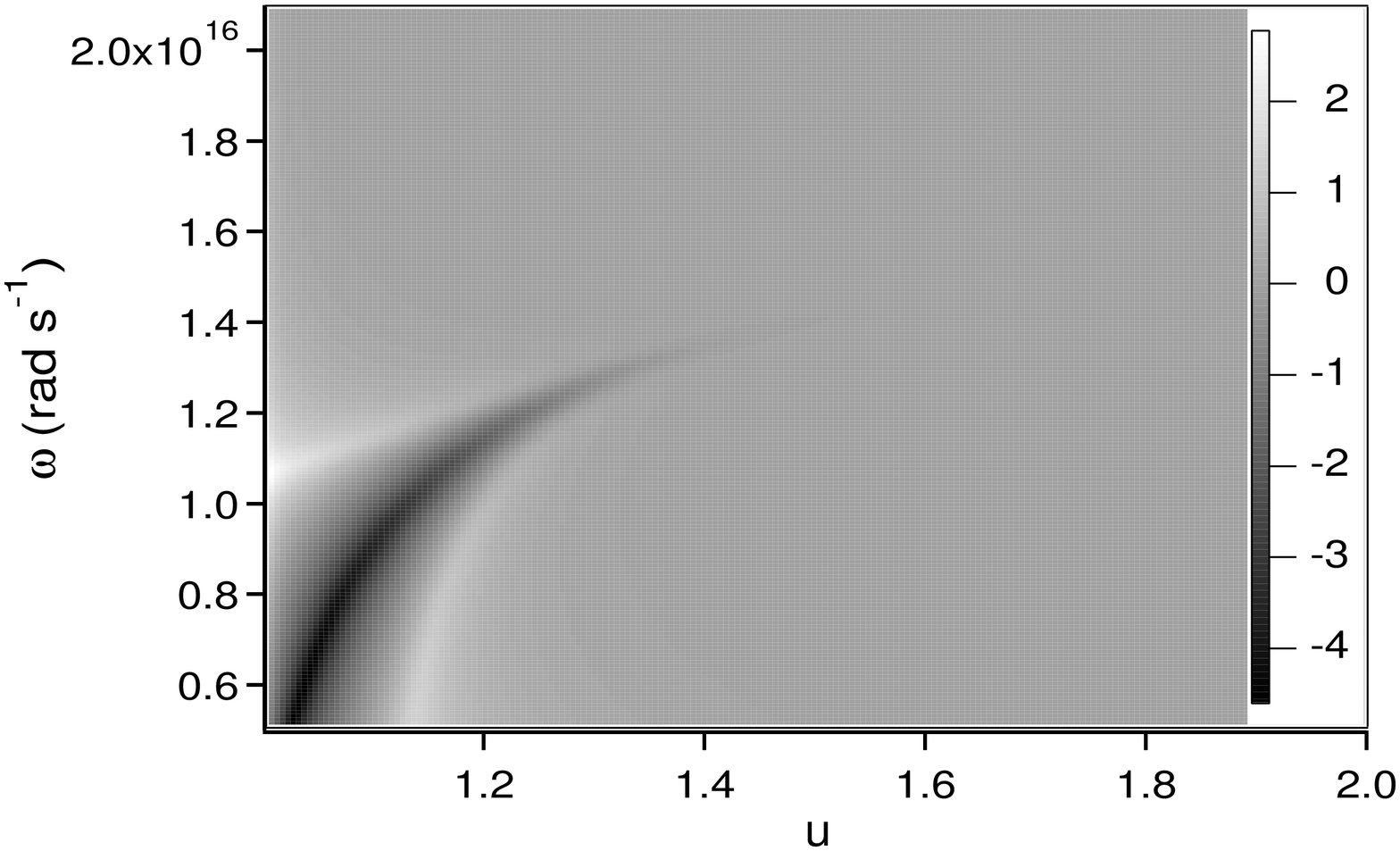}
     \caption{Same as Fig.\ref{fig:4a} but for $d=100$ nm. From  \cite{Henkel04}}
     \label{fig:5a}
\end{center}
\end{figure}

The influence of the dispersion relation on the force is illustrated in Fig.~\ref{fig:4a}b. In this figure, the quantity $1/|1-r_{\rm p}^{2}e^{-2
\frac{\omega}{c} v d}|^{2}$ is plotted in the
$(u,\omega)$ plane. Upon comparison between Fig.~\ref{fig:4a}b and
Fig.~\ref{fig:4a}a, it is clearly seen that the main contribution to
the force can be attributed to the SPP.  In addition, we observe on
Fig.~\ref{fig:4a}b a dark line which corresponds to minima of
$1/|1-r_{\rm p}^{2}e^{-2 \frac{\omega}{c} v d}|^{2}$.  The minima can
be attributed to very large values of the reflection factor of
a plane interface $r_{\rm p}$.  Thus, the dark line is
the dispersion relation of a single SPP on a flat interface.
In Fig.\ref{fig:5a}, the integrand is plotted for two aluminum half-spaces separated by a distance $d=100$ nm : the two branches tend to merge with the flat interface dispersion relation. One can thus propose the following interpretation : when the surfaces approach each other, the overlapping of the two SPP leads to two coupled modes and to a splitting of the polaritons frequencies \cite{Marcuse,Sarid}. The frequency splitting can be found from the solutions of (\ref{dispslab}) which are implicitely defined by
$$
r_{\rm p}(u,\om)=\pm e^{\om v d/c}.
$$
The signs correspond to either symmetric or antisymmetric mode functions (for the magnetic field), as sketched in the inset of Fig.\ref{fig:5a}. The symmetric (antisymmetric)
branch corresponds to a lower (higher) resonance frequency,
respectively, similar to molecular orbitals and tunneling doublets.
These branches contribute with opposite signs to the Casimir force,
due to the following identity:
\begin{eqnarray}
     &&\frac{ 2 \,r_{\rm p}^2( \omega, u ) \, {\rm e}^{ -2
     \omega v d /c} }{
     1 - r_{\rm p}^2( \omega, u ) \, {\rm e}^{ -2
     \omega v d/c } }
     =
     \nonumber\\
     &&
     \frac{ r_{\rm p}( \omega, u ) \, {\rm e}^{ -
     \omega v d /c} }{
     1 - r_{\rm p}( \omega, u ) \, {\rm e}^{ -
     \omega v d/c } }
-
     \frac{ r_{\rm p}( \omega, u ) \, {\rm e}^{ -
     \omega v d /c} }{
     1 + r_{\rm p}( \omega, u ) \, {\rm e}^{ -
     \omega v d /c} }
     ,
\label{eq:4l}
\end{eqnarray}
where the first (second) term is peaked at the symmetric
(antisymmetric) cavity mode. The symmetry of the resonance mode
function hence determines the attractive or repulsive character of its
contribution to the Casimir force. 

\subsection{Analytical formulation of the short-distance limit}

Using a simple Lorentz-Drude model for the dielectric function, 
\begin{equation}
\label{Drude}
\epsilon(\om)=1+\frac{2(\Omega^2-\omega_0^2)}{\Omega^2-i\gamma\om-\om^2},
\end{equation}
 we can derive a simple analytic form of the force in the near-field limit that accounts for the modification of the local density of states due to the surface waves and that takes into account absorption effects.  This model describes either dielectric or metals depending on the value of $\om_0$. The corresponding plasma frequency is $\sqrt{2(\Omega^2-\omega_0^2)}$. With this convention, the large $u$ asymptote of the SPP dispersion (\ref{dispvac}) occurs at  $\om\approx\Omega$.  As the distance $d$ reduces to a quantity small compared to the wavelength, the evanescent contribution to the force comes from higher and higher parrallel wave vector. When $u\gg 1$, the reflection coefficient in $p$-polarisation can be approximated by:
\begin{equation}
     u \gg 1: \qquad
r_{\rm p}( \omega, u ) \approx
\frac{ \Omega^2-\omega_0^2}{ \Omega^2 - {\rm i} \gamma \omega - \omega^2 }.
\label{eq:rp-estat}
\end{equation}
It is seen that the reflection factor has a complex pole. From (\ref{dispslab}), we thus get the following dispersion relation
for the (anti)symmetric surface-plasmon resonances, neglecting for the
moment the damping coefficient~$\gamma$:
\begin{equation}
     \omega_{\pm}^2 \approx \Omega^2  \mp
     {\rm e}^{ - \omega_{\pm} u d/c}\left(\Omega^2-\om_0^2 \right),
\end{equation}
where we have used $v \approx u$ for $u \gg 1$. For large $u$, we solve by
iteration and find that $\omega_{\pm}\,
\raisebox{0.65ex}{$<$}\kern-1.7ex\raisebox{-0.65ex}{$>$}\,
\Omega$. As announced above, the symmetric mode thus occurs at a
lower resonance frequency.

To derive an analytical estimate for the Casimir force, we retain in
~(\ref{eq:1a}) only the contribution of $p$-polarized, evanescent
waves. The final result\cite{Henkel04} contains two contributions:
\begin{equation}
     F =  \frac{ \hbar \Omega }{ 4\pi d^3 }
\left( \alpha(z) -
     \frac{ \gamma \,\textnormal{Li}_3(z^2) }{ 4\pi\Omega }
     \right),
     \label{eq:7a}
\end{equation}
where
\begin{equation}
     \alpha(z) = \frac14 \sum_{n=1}^{\infty}z^{2n}
     \frac{(4n-3)!!}{n^3 (4n-2)!!} 
\end{equation}
and
\begin{equation}
     \textnormal{Li}_3(z^2) = \sum_{n=1}^{\infty} \frac{ z^{2n} }{ n^3 } 
\end{equation}
When $(z\rightarrow1)$, we get the following asymptotics
\begin{equation}
\label{ }
\alpha(z)\approx 0.1388-0.32(1-z)+0.4(1-z)^2
\end{equation}
\begin{equation}
\label{ }
\textnormal{Li}_3(z^2)\approx\zeta(3)-\frac{\pi^2}{3}(1-z)+\left[3-\frac{\pi^2}{6}-2\ln[2(1-z)]\right]\times(1-z)^2
\end{equation}
with $\zeta(3)\approx1.202$. For a metal, the parameter $z$ takes the value $1$ so that we have $\alpha(1)\approx 0.1388$ and $\textnormal{Li}_3\approx 1.202$.
\begin{figure}
\begin{center}
\includegraphics[width=6cm]{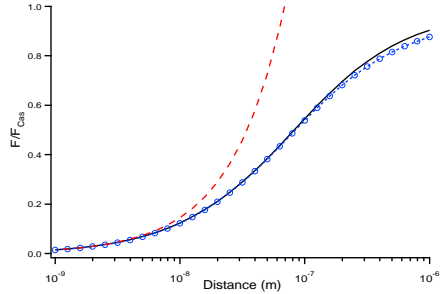}
\caption{Comparison of different expressions for the Casimir force between aluminum surfaces.
We plot the ratio $F(d)/F_{Cas}(d)$, where $F_{Cas}(d)=\hbar c \pi^2/(240 d^4)$ is the Casimir Force for perfect mirrors. Solid line: numerical integration of (\ref{eq:1a}), using tabulated data \cite{Palik}. Short-dashed line with circles : same, with a model dielectric function of Drude form (\ref{Drude}) with $\om_0=0$, $\Omega=1.66\times10^{16}$ s$^{-1}$, and $\gamma/\Omega=0.036$. These parameters have been derived by fitting the value of the reflectivity. Long-dashed line: short distance asymptotics (\ref{eq:7a}) with the same values for $\om_0$, $\Omega$ and $\gamma$. From \cite{Henkel04}}
\label{fig:6a}
\end{center}
\end{figure}
We compare (\ref{eq:7a}) in Fig.\ref{fig:6a} to the full integral (\ref{eq:1a}) for the case of aluminum : the asymptotic estimation turns out to be quite accurate for distances $d\leq 0.1\lambda_{SPP}$ where $\lambda_{SPP}$ is the wavelength of the SPP with the largest frequency. In the case of aluminum,  the first order correction in $\gamma/\Omega$ is 2.5\% of the zeroth order value of the force. The plot also shows that for numerical integration, the tabulated data\cite{Palik} and the Lorentz-Drude model (\ref{Drude}) with parameters fitted around the surface resonance give very close results over a large range of distances. This is another indication that the short range Casimir force is dominated by a narrow frequency range. 

\subsection{Friction forces}
The Casimir force that has been treated in the previous section is not the only one  that occurs when two bodies are approached one to each other. A particle moving in a vacuum experiences a friction force proportional to its velocity as first discussed by Einstein \cite{Milonnibook,Einstein1,Einstein2,Milonni,Saslow}. Again, the friction force arises from the fluctuations of the electromagnetic field. A major difference between friction forces and Casimir force is that the leading contributions come from the low frequencies for friction. By contrast, high frequencies contribution gives the largest contribution to vacuum energy and therefore gives the leading contribution to the Casimir force. The friction force has been studied recently for a particle in close proximity to an interface. This force has been used to develop a near field imagery technique called shear-force microscopy \cite{Betzig,Vaez-Iravani,Karrai1}.  The exact origin of these forces is still unclear. Recently, experiments on shear force \cite{Karrai,Stipe} have been conducted in ultra-high vacuum in order to eliminate forces due to a water monolayer for instance.  
Different models have been developped to explain the origin of the friction force between two parallel surfaces \cite{Pendry97,Teodorovitch,Levitov,Barton,Volokitin} in the framework of fluctuational electrodynamic fields. The case of a particle moving parallel to a surface has been considered in \cite{Schaich,Tomassone,Kyasov}. The friction force for particles moving perpendicular to the surface has also been considered experimentally \cite{Dorofeyev99} and theoretically \cite{Dorofeyev01}. Volokitin and Persson \cite{Volokitin03PRL,Volokitin03} have shown that the friction force is  much larger for the perpendicular case. The problem is still open because the discrepancy between theory and experiments is very large. Recently, it has been proposed \cite{Zurita} that the friction observed in the experiment by Stipe \textit{et al.} could be due to the dielectric located below the gold film. Volokitin and Persson \cite{Volokitin03PRL} have suggested that adsorbates might produce a very large enhancement of the friction forces. Clearly, more experiments are needed to clarify this issue.

\section{Concluding remarks}

Many years after the discovery of surface polaritons, new discoveries and effects are still being reported. A major reason is the development of near-field techniques that allows to probe the properties of surfaces with a nanometric resolution and motivates further work. Although thermal excitation of surface waves had been studied in the past, their major role in many phenomena has been realized only recently. It has been shown that heat transfer is dramatically enhanced in the near field due to the resonant contribution of surface waves. It has been shown that the electromagnetic field emitted by a thermal source may be partially temporally coherent and partially spatially coherent in the near field. It has also been shown that the far field emission properties of surfaces can be engineered by exciting and coupling efficiently surface waves. Partially spatially coherent thermal sources have been realized. Applications to emitting light devices are under study. The role of surface excitations in the Casimir force has been clarified and the agreement between experiments and models is now satisfactory. The situation for friction forces is far less clear. There are still large discrepancies between published data and models. 

{\bf Acknowledgement}

The authors are very grateful to Carsten Henkel and Lukas Novotny for many fruitful discussions. 

\appendix{}
\section{Green's tensor }\label{AppGT}

\subsection{Green's tensor in a vacuum}
The Green's tensor in a vacuum is defined by the relation:
\begin{equation}
\mathbf{E}(\rv)=\mu_0 \om^2 \G(\rv,\rv') \mathbf{p}.
\end{equation}
Its explicit form expression is given by:
\begin{eqnarray}
\label{Greentensor}
\G(\rv,\rv ',\om)&=&\frac{ke^{ikR}}{4\pi}\left[\left(\frac{1}{kR}+\frac{i}{(kR)^2}-\frac{1}{(kR)^3}\right)
\Id\right.\nonumber\\
&+&\left.(\uv_r\ \uv_r)\left(\frac{3}{(kR)^3}-\frac{3i}{(kR)^2}-\frac{1}{kR}\right)\right]
\end{eqnarray}
where $R=|\rv-\rv'|$, $\Id$ is the identity tensor and $\uv_r \uv_r$ is a tensor in dydadic notation such that $(\uv_r\ \uv_r) \mathbf{A}=\uv_r (\uv_r\cdot\mathbf{A})$.

\subsection{Green's tensor above an interface}
For the plane interface system, it is convenient to use the representation due to Sipe \cite{Sipe} that consists in a decomposition over elementary plane waves. We use again the dyadic notation for the tensors. For instance, the s-component of the electric field is given by $\shat \shat \mathbf{E}=\shat (\shat\cdot\mathbf{E})$.  In the case of a Green tensor relating currents in the lower half space ($\rv'$ in medium 2) to a field in the upper half space ($\rv$ in medium 1), one has:
\begin{equation}
\label{gelec}
\G^{EE}(\rv,\rv',\om)=\frac{i}{2}\int \frac{d^2\K}{4\pi^2}\frac{1}{\gamma_2}\left[\shat t_{21}^s\shat
+\pup t_{21}^p\pdp\right]e^{i\K.(\R-\R')}e^{i\gamma_1z-i\gamma_2z'}
\end{equation}
and
\begin{equation}
\label{gmagn}
\G^{HE}(\rv,\rv',\om)=\frac{k_0n_1}{2}\int \frac{d^2\K}{4\pi^2}\frac{1}{\gamma_2}\left[\pup t_{21}^s\shat
-\shat t_{21}^p\pdp\right]e^{i\K.(\R-\R')}e^{i\gamma_1z-i\gamma_2z'}
\end{equation}
In the expression $\shat=\K\times\zu/|K|$ and $\hat{p}_i^\pm=-[\gamma_i\K/|K|\mp K\zu]/(n_ik_0)$. The transmission factors are defined by: 
\begin{equation}
t_{21}^p=\frac{2 n_1n_2\gamma_2}{\epsilon_1\gamma_2+\epsilon_2 \gamma_1}\quad ; \quad t_{21}^s=\frac{2 \gamma_2}{\gamma_1 +\gamma_2}
\end{equation}
Note that the transmission factor for $p$-polarization has a pole that corresponds to the surface wave. Thus the Green's tensor contains all the information on surface waves. 

In the case of two points lying above the interface in medium 1, the tensor can be cast in the form:

\begin{eqnarray}
\G^{EE}(\rv,\rv',\om) & = & \frac{i}{2}\int\frac{d^2\K}{4\pi^2\gamma_1}\left[\shat r_{12}^s\shat
+\pup r_{12}^p\pum\right] e^{i\K.(\R-\R')}e^{i\gamma_1(z+z')} \\
\G^{HE}(\rv,\rv',\om) & = & \frac{k_0n_1}{2}\int\frac{d^2\K}{4\pi^2} \frac{1}{\gamma_1} \left[\pup r_{12}^s \shat
-\shat r_{12}^p\pum\right] e^{i\K.(\R-\R')}e^{i\gamma_1(z+z')}.
\end{eqnarray}

where the reflection factors are :
 \begin{equation}
r_{12}^p=\frac{-\epsilon_1\gamma_2+\epsilon_2\gamma_1}{\epsilon_1\gamma_2+\epsilon_2 \gamma_1}\quad ; \quad r_{12}^s=\frac{ \gamma_1-\gamma_2}{\gamma_1 +\gamma_2}
\end{equation}

\subsection{Green's tensor for a two-interfaces system}

We now consider a two layers system. The upper medium is denoted 1 and lies above $z=d$. The lower medium ($z<0$) is medium 2.  Medium 3 is defined by $d>z>0$. The Green's tensor relating the currents in medium 2 to the field in medium 1 with a film of medium 3 between media 1 and 2 are given by:
\begin{equation}
\label{ }
\G^{EE}(\rv,\rv',\om)=\frac{i}{2}\int\frac{d^2\K}{4\pi^2}\frac{1}{\gamma_2}\left(\shat t_{12}^s\shat+\pup t_{12}^p\pdp\right)
e^{i[\K.(\R-\R')}e^{i[\gamma_1( z-d) -\gamma_2 z']},
\end{equation}
and
\begin{equation}
\label{ }
\G^{HE}(\rv,\rv',\om)=\frac{k_0n_2}{2}\int\frac{d^2\K}{4\pi^2}\frac{1}{\gd}\left(\pup t_{21}^s\shat-\shat t_{21}^p\pdp\right)
e^{i[\K.(\R-\R')]}\e^{i[\gu (z-d)-i\gd z']},
\end{equation}
where
\begin{equation}
\label{ }
t_{21}^{s,p}=\frac{t_{31}^{s,p}t_{23}^{s,p}e^{i\gt d}}{1-r_{31}^{s,p}r_{32}^{s,p}e^{2i\gt d}}.
\end{equation}
Note that the two-interfaces Green's tensor is very similar to the single interface one, except that the single interface transmission coefficient has to be replaced by a generalised transmission coefficient taking into account the multiple reflections.

When $\rv$ and  $\rv'$ are in the film (medium 3), the Green's tensor can be cast in the form :
\begin{equation}
\G^{EE}(\rv,\rv',\om)=\int\frac{d^2\K}{4\pi^2} \g^{EE}(\K,z,z')\exp[i\K(\R-\R')],
\end{equation}

where $\g^{EE}(\K,z,z')$ is the sum of three contributions :
\begin{eqnarray}
\g^{EE}(\K,z,z')=&\frac{i}{2\gamma_3}[\shat\shat+\hat{p}\hat{p}]\exp[i\gamma_3\vert z-z'\vert]-\frac{1}{k_0^2\epsilon_3}\delta(z-z')\zu\zu \nonumber \\
&+ \frac{i}{2\gamma_3}[\shat \rho^s_{32}\shat+\hat{p}_3^{+}\rho_{32}^p\hat{p}_3^{-}]\exp[i\gamma_3(z+z')] \nonumber \\
&+ \frac{i}{2\gamma_3}[\shat \rho^s_{31}\shat+\hat{p}_3^{-}\rho_{31}^p\hat{p}_3^{+}]\exp[i\gamma_3(d-z+d-z')],
\end{eqnarray}
where 
\begin{equation}
\rho_{31}^{s,p}=\frac{r_{31}^{s,p}}{1-r_{31}^{s,p}r_{32}^{s,p}\e^{i2\gamma_3d}} \quad ; \quad
\rho_{32}^{s,p}=\frac{r_{32}^{s,p}}{1-r_{31}^{s,p}r_{32}^{s,p}\e^{i2\gamma_3d}}.
\end{equation}

\section{Fluctuation-dissipation theorem}\label{AppFD}
In this section, we derive the apropriate spectral density for an absorption measurement and for a quantum counter measurement. The operator needed to describe the absorption of a photon involves the normally ordered correlation function of the field operator $<E_k^{(-)}E_l^{(+)}>$ where $E^{(+/-)}$ is defined using only the positive or negative frequencies of the spectrum. We quote from \cite{Agarwal} the cross-spectral density of the normally and antinormally ordered electric fields:
\begin{eqnarray}
\label{FDTEapp}
\mathcal{E}^{(N)}_{kl}(\rv,\rv',\omega)&=&\eta(-\om)\mu_0\hbar\om^2[1+\coth(\hbar\omega/2k_{B}T)] Im[G_{kl}^{EE}(\rv,\rv',\om)] , \\
\mathcal{E}^{(A)}_{kl}(\rv,\rv',\omega)&=&\eta(\om)\mu_0\hbar\om^2[1+\coth(\hbar\omega/2k_{B}T)] Im[G_{kl}^{EE}(\rv,\rv',\om)],
\end{eqnarray}
where $\eta(\om)$ is the Heaviside function. It is seen that the normally ordered cross-spectral density involves only negative frequencies whereas the antinormally ordered correlation involves only positive frequencies. They can be viewed as an analytic signal. Despite the apparent symmetry of the above equations, there is a critical difference as will be seen in the next subsection.

\subsection{Absorption measurement}

Let us now derive the time correlation function for the electric field. From \cite{Agarwal}, we have
\begin{equation}
\left<E_k(\rv,t+\tau)E_l(\rv',t)\right>=2Re[\left<E^{(-)}_k(\rv,t+\tau)E^{(+)}_l(\rv',t)\right>]
\end{equation}
Using the cross-spectral density (\ref{FDTEapp}), we get
\begin{equation}
\left<E_k(\rv,t+\tau)E_l(\rv',t)\right>=2Re\left[ \int_{-\infty}^0\frac{d\om}{2\pi}\exp(-i\om\tau)\mu_0\hbar\om^2 Im[G^E_{kl}] [1+\coth(\hbar\om/2k_{B}T)] \right].
\end{equation}
We note that 
\begin{equation}
Im[G_{kl}^E(\rv,\rv',-\om)]=-Im[G^E_{kl}(\rv,\rv',\om)],
\end{equation}
and 
\begin{equation}
 [1+\coth(\hbar\omega/2k_{B}T)]=-2/[\exp(\hbar\vert\om\vert/k_{B}T)-1]
 \end{equation}
so that the correlation function can be cast in the form:
\begin{equation}
\left<E_k(\rv,t+\tau)E_l(\rv',t)\right>=2Re\left[ \int_{0}^{\infty}\frac{d\om}{2\pi}\exp(i\om\tau)2\mu_0\om Im[G^{EE}_{kl}]\Theta(\om,T) \right].
\end{equation}

\subsection{Quantum counter measurement}

We derive now the time correlation starting from the antinormally ordered correlation function. This choice is apropriate for a quantum counter experiment. The Casimir effect that depends on the total energy of the system pertains to this category.  From \cite{Agarwal}, we have
\begin{equation}
\left<E_k(\rv,t+\tau)E_l(\rv',t)\right>=2Re[\left<E^{(+)}_k(\rv,t+\tau)E^{(-)}_l(\rv',t)\right>]
\end{equation}
Using the cross-spectral density (\ref{FDTEapp}), we get
\begin{equation}
\left<E_k(\rv,t+\tau)E_l(\rv',t)\right>=2Re\left[ \int_{0}^{\infty}\frac{d\om}{2\pi}\exp(-i\om\tau)\mu_0\hbar \om^2Im[G^{EE}_{kl}][1+\coth(\hbar\om/2k_{B}T)] \right].
\end{equation}
We note that 
\begin{equation}
 [1+\coth(\hbar\omega/2k_{B}T)]=2+\frac{2}{\exp(\hbar\om/k_{B}T)-1},
 \end{equation}
so that the correlation function can be cast in the form:
\begin{equation}
\left<E_k(\rv,t+\tau)E_l(\rv',t)\right>=2Re\left[ \int_{0}^{\infty}\frac{d\om}{2\pi}\exp(-i\om\tau)2\mu_0\om Im[G^{EE}_{kl}][\hbar\om+\Theta(\om,T)] \right].
\end{equation}

We conclude by noting that both results have a similar structure and can be described by an effective  spectrum defined for positive frequencies only.  This analysis provides a justification for the heuristic argument often used to drop the vacuum energy fluctuation.

\bibliographystyle{elsart-num}

\bibliography{Article291204}

\begin{thebibliography}{100}
\expandafter\ifx\csname url\endcsname\relax
  \def\url#1{\texttt{#1}}\fi
\expandafter\ifx\csname urlprefix\endcsname\relax\def\urlprefix{URL }\fi

\bibitem{Chance}
R.~Chance, A.~Prock, R.~Silbey, Adv.Chem.Phys 37 (1978) 1.

\bibitem{Bloch}
H.~Failache, S.~Saltiel, M.~Fichet, D.~Bloch, M.~Ducloy, Appl.Phys. Lett 60
  (1992) 2484.

\bibitem{Keilmannature}
R.~Hillenbrand, T.~Taubner, F.~Keilmann, Nature 418 (2002) 159.

\bibitem{Ebbesen}
T.~Ebbesen, H.~Lezec, H.~Ghaemi, T.~Thio, P.~Wolff, Nature 391 (1998) 667.

\bibitem{PendryOC}
A.~Krishnan, T.~Thio, T.~Kim, H.~Lezec, T.~Ebbesen, P.~Wolff, J.~Pendry,
  L.~Martin-Moreno, F.~Garcia-Vidal, Opt. Commun. 200 (2001) 1.

\bibitem{PendryPRL}
J.~Pendry, Phys. Rev. Lett. 85 (2000) 3966.

\bibitem{Shchegrov}
A.~Shchegrov, K.~Joulain, R.~Carminati, J.-J. Greffet, Phys. Rev. Lett. 85
  (2000) 1548.

\bibitem{Carminati}
R.~Carminati, J.-J. Greffet, Phys. Rev. Lett. 82 (1999) 1660.

\bibitem{Henkel_euro}
C.~Henkel, M.~Wilkens, Europhys. Lett. 47 (1999) 414.

\bibitem{Greffet_nat}
J.-J. Greffet, R.~Carminati, K.~Joulain, J.-P. Mulet, S.~Mainguy, Y.~Chen,
  Nature 416 (2002) 61.

\bibitem{Kittel}
C.~Kittel, Introduction to Solid State Physics, 7th Edition, John Wiley and
  sons, New-York, 1996.

\bibitem{Ashcroft}
N.~Ashcroft, D.~Mermin, Solid State Physics, international Edition, Saunders,
  Philiadelphia, 1976.

\bibitem{Ziman}
J.~Ziman, Electrons and Phonons, Oxford University Press, Oxford, 1960.

\bibitem{Raether}
H.~Raether, Surface Plasmons, Springer-Verlag, Berlin, 1988.

\bibitem{Agranovitch}
V.~Agranovitch, D.~Mills, Surface Polaritons, North-Holland, Amsterdam, 1982.

\bibitem{Economou1}
E.~Economou, K.~Ngai, Adv.Chem.Phys. 27 (1974) 265.

\bibitem{Boardman}
E.~A. Boardman, Electromagnetic Surface Modes, John Wiley, New York, 1982.

\bibitem{Legall}
J.~L. Gall, M.~Olivier, J.-J. Greffet, Phys. Rev. B 55 (1997) 10105.

\bibitem{Halevi}
P.~Halevi, Electromagnetic Surface Modes, Wiley, New-York, 1982.

\bibitem{Arakawa}
E.~Arakawa, M.~Williams, R.~Ritchie, Phys. Rev. Lett. 31 (1973) 1127.

\bibitem{Kliever}
K.~Kliewer, R.~Fuchs, Adv. Chem. Phys. 27 (1974) 355.

\bibitem{Otto}
A.~Otto, Z. Phys. 241 (1971) 398.

\bibitem{Kretschmann}
E.~Kretschmann, Z. Phys. 216 (1971) 398.

\bibitem{Vinogradov}
E.A.Vinogradov, G.N.Zhizhin, V.I.Yudson, North-Holland, 1982.

\bibitem{Spitzer}
W.~Spitzer, D.~Kleinman, D.~Walsh, Phys. Rev. 113 (1959) 127.

\bibitem{Rytov1}
S.~Rytov, Sov. Phys. JETP 6 (1958) 130.

\bibitem{Rytov2}
S.~Rytov, Y.~Kravtsov, V.~Tatarskii, Principles of Statistical Radiophysics,
  Vol.~3, Springer-Verlag, Berlin, 1989.

\bibitem{Langevin08}
P.~Langevin, C. R. Acad. Sci. (Paris) 146 (1908) 530.

\bibitem{Mandel}
L.~Mandel, E.~Wolf, Optical Coherence and Quantum Optics, Cambridge University
  Press, Cambridge, 1995.

\bibitem{Callen}
H.B.Callen, T.A.Welton, Phys. Rev. 83 (1951) 34.

\bibitem{Agarwal}
G.~Agarwal, Phys. Rev. A 11 (1975) 230.

\bibitem{Tai}
C.-T. Tai, Dyadic Green's functions in electromagnetic theory, 2nd Edition,
  Oxford University Press, Oxford, 1996.

\bibitem{Agarwal3}
G.~Agarwal, Phys. Rev. A 11 (1975) 253.

\bibitem{Jackson}
J.~Jackson, Classical Electrodynamis, 2nd Edition, John Wiley and sons,
  New-York, 1975.

\bibitem{Henkel00}
C.~Henkel, K.~Joulain, R.~Carminati, J.-J. Greffet, Opt. Commun. 186 (2000) 57.

\bibitem{Tersoff}
J.~Tersoff, D.~Hamann, Phys.Rev.B 31 (1985) 805.

\bibitem{Economou2}
E.~Economou, Green's Functions in Quantum Physics, Springer, Berlin, 1983.

\bibitem{vankampen}
N.~G.~V. Kampen, B.~Nijboer, K.~Schram, Phys. Lett. A 26 (1968) 307.

\bibitem{Gerlach}
E.~Gerlach, Phys. Rev. B 4 (1971) 393.

\bibitem{Pendry97}
J.~Pendry, J. Phys.: Condens. Matter 9 (1997) 10301.

\bibitem{Pendryldos}
F.Wijnands, J.~Pendry, F.J.Garcia-Vidal, P.J.Roberts, L.~Martin-Moreno,
  Opt.Quantum Electron. 29 (1997) 199.

\bibitem{Chicanne}
C.~Chicanne, T.~David, R.~Quidant, J.~Weeber, Y.~Lacroute, E.~Bourillot,
  A.Dereux, G.~{Colas des Francs}, C.~Girard, Phys. Rev. Lett. 88 (2002)
  097402.

\bibitem{Colas}
G.~{Colas des Francs}, C.~Girard, J.-C. Weeber, C.~Chicanne, T.~David,
  A.~Dereux, D.Peyrade, Phys. Rev. Lett. 86 (2001) 4950.

\bibitem{Joulain}
K.~Joulain, R.~Carminati, J.-P. Mulet, J.-J. Greffet, Phys. Rev. B 68 (2003)
  245405.

\bibitem{Sipe}
J.~Sipe, J. Opt. Soc. Am. B 4 (1987) 481.

\bibitem{Palik}
E.~Palik, Handbook of Optical constants of Solids, Academic Press, San Diego,
  1991.

\bibitem{Harrisson}
W.~Harrisson, Solid State Theory, Dover, New-York, 1980.

\bibitem{Goodman}
J.~Goodman, Statistical Optics, Wiley, New-York, 1985.

\bibitem{Friberg}
T.~Setälä, M.~Kaivola, A.~Friberg, Phys.Rev.Lett. 88 (2002) 123902.

\bibitem{Nietolivre}
M.~Nieto-Vesperinas, Scattering and Diffraction in Physical Optics, Wiley, New
  York, 1991.

\bibitem{Heskethnature}
P.~Hesketh, J.N.Zemel, B.~Gebhart, Nature 324 (1986) 549.

\bibitem{HeskethPRB1}
P.~Hesketh, J.N.Zemel, B.~Gebhart, Phys.Rev.B 37 (1988) 10795.

\bibitem{HeskethPRB2}
P.~Hesketh, J.N.Zemel, B.~Gebhart, Phys.Rev.B 37 (1988) 10803.

\bibitem{Kreiter}
M.~Kreiter, J.~Auster, R.~Sambles, S.~Herminghaus, S.~Mittler-Neher, W.~Knoll,
  Opt.Commun. 168 (1999) 117.

\bibitem{Heinzel}
A.Heinzel, V.~Boerner, A.~Gombert, B.~Bläsi, V.~Wittwer, J.~Luther, J. Modern
  Optics 47 (2000) 2399.

\bibitem{FrancoisPRB}
F.Marquier, K.~Joulain, J.~Mulet, R.~Carminati, Y.~Chen, Phys.Rev.B 69 (2004)
  155412.

\bibitem{Kollyukh}
O.G.Kollyukh, A.~Liptuga, V.~Morozhenko, V.~Pipa, Opt.Commun. 225 (2003) 349.

\bibitem{Ben-abd}
P.~Ben-Abdallah, J.Opt.Soc.Am.A 21 (2004) 1368.

\bibitem{Zhang}
Z.M.Zhang, C.J.Fu, Q.Z.Zhu, Advances in Heat Transfer 37 (2003) 179.

\bibitem{Ghmari}
F.~Ghmari, T.~Ghbara, M.~Laroche, R.~Carminati, J.~Greffet, J. Appl.Phys. 96
  (2004) 2656.

\bibitem{Beckmann}
P.Beckmann, A.~Spizzichino, The Scattering of Electromagnetic Waves, Pergamon
  Press, Oxford, 1963.

\bibitem{Desanto}
J.A.Desanto, G.S.Brown, Progress in Optics XIII, North-Holland, Amsterdam,
  1986.

\bibitem{Ogilvy}
J.Ogilvy, Theory of wave scattering from random rough surfaces, Adam Hilger,
  Bristol, 1991.

\bibitem{Tsang}
L.Tsang, J.~Kong, Scattering of Electromagnetic Waves. Advanced Topics, Wiley,
  New York, 2001.

\bibitem{Voronovich}
A.G.Voronovich, Wave Scattering from Rough Surfaces, Springer-Verlag, Berlin,
  1994.

\bibitem{Baylard}
C.Baylard, J.~Greffet, A.A.Maradudin, J.Opt.Soc.Am.A 10 (1993) 2637.

\bibitem{Hava1}
M.Auslander, S.~Hava, Infrared Phys.Technol. 36 (1999) 414.

\bibitem{Hava2}
M.~Auslander, D.~Levy, S.~Hava, Appl.Opt. 37 (1998) 369.

\bibitem{Hava3}
S.~Hava, J.~Ivri, M.~Auslander, J. Appl. Phys. 85 (1999) 7893.

\bibitem{Hava4}
M.~Auslander, D.~Levy, S.~Hava, Appl.Opt. 37 (1998) 369.

\bibitem{Sai}
H.~Sai, H.~Yugami, Y.~Akiyama, Y.~Kanamori, K.~Hane, J.Opt. Soc.Am. A 18 (2001)
  1471.

\bibitem{MarquierOC}
F.Marquier, K.~Joulain, J.~Mulet, R.~Carminati, J.-J. Greffet, Opt.Commun. 237
  (2004) 379.

\bibitem{Kusunoki}
F.~Kusunoki, J.~Takahara, T.~Kobayashi, Electron. Lett. 39 (2003) 23.

\bibitem{Pralle}
M.U.Pralle, N.~Moelders, M.~McNeal, I.~Puscasu, A.C.Greenwald, J.T.Daly,
  E.A.Johnson, T.George, D.S.Choi, I.~ElKady, R.~Biswas, Appl.Phys.Lett. 81
  (2002) 4685.

\bibitem{BarnesNatMat}
W.L.Barnes, Nature Materials 3 (2004) 588.

\bibitem{Okamoto}
K.Okamoto, I.~Niki, A.~Shvartser, Y.~Narukawa, T.~Mukai, A.~Scherer, Nature
  Mater. 3 (2004) 601.

\bibitem{Cravalho}
E.~Cravalho, C.~Tien, R.~Caren, J. Heat Trans. 89 (1967) 351.

\bibitem{Boehm}
R.~Boehm, C.~Tien, J. Heat Trans. 92 (1970) 405.

\bibitem{Polder}
D.~Polder, D.~V. Hove, Phys. Rev. B 4 (1971) 3303.

\bibitem{Levin}
M.L.Levin, V.~Polevai, Sov.Phys.JETP 52 (1980) 1054.

\bibitem{loom}
J.~Loomis, H.~Maris, Phys. Rev. B 50 (1994) 18517.

\bibitem{Pendry99}
J.~Pendry, J. Phys.: Condens. Matter 11 (1999) 6621.

\bibitem{Volokitin01}
A.~Volokitin, B.~Persson, Phys.Rev.B 63 (2001) 205404.

\bibitem{Chen1}
G.Chen, Microscale Thermophysical Engineering 1 (1997) 215.

\bibitem{Mulet02}
J.~Mulet, K.~Joulain, R.~Carminati, J.-J. Greffet, Micr. Thermophys. Eng. 6
  (2002) 209.

\bibitem{Pan}
J.L.Pan, Opt.Lett. 25 (2001) 369.

\bibitem{Pan2}
J.L.Pan, Opt.Lett. 26 (2001) 482.

\bibitem{Maradudin}
A.A.Maradudin, Opt.Lett. 26 (2001) 479.

\bibitem{Mulet01b}
J.~Mulet, K.~Joulain, R.~Carminati, J.-J. Greffet, Opt. Lett. 26 (2001) 480.

\bibitem{Heargraves1}
C.M.Heargraves, Phys.Rev.Lett. 30A (1969) 491.

\bibitem{Dransfeld}
K.Dransfeld, J.~Xu, J. of Microscopy 152 (1988) 35.

\bibitem{Xu}
J.B.Xu, K.Lauger, R.Moller, K.~Dransfeld, I.H.Wilson, J.Appl.Phys. 76 (1994)
  7209.

\bibitem{Dorofeyev98}
I.Dorofeyev, J.Phys.D:Appl.Phys. 31 (1998) 600.

\bibitem{Mulet01}
J.~Mulet, K.~Joulain, R.~Carminati, J.-J. Greffet, Appl. Phys. Lett. 78 (2001)
  2931.

\bibitem{ChenAPL}
A.Narayanaswamy, G.Chen, Appl.Phys.Lett. 82 (2003) 3544.

\bibitem{Forster}
TH.Forster, Ann. Phys.(Leipzig) 6 (1948) 55.

\bibitem{Bohren}
C.F.Bohren, D.~Huffman, Absorption and Scattering of light by small particles,
  Wiley, New York, 1983.

\bibitem{CarminatiPRA}
R.~Carminati, J.~Saenz, J.~Greffet, M.~Nieto-Vesperinas, Phys.Rev.A. 62 (2000)
  012712.

\bibitem{Casimir}
H.~Casimir, Proc. Konikl. Ned. Akad. Wetenschap. 51 (1948) 793.

\bibitem{Caspol}
H.~Casimir, D.~Polder, Phys. Rev. 73 (1948) 360.

\bibitem{Mostepanenko}
V.~Mostepanenko, N.N.Trunov, The Casimir Effet and its Applications, Oxford
  Science Publications, Oxford, 1997.

\bibitem{Milonnibook}
P.W.Milonni, The Quantum Vacuum. An introduction to quantum Electrodynamics,
  Academic Press, San Diego, 1994.

\bibitem{Plunien}
G.Plunien, B.~Muller, W.~Greiner, Phys.Rep. 134 (1986) 87.

\bibitem{Bordag}
M.Bordag, U.Mohideen, V.M.Mostepanenko, Phys.Rep. 53 (2001) 1.

\bibitem{Lamoreaux}
S.~Lamoreaux, Am. J. Phys. 67 (1999) 850.

\bibitem{Milton}
K.~Milton, [ArXiv:hep-th/0406024].

\bibitem{Lifshitz56}
E.~Lifshitz, Sov. Phys. JETP 2 (1956) 73.

\bibitem{Agarwal2}
G.~Agarwal, Phys. Rev. A 11 (1975) 243.

\bibitem{Lamoreaux97}
S.~Lamoreaux, Phys. Rev. Lett. 78 (1997) 5.

\bibitem{Mohideen}
U.Mohideen, A.~Roy, Phys.Rev. Lett. 81 (1998) 4549.

\bibitem{Harris}
B.W.Harris, F.~Chen, U.~Mohideen, Phys.Rev.A 62 (2000) 052109.

\bibitem{Ederth}
T.~Ederth, Phys.Rev.A 62 (2000) 062104.

\bibitem{Bressi}
G.~Bressi, G.~Carugno, R.~Onofrio, G.~Ruos, Phys.Rev.Lett. 88 (2002) 041804.

\bibitem{Decca}
R.~Decca, D.~Lopez, E.~Fischbach, D.~Krause, Phys.Rev.Lett. 91 (2003) 050402.

\bibitem{Chan01a}
H.~Chan, V.~Aksyuk, R.~Kleinman, D.~Bishop, F.~Capasso, Phys. Rev. Lett. 87
  (2001) 211801.

\bibitem{Chan01b}
H.~Chan, V.~Aksyuk, R.~Kleinman, D.~Bishop, F.~Capasso, Science 291 (2001)
  1941.

\bibitem{Lambrecht00}
A.~Lambrecht, S.~Reynaud, Eur. Phys. J. D. 8 (2000) 309.

\bibitem{Genet00}
C.~Genet, A.~Lambrecht, S.~Reynaud, Phys. Rev. A 62 (2000) 012110.

\bibitem{Klim01}
G.~Klimchitskaya, V.~Mostepanenko, Phys. Rev. A 63 (2001) 062108.

\bibitem{Henkel04}
C.~Henkel, K.~Joulain, J.-P. Mulet, J.-J. Greffet, Phys. Rev. A 69 (2004)
  023808.

\bibitem{Schwinger78}
J.~Schwinger, L.~DeRaad, K.~Milton, Annals of Physics 115 (1978) 1--23.

\bibitem{Milonni88}
P.~Milonni, R.~Cook, M.~Goggin, Phys. Rev. A 38 (1988) 1621.

\bibitem{Hushwater}
V.~Hushwater, Am. J. Phys. 65 (1997) 381.

\bibitem{Genet03b}
C.~Genet, A.~Lambrecht, S.~Reynaud, Phys. Rev. A 67 (2003) 043811.

\bibitem{Kenneth}
O.~Kenneth, I.~Klich, A.~Mann, M.~Revzen, Phys. Rev. Lett. 89 (2002) 033001.

\bibitem{Buks02}
E.Buks, M.~Roukes, Nature 419 (2002) 119.

\bibitem{Marcuse}
D.~Marcuse, Theory of Dielectric Optical Waveguides, 2nd Edition, Academic
  Press, San Diego, 1991.

\bibitem{Sarid}
D.Sarid, Phys.Rev.Lett. 47 (1981) 1927.

\bibitem{Einstein1}
A.Einstein, L.Hopf, Ann. Phys.(Leipzig) 33 (1910) 1105.

\bibitem{Einstein2}
A.Einstein, O.Stern, Ann. Phys.(Leipzig) 40 (1913) 551.

\bibitem{Milonni}
P.W.Milonni, M.~Shih, Am.J.Phys. 59 (1991) 684.

\bibitem{Saslow}
V.~Mkrtchian, V.~Parsegian, R.~Podgornik, W.~M. Saslow, Phys.Rev.Lett. 91
  (2004) 220801.

\bibitem{Betzig}
E.Betzig, P.~Finn, J.~Weiner, Phys. Rev. B 62 (2000) 13174.

\bibitem{Vaez-Iravani}
R.~Toledo-Crow, P.~Yang, Y.~Chen, M.~Vaez-Iravani, Appl.Phys.Lett. 60 (1992)
  2957.

\bibitem{Karrai1}
K.~Karrai, R.~Grober, Appl.Phys.Lett. 66 (1995) 1842.

\bibitem{Karrai}
K.~Karrai, I.~Tiemann, Phys. Rev. B 62 (2000) 13174.

\bibitem{Stipe}
B.~Stipe, H.~Mamin, T.~Stowe, T.~Kenny, D.~Rugar, Phys. Rev. Lett. 87 (2001)
  096801.

\bibitem{Teodorovitch}
E.V.Teodorovitch, Proc.Roy.Soc. London A 362 (1978) 71.

\bibitem{Levitov}
L.S.Levitov, Europhys.Lett. 8 (1989) 499.

\bibitem{Barton}
G.Barton, Ann.Phys. NY 245 (1996) 361.

\bibitem{Volokitin}
A.~Volokitin, B.~Persson, J. Phys. : Condens Matter 11 (1999) 345.

\bibitem{Schaich}
W.L.Schaich, J.~Harris, J.Phys.F : Met.Phys. 11 (1981) 65.

\bibitem{Tomassone}
M.S.Tomassone, A.~Widom, Phys.Rev.B 56 (1997) 4938.

\bibitem{Kyasov}
A.~Kyasov, G.~Dedkov, Surface Science 463 (2000) 11.

\bibitem{Dorofeyev99}
I.Dorofeyev, H.Fuchs, G.~Wenning, B.~Gotsmann, Phys.Rev.Lett. 83 (1999) 848.

\bibitem{Dorofeyev01}
I.Dorofeyev, H.~Fuchs, B.~Gotsmann, J.~Jersch, Phys.Rev.B 64 (2001) 035403.

\bibitem{Volokitin03PRL}
A.~Volokitin, B.~Persson, Phys.Rev.Lett. 91 (2003) 106101.

\bibitem{Volokitin03}
A.~Volokitin, B.~Persson, Phys.Rev.B 68 (2003) 155420.

\bibitem{Zurita}
J.~Zurita-Sanchez, J.-J. Greffet, L.~Novotny, Phys. Rev. A 69 (2004) 022902.

\end{thebibliography}









\end{document}